\pdfoutput=1

\documentclass[11pt,twoside,a4paper,cmspaper,final,collab]{cms-tdr}
\def\svnVersion{373476}\def\cmsCernNoTag{CERN-PH-EP/2015-074}\def\cmsCernDate{\today}\def\cmsMessage{Published in the Journal of High Energy Physics as \href{http://dx.doi.org/10.1007/JHEP10(2015)144}{\doi{10.1007/JHEP10(2015)144.}}}

\begin{document}\cmsNoteHeader{HIG-13-031}

\hyphenation{had-ron-i-za-tion}
\hyphenation{cal-or-i-me-ter}
\hyphenation{de-vices}
\RCS$Revision: 294036 $
\RCS$HeadURL: svn+ssh://svn.cern.ch/reps/tdr2/papers/HIG-13-031/trunk/HIG-13-031.tex $
\RCS$Id: HIG-13-031.tex 294036 2015-06-24 21:42:15Z ntran $

\newcommand{\Hww}{\PH\to\WW}
\newcommand{\METpr}{\ensuremath{E_{\text{T, pr}}^{\text{miss}}}}
\newcommand{\Mm}{\Pgmm}
\newcommand{\WW}{\ensuremath{\PW\PW}}
\newcommand{\WZ}{\ensuremath{\PW\Z}}
\newcommand{\Wp}{\ensuremath{\PWp}}
\newcommand{\ZZ}{\ensuremath{\cPZ\cPZ}}
\newcommand{\dyll}{\ensuremath{\cPZ/\gamma^*\to \ell^+\ell^-}}
\newcommand{\dytt}{\ensuremath{\cPZ/\gamma^* \to\tau^+\tau^-}}
\newcommand{\mH}{\ensuremath{m_{\PH}}}
\newcommand{\mZZ}{\ensuremath{m_{\cPZ\cPZ}}}
\newcommand{\mll}{\ensuremath{m_{\ell \ell}}}
\newcommand{\mumu}{\ensuremath{\Pgmp\Pgmm}}
\newcommand{\ptlmax}{\ensuremath{p_{\mathrm{T}}^{\Lep,\text{max}}}}
\newcommand{\ptlmin}{\ensuremath{p_{\mathrm{T}}^{\Lep,\text{min}}}}
\newcommand{\taue}{\ensuremath{\Pgt_\Pe}}
\newcommand{\tauh}{\ensuremath{\Pgt_\mathrm{h}}}
\newcommand{\taumu}{\ensuremath{\Pgt_\Pgm}}
\newcommand{\wgamma}{\ensuremath{\PW\gamma}}
\providecommand{\mjj}{\ensuremath{m_\mathrm{jj}}\xspace}
\providecommand{\mt}{\ensuremath{m_\mathrm{T}}\xspace}
\providecommand{\MT}{\ensuremath{M_{\mathrm{T}}\xspace}}
\providecommand{\vecEtm}{\ensuremath{\vec{E}_\mathrm{T}^{\text{miss}}}\xspace}
\providecommand{\vecPtell}{\ensuremath{\vec{p}_\mathrm{T}^{\ell\ell}}\xspace}
\providecommand{\vecPtel}{\ensuremath{\vec{p}_\mathrm{T}^{\ell}}\xspace}

\cmsNoteHeader{HIG-13-031}
\title{Search for a Higgs boson in the mass range from 145 to 1000\GeV decaying to a pair of W or Z bosons}

\date{\today}

\abstract{
A search for a heavy Higgs boson in the $\PH \to \PW\PW$ and $\PH \to \Z\Z$ decay channels is reported.
The search is based upon proton-proton collision data samples corresponding to an integrated luminosity
of up to 5.1\fbinv at $\sqrt{s} = 7$\TeV and up to 19.7\fbinv at $\sqrt{s} = 8$\TeV, recorded by the CMS experiment at the CERN LHC.
Several final states of the $\PH \to \PW\PW$ and $\PH\to \Z\Z$  decays are analyzed.
The combined upper limit at the 95\% confidence level on the product of the cross section and branching
fraction exclude a Higgs boson with standard model-like couplings and decays in the range $145 < m_{\PH} < 1000$\GeV.
We also interpret the results in the context of an electroweak singlet extension of the standard model.
}

\hypersetup{%
pdfauthor={CMS Collaboration},%
pdftitle={Search for a Higgs boson in the mass range from 145 to 1000 GeV decaying to a pair of W or Z bosons},%
pdfsubject={CMS},%
pdfkeywords={CMS, Higgs, heavy Higgs, BSM}}

\maketitle

\section{Introduction}
\label{sec:introduction}

The standard model (SM) of electroweak (EW) interactions~\cite{StandardModel67_1,StandardModel67_2,StandardModel67_3}
posits the existence of the Higgs boson, a scalar particle associated with the field responsible for
spontaneous EW symmetry breaking~\cite{Englert:1964et,Higgs:1964ia,Higgs:1964pj,Guralnik:1964eu,Higgs:1966ev,Kibble:1967sv}.
The mass of the boson is not predicted by the theory.
In 2012 the ATLAS and CMS collaborations at the CERN LHC reported the observation of a new boson with a mass of about
125\GeV~\cite{ATLASobservation125,CMSobservation125,CMSlongpaper,ATLASlongpaper,LHCmasscombination:2015}.
Throughout this paper, we denote the observed Higgs boson as \Ph{}(125).
Subsequent studies of the production and decay
rates~\cite{CMSttHbb7TeV,ATLASDiboson2013,Chatrchyan:2013zna,CMSHWWlvlv2014,Chatrchyan:2013mxa,Chatrchyan:2014nva,CMSHggLegacyRun1-arxiv,CMSHzg2013,ATLASHzg2014,ATLASInvisible2014,CMSInvisible2014}
and of the spin-parity quantum numbers~\cite{CMSHWWlvlv2014,Chatrchyan:2013mxa,CMSMassParity2012,Aad:2013xqa}
of the new boson show that its properties are compatible with those expected for the SM Higgs boson.

The observation of a Higgs boson with a mass of 125\GeV is also
consistent with the unitarity constraints on diboson scattering at high
energies~\cite{Dicus:1992vj,Veltman:1976rt,Lee:1977eg,Lee:1977yc,Passarino:1990hk,Chanowitz:1985hj,Duncan:1985vj,Dicus:1986jg,Bagger:1995mk,Ballestrero:2009vw}.
Nevertheless, there is a possibility that the newly discovered particle
is part of a larger Higgs boson sector and thus only partially responsible for EW symmetry breaking.
This can be realized in several scenarios, such as two-Higgs-doublet models~\cite{Branco:2011iw,craig},
or models in which the SM Higgs boson mixes with a heavy EW
singlet~\cite{Chacko:2005pe,Patt:2006fw,Barger:2007,Bowen:2007ia,Bhattacharyya:2007,Profumo:2007wc,Dawson:2009,Bock:2010,Baek:2011aa,Fox:2011,Englert:2011,Englert:2012,Batell:2012,Englert:2012a,Gupta:2012,Dolan:2012,Bertolini:2012,Batell:2012a,Chpoi:2013wga,Hambye:2013dgv,Craig:2014lda,Curtin:2014jma,Lopez-Val:2014jva,Robens:2015gla},
which predict the existence
of additional resonances at high mass, with couplings similar to those of the SM Higgs boson.

Previous searches at the LHC for heavy SM-like Higgs bosons have been reported by ATLAS and CMS.
Based on a dataset of 4.7\fbinv at $\sqrt{s} = 7$\TeV and up to 5.9\fbinv at $\sqrt{s} = 8$\TeV, and combining all channels listed in Ref.~\cite{ATLASobservation125},
ATLAS excludes a SM-like heavy Higgs boson of $131 < \mH < 559$\GeV at 95\% CL~\cite{ATLASobservation125}.
The CMS collaboration reported a search in the $\WW$ and $\ZZ$ decay channels using an initial dataset of 5.1\fbinv at
$\sqrt{s} = 7$\TeV and 5.3\fbinv at $\sqrt{s} = 8$\TeV, searching in the mass range $145 < \mH < 1000$\GeV, and excluding Higgs boson masses
up to 710\GeV at 95\% CL~\cite{CMShighmass}. In this paper, we report on an extension of this search using the complete 7 and 8\TeV dataset.
In addition, the search is interpreted in the context of the SM expanded by an additional EW singlet.
Both the SM-like heavy Higgs boson as well as the EW singlet are denoted as H here.

The analysis uses proton-proton collision data recorded with the CMS detector, corresponding to integrated luminosities of up
to 5.1\fbinv at $\sqrt{s}=7$\TeV
and up to 19.7\fbinv at $\sqrt{s}=8$\TeV.
The analysis is performed in a mass range $145 < \mH < 1000$\GeV exploiting both the $\PH \to \WW$ and $\PH \to \ZZ$
decay channels, with the lower boundary being chosen to reduce contamination from \Ph{}(125).
In the case of a Higgs boson decaying into a pair of $\PW$ bosons, the
fully leptonic ($\PH \to \WW \to \ell\nu\ell\nu$) and semileptonic ($\PH \to \WW \to \ell \nu
\cPq\cPq$) final states are considered in the analysis. For a Higgs boson decaying into two $\cPZ$
bosons, final states containing four charged leptons ($\PH \to \ZZ \to 2\ell 2\ell'$), two charged leptons and two
quarks  ($\PH \to \ZZ \to 2\ell 2\cPq$), and two charged leptons and two neutrinos
($\PH \to \ZZ \to 2\ell 2\nu$) are considered, where $\ell = \Pe$ or $\Pgm$ and $\ell' = \Pe$, $\Pgm$,
or $\Pgt$.

\section{CMS detector}
\label{sec:cms}

The central feature of the CMS apparatus is a superconducting solenoid of 6\unit{m} internal diameter,
providing a magnetic field of 3.8\unit{T}. Within the solenoid volume are a silicon pixel and strip tracker,
a lead tungstate crystal electromagnetic calorimeter (ECAL), and a brass and scintillator hadron calorimeter, each composed of a barrel and
two endcap sections.
Muons are measured in
gas-ionization detectors embedded in the steel flux-return yoke outside the solenoid. Extensive forward calorimetry complements the coverage
provided by the barrel and endcap detectors. The first level of the CMS trigger system,
composed of custom hardware processors, uses information from the calorimeters and the muon detectors to select
the most interesting events in a fixed time interval of less than $4\mus$. The high level trigger processor farm
further decreases the event rate from around 100\unit{kHz} to less than 1\unit{kHz}, before data storage.
A more detailed description of the detector as well as the definition of the coordinate system and relevant kinematic variables can be
found in Ref.~\cite{Chatrchyan:2008zzk}.

\section{Signal model and simulations}
\label{sec:sim}
Several Monte Carlo event generators are used to simulate the signal and background event samples.
The Higgs boson signal samples from gluon fusion (ggF, $\Pg\Pg \to \PH$),
and vector boson fusion (VBF, $\Pq\Pq \to \Pq\Pq \PH$), are generated with {\POWHEG} 1.0~\cite{Bagnaschi:2012,Nason:2004rx,Frixione:2007vw,Alioli:2010xd,Nason:2010} at next-to-leading order (NLO)
and a dedicated program~\cite{Gao:2010qx} used for angular correlations. Samples of $\PW\PH$, $\cPZ\PH$, and $\ttbar\PH$ events are
generated using the leading-order (LO) \PYTHIA 6.4~\cite{Sjostrand:2006za} program.
At the generator level, events are weighted according to the total cross section $\sigma(\Pp\Pp\to \PH)$~\cite{Heinemeyer:2013tqa},
which contains contributions from ggF computed to next-to-next-to-leading order (NNLO)
and next-to-next-to-leading-log  (NNLL), and from weak-boson fusion computed at NNLO.
The \WW (\ZZ) invariant mass, $m_{\WW}$ ($m_{\ZZ}$), lineshape is affected by interference between signal and SM background
processes. The simulated $\mH$ lineshape is therefore corrected to match
theoretical predictions~\cite{Passarino:2010qk,Goria:2011wa,Kauer:2012hd} using the complex-pole scheme
for the Higgs boson propagator.
The procedure for including lineshape corrections and uncertainties from interference of the signal with background processes for both
ggF and VBF production are described below in the discussion of the lineshape corrections applied for the EW singlet interpretation.

The background contribution from $\cPq\cPaq \to \WW$ production is generated using \MADGRAPH 5.1~\cite{Alwall:2007st},
and the subdominant $\Pg\Pg \to \WW$ process is generated at LO with \textsc{gg2ww} 3.1~\cite{ggww}.
The $\cPq\cPaq \to \ZZ$ production process is simulated at NLO
with \POWHEG, and the $\mathrm{gg} \to \ZZ$ process is simulated at LO using
\textsc{gg2zz} 3.1~\cite{Binoth:2008pr}. Other diboson processes ($\PW\cPZ$, $\cPZ\gamma^{(*)}$, $\PW\gamma^{(*)}$), $\cPZ+$jets, and
$\PW+$jets are generated
with \PYTHIA and \MADGRAPH.
The $\ttbar$ and $\cPqt\PW$ events are generated at NLO with \POWHEG.
For all samples, \PYTHIA is used for parton showering, hadronization, and underlying event simulation.
For LO generators, the default set of parton distribution functions (PDF) used to produce these samples is
CTEQ6L~\cite{CTEQ66}, while CT10~\cite{Lai:2010vv} is used for NLO generators. The tau lepton decays are simulated
with \TAUOLA~\cite{Jadach:1993hs}. The detector response is simulated using a detailed description of the CMS detector, based on
the \GEANTfour package~\cite{GEANT}, with event reconstruction performed identically to that of recorded data.
The simulated samples include the effect of multiple $\Pp\Pp$ interactions per bunch crossing (pileup).
The \PYTHIA parameters for the underlying events and pileup interactions
are set to the Z2 ($\cPZ 2^*$) tune~\cite{Chatrchyan:2011id} for the 7\,(8)\TeV simulated data sample,
with the pileup multiplicity distribution matching the one observed in data.

The data are analysed to search for both a beyond the standard model (BSM) case in the form of an EW singlet scalar
mixed with the recently discovered Higgs boson, \Ph{}(125),
as well as a heavy Higgs boson with SM-like couplings.
The couplings of the two gauge eigenstates (\Ph{}(125) and EW singlet)
are phenomenologically constrained by unitarity and the coupling strength of the
light Higgs boson is therefore reduced with respect to the SM case.
The unitarity constraint is ensured by enforcing $C^2+C^{'2}=1$, where $C$ and $C'$ are defined as the scale
factors of the couplings with respect to the SM of the low- and high-mass Higgs boson, respectively.
The EW singlet production cross section is also modified by a factor $\mu'$
and the modified width is $\Gamma'$; they are defined as

\begin{equation}
\mu'=C'^2\, (1-\mathcal{B}_\text{new}),
\label{eq:bsmstrength}
\end{equation}

\begin{equation}
\Gamma'=\Gamma_\mathrm{SM} \, \frac{C'^2}{1-\mathcal{B}_\text{new}},
\end{equation}

where $\mathcal{B}_\text{new}$ is the branching fraction of the EW singlet to non-SM decay modes.
An upper limit at 95\% CL can be set indirectly as $C^{'2}<0.28$ using the signal strength
fits to the \Ph{}(125) boson as obtained in Ref.~\cite{CMS-PAS-HIG-14-009}.

This paper focuses on the case where $C'^2\leq (1-\mathcal{B}_\text{new})$.
In this regime the new state is expected to have an equal or narrower width with respect to the SM case.
Results are presented distinguishing between the $\mathcal{B}_\text{new}=0$ and $\mathcal{B}_\text{new}>0$ cases.
Under this hypothesis, signal samples with different Higgs boson widths are generated, scanning the $C^{'2}$ and
$\mathcal{B}_\text{new}$ space. We follow the recommendations of the ``LHC Higgs Cross Section Working
Group''~\cite{Heinemeyer:2013tqa} described below.

The SM signal mass lineshape generated with \POWHEG is weighted in order to simulate the
narrow scalar singlet lineshape. For the ggF production mode,
the weights are calculated using either the \textsc{gg2zz} generator for the ZZ channel, or the \POWHEG and
\MCFM 6.2~\cite{MCFM} generators for the interference calculation for the WW channel.
For the VBF production mode, the interference weights are computed using the \textsc{Phantom} 1.2~\cite{Ballestrero:2007xq} generator,
where the signal-only lineshape at LO is weighted based on results obtained with \MADGRAPH generator predictions.
The weights are defined as the ratio
of the sum of a narrow resonance
signal plus interference and the standard model signal lineshape as generated.
The contribution from the interference term between the BSM Higgs boson and
the background is furthermore assumed to scale according to the modified coupling of the Higgs boson as
$(\mu+I)_\mathrm{BSM}=\mu_\mathrm{SM}C'^2+I_\mathrm{SM}C'$, where $\mu(I)$
 is the signal strength (interference) in the BSM or SM cases.
This assumption is based on the hypothesis that the couplings are
similar to the SM case and simply rescale due to unitarity constraints.
Systematic uncertainties considered for this procedure are detailed later.

If the new resonance has a
very small width, its production will tend to interfere less with the
background continuum.
Thus in the most interesting region of low-$C'^2$, the effect of the
interference and its exact modeling is of limited importance. Any possible interference between \Ph{}(125) and its EW singlet
partner~\cite{Englert:2014,Maina:2015ela} is assumed to be covered by a conservative systematic uncertainty.
In addition to the EW singlet, the analysis searches for a heavy Higgs
boson that gets produced and decays like the SM Higgs boson, but has a higher mass and interferes with \Ph{}(125).

\section{Event reconstruction}
\label{sec:reconstruction}
CMS uses a particle-flow (PF) reconstruction algorithm~\cite{CMS-PAS-PFT-09-001,Chatrchyan:2011tn} to provide an event description in the form of
particle candidates, which are then used to build higher-level objects, such as jets and missing transverse energy, as well as
lepton isolation quantities. Not all the channels considered here use the same selection criteria for their objects,
but the common reconstruction methods are listed below.

The high instantaneous luminosity delivered by the LHC
provides an average of about 9 (21) pileup interactions
per bunch crossing in 7 (8)\TeV data, leading to events with several possible primary vertices.
The vertex with largest value of the sum of the square of the transverse momenta ($\pt$) for the associated
tracks is chosen to be the reference vertex.
According to simulation, this requirement provides the correct primary
vertex in more than 99\% of both signal and background events.

Muon candidates are reconstructed by using one of two
algorithms: one in which tracks in the silicon tracker are matched to hits in
the muon detectors, and another in which a combined fit is performed to
signals in both the silicon tracker and the muon
system.
Other identification criteria based on the number of
measurements in the tracker and in the muon system, the fit quality of
the muon track, and its consistency with its origin from the primary vertex are
also imposed on the muon candidates to reduce the misidentification rate~\cite{CMS-PAS-MUO-10-002}.
In some of the channels, vetos are imposed on additional low-$\pt$ muons in the event,
whose tracks in the silicon tracker fulfill more stringent requirements, but whose primary vertex association can be more relaxed.
We call these muons \textit{soft} muons.

Electron candidates are reconstructed from superclusters, which are arrays of energy
clusters along the $\phi$ direction in the ECAL, matched to tracks in the silicon tracker. A complementary
algorithm reconstructs electron candidates by extrapolating measurements in the innermost tracker layers
outward to the ECAL. Trajectories from both algorithms are reconstructed using the Gaussian sum filter
algorithm~\cite{CMSTracking}, which accounts for the electron energy loss by bremsstrahlung. Additional requirements are applied
to reject electrons originating from photon conversions in the tracker material. Electron identification
relies on a multivariate discriminant that combines observables sensitive to the bremsstrahlung along the electron
trajectory, the geometrical and momentum-energy matching between the electron trajectory and the associated supercluster,
as well as ECAL shower shape observables. Electron candidates with a pseudorapidity $\eta$ of their ECAL supercluster 
in the transition region
between ECAL barrel and endcap ($1.4442 < \abs{\eta} < 1.5660$) are rejected, 
because the reconstruction of electrons in
this region is compromised~\cite{ElePerformance:2015}.

The PF candidates are used to reconstruct hadronic tau candidates, $\tauh$, with the ``hadron plus strip"
algorithm~\cite{Chatrchyan:2011xq} that is designed to optimize the performance of $\tauh$ identification and reconstruction
by considering specific $\tauh$ decay modes. The neutrinos produced in all $\tau$ decays escape detection and are ignored in the
$\tauh$ reconstruction. The algorithm provides high $\tauh$ identification efficiency, approximately 50\% for the range of $\tauh$
energies relevant for this analysis, while keeping the misidentification rate for jets at the level of $\sim1\%$.

Leptons produced in the decay
of $\PW$ and $\cPZ$ bosons are expected to be isolated from hadronic activity in
the event. Isolation is defined as the scalar sum of the transverse momenta of the PF candidates found (excluding the selected leptons in the
event themselves)
in a cone of radius $R=\sqrt{\smash[b]{\Delta\eta^2+\Delta\phi^2}} =  0.4$, with $\phi$ being measured in radians, built around
each lepton. We require that the isolation sum is smaller than 20\% (15\%) of the muon (electron)
transverse momentum. To account for the contribution to the isolation sum from pileup interactions
 and the underlying event, a median energy density is determined on an event-by-event basis using
the method described in Ref.~\cite{FASTJET}.
For electron candidates, an effective area that is proportional to the
isolation cone, is derived to renormalize this density estimate to the
number of pileup interactions, and is subtracted from the isolation sum.
For muon and tau candidates, the correction is done by subtracting
half the sum of the transverse momenta of the charged particles not associated to the primary vertex in the cone of
interest. Soft muons do not need to fulfill isolation requirements.

The combined efficiency of lepton reconstruction, identification, and isolation
is measured using observed $\cPZ$ decays and ranges between 90\% and 97\%
for muons, between 70\% and 90\% for electrons, and approximately 50\% for hadronic taus depending on the
$\pt$ and $\eta$ of the leptons.

Jets are reconstructed from PF candidates~\cite{CMS-PAS-JME-10-003} using the anti-\kt
clustering algorithm~\cite{Cacciari:2008gp} with a distance parameter of 0.5 (called AK5 jets), as implemented
in the \textsc{fastjet} package~\cite{fastjetmanual}, and with the Cambridge--Aachen
algorithm~\cite{Dokshitzer:1997gp} with a distance parameter of 0.8
(called CA8 jets); when not otherwise specified we use the term \textit{jets} to mean AK5 jets throughout this
paper. Any reconstructed jet overlapping with isolated leptons within a distance of
0.5 (0.8) for AK5 (CA8) jets is removed in order to avoid double counting the lepton as a jet.
AK5 (CA8) jets are required to have $\abs{\eta}< 4.7\,(2.4)$.
At hadron level, the jet momentum is defined as the vectorial sum of all particle
momenta in the jet,  and is found in the simulation to be within 5--10\% of the true momentum over the whole \pt range
and detector acceptance.
A correction is applied to the jet $\pt$ to take into account the extra energy clustered in jets
due to additional proton-proton interactions within the same bunch
crossing~\cite{FASTJET,Cacciari:2008gn}.
Other jet energy scale (JES) corrections applied are derived from the simulation,
and are calibrated with in situ measurements of the energy balance of
dijet and $\cPZ/\gamma$+jets events.
For some of the channels in this analysis, a combinatorial background arises from low-\pt\ jets from pileup interactions,
which get clustered into high-\pt\ jets.
At $\sqrt{s}$ = 8\TeV the number of pileup events is larger than at $\sqrt{s}= 7$\TeV,
and a multivariate selection is applied to separate jets stemming from the primary interaction and those
reconstructed due to energy deposits associated with pileup interactions~\cite{CMS-PAS-JME-13-005}. The discrimination
is based on the differences in the jet shapes, on the relative multiplicity of charged and
neutral components, and on the different fraction of transverse momentum, which is carried
by the hardest components. Within the tracker acceptance, the tracks belonging to each jet are
also required to be compatible with the primary vertex.

At high Higgs boson masses, the $\pt$ of the $\cPZ$ or $\PW$ boson is high enough that the two quarks from the vector boson
decay are expected to be reconstructed as a single CA8 jet, or \textit{merged} jet. To improve background rejection and jet mass
resolution, we apply a jet pruning algorithm~\cite{CMS-PAS-JME-13-006}. Additionally, the ``N-subjettiness ratio" variable
$\tau_2/\tau_1$, a measure of the compatibility of a jet having $N = 2$ subjets, is used to reduce the backgrounds~\cite{Thaler:2011,CMS-PAS-JME-13-006}.
We require $\tau_2/\tau_1 < 0.5$ to avoid contamination from jets originating from the hadronization of gluons and single quarks.

In some of the channels included in this analysis, the identification of jets originating from \cPqb\ quarks is important.
These b jets are tagged with dedicated algorithms~\cite{BTV},
which are applied either directly to the AK5 jets, or to the subjets
of the merged jets.
The b-tagging algorithm used is either the Combined Secondary Vertex algorithm, Jet Probability algorithm, or Track Counting High Efficiency algorithm depending on the channel.
Tagged b-jet candidates are required to have $\pt > 30$\GeV and to be
within the tracker acceptance ($\abs{\eta} < 2.4$).
For jets in this kinematic range with a b-tagging efficiency of 70\%, the misidentification probability from light quark and gluon jets is approximately 1\%.

The missing transverse momentum vector, \vecEtm, is defined as the projection on the plane perpendicular to the beams
of the negative vector sum of the momenta of all reconstructed particles in an event~\cite{CMS-PAS-JME-10-005,Chatrchyan:2011tn}.
Its magnitude is referred to as \MET.

\section{Data analysis}
\label{sec:analyses}

The results presented in this paper are obtained by combining searches exploiting different Higgs boson production and decay modes
as detailed in Table~\ref{tab:channels}.
All final states are exclusive, without overlap between channels. For the $\PW\PW \to \ell\nu\ell\nu$ and
$\cPZ\cPZ \to 2\ell 2\ell'$ channels a mass range starting from 110\GeV has been analyzed in other contexts
(e.g.~\cite{CMSlongpaper}), but in this paper
both analyses are restricted to searches above 145\GeV.
In the rest of this section, the individual analysis strategies and details are defined including
a discussion of the leading systematic uncertainties.
In the next section, statistical interpretations of the individual and combined searches are given for the Higgs boson with SM-like couplings hypothesis
as well as the EW singlet extension of the SM.

\begin{table}[ht]
\centering
\begin{minipage}{\textwidth}
\topcaption{A summary of the analyses included in this paper. The column ``\PH production''
indicates the production mechanism targeted by an analysis; it does not imply 100\% purity in the selected sample.
The main contribution in the untagged categories is always ggF.
The $(\mathrm{jj})_\mathrm{VBF}$ refers to a dijet pair consistent with the VBF topology.
$(\mathrm{jj})_{\PW(\cPZ)}$ and $(\mathrm{J})_{\PW(\cPZ)}$ refer to a dijet pair and
single merged jet from a Lorentz-boosted $\PW$ ($\cPZ$) with
an invariant mass consistent with a $\PW$ ($\cPZ$) dijet decay, respectively. The superscript ``0,1,2 b tags'' refers to the three possible categories of b tag multiplicities.
Exclusive final states are selected according to the lepton and reconstructed jet content of the event.
The mass range under investigation and the mass resolution are also listed. Mass ranges differ with the sensitivities of each channel.
}
\label{tab:channels}
\small
\begin{tabular}{cccccc}
  \hline
 \PH & \PH & Exclusive & No. of & $m_{\PH}$ range & $m_{\PH}$ \\
 decay mode & production & final states & channels & [\GeVns{}] & resolution \\
\hline
$\PW\PW \to \ell\nu\ell\nu$      & untagged  &  ((\Pe\Pe, \Pgm\Pgm), \Pe\Pgm) + (0 or 1 jets)                           & 4  & 145--1000~\footnote{\label{note:ewks}EW singlet model interpretation starts at 200\GeV to avoid contamination from \Ph{}(125).}\footnote{\label{note:8tev}600-1000\GeV for $\sqrt{s} = 8\TeV$ only.}  & 20\%  \\
                                 & VBF tag   &  ((\Pe\Pe, \Pgm\Pgm), \Pe\Pgm) + $(\mathrm{jj})_\mathrm{VBF}$            & 2  & 145--1000 $^{ab}$         & 20\%  \\
\hline
$\PW\PW \to \ell\nu\cPq\cPq$     & untagged  &  ($\Pe\nu$, $\Pgm\nu$) + (jj)$_{\PW}$                                    & 2  & 180--600                  & 5--15\%  \\
                                 & untagged  &  ($\Pe\nu$, $\Pgm\nu$) + (J)$_{\PW}$ + (0+1-jets)                     & 2  & 600--1000 $^{b}$          & 5--15\%  \\
                                 & VBF tag   &  ($\Pe\nu$, $\Pgm\nu$) + (J)$_{\PW}$ + $(\mathrm{jj})_\mathrm{VBF}$      & 1  & 600--1000 $^{b}$          & 5--15\%  \\
\hline
$\cPZ\cPZ \to 2\ell 2\ell'$      & untagged  &  4\Pe, 4\Pgm, 2\Pe2\Pgm                                                  & 3  & 145--1000         & 1--2\%   \\
                                 & VBF tag   &  (4\Pe, 4\Pgm, 2\Pe2\Pgm) + (jj)$_{\mathrm{VBF}}$                        & 3  & 145--1000         & 1--2\%   \\
                                 & untagged  &  (\Pe\Pe, \Pgm\Pgm) + (\tauh\tauh, \taue\tauh, \taumu\tauh, \taue\taumu) & 8  & 200--1000         & 10--15\%  \\
\hline
$\cPZ\cPZ \to 2\ell 2\nu$        & untagged  &  (\Pe\Pe, \Pgm\Pgm) + (0 or $\geq$ 1 jets)                               & 4  & 200--1000         & 7\%   \\
                                 & VBF tag   &  (\Pe\Pe, \Pgm\Pgm) + $(\mathrm{jj})_\mathrm{VBF}$                       & 2  & 200--1000         & 7\%   \\
\hline
$\cPZ\cPZ \to 2\ell 2\cPq$       & untagged  &  (\Pe\Pe, \Pgm\Pgm) + $(\mathrm{jj})_\cPZ^{0,1,2 \,\text{b tags}}$          & 6  & 230--1000~\footnote{\label{note:8tev2}For $\sqrt{s} = 8\TeV$ only.} & 3\% \\
                                 & untagged  &  (\Pe\Pe, \Pgm\Pgm) + $(\mathrm{J})_\cPZ^{0,1,2 \,\text{b tags}}$          & 6  & 230--1000 $^{c}$  & 3\% \\
                                 & VBF tag   &  (\Pe\Pe, \Pgm\Pgm) + $(\mathrm{jj})_\cPZ^{0,1,2 \,\text{b tags}}$ + $(\mathrm{jj})_\mathrm{VBF}$         & 6  & 230--1000 $^{c}$ & 3\% \\
                                 & VBF tag   &  (\Pe\Pe, \Pgm\Pgm) + $(\mathrm{J})_\cPZ^{0,1,2 \,\text{b tags}}$ + $(\mathrm{jj})_\mathrm{VBF}$          & 6  & 230--1000 $^{c}$ & 3\% \\
  \hline
\end{tabular}
\end{minipage}

\end{table}

\subsection{\texorpdfstring{$\PH \to \PW\PW \to \ell\nu\ell\nu$}{H to WW to l nu l nu}}
\label{sec:hwwlnulnu}
In the $\PH \to \PW\PW \to \ell\nu\ell\nu$ channel the Higgs boson decays to two $\PW$ bosons, both of which decay leptonically,
resulting in a signature with two isolated, oppositely charged, high-$\pt$ leptons (muons or electrons)
and large $\MET$ due to the undetected neutrinos.
A complete description of the analysis strategy is given in Ref.~\cite{CMSHWWlvlv2014}.
For this analysis, we require triggers with either one or two high-\pt muons or electrons.
The single muon or electron triggers are based on relatively tight lepton identification with $\pt$ thresholds from 17 to 25\GeV (17 to 27\GeV) in the muon (electron) channel.
The higher thresholds are used for periods of higher instantaneous luminosity.
The dilepton trigger $\pt$ thresholds for the leading and trailing leptons were 17 and 8\GeV, respectively.

Candidate events must contain two reconstructed leptons with opposite charge, $\pt>20$\GeV
for the leading lepton, and $\pt>10$\GeV for the subleading one.
Only muons (electrons) with $\abs{\eta} < 2.4\,(2.5)$ are considered in this channel.
The analysis is very similar to that in the Higgs boson discovery~\cite{CMSobservation125,CMSlongpaper},
but additionally uses an improved Higgs boson lineshape model.

Events are classified into three mutually exclusive categories, according to the number of reconstructed jets
with $\pt>30$\GeV.
The categories are characterized by different signal yields and signal-to-background ratios.
In the following, these are referred to as 0-jet, 1-jet, and 2-jet multiplicity categories.
Events with more than two jets are considered only if they are consistent with the VBF signature:
a dijet mass of the two highest $\pt$ jets greater than 500\GeV, $\abs{\Delta\eta_{\mathrm{jj}}}> 3.5$, and no additional jets with $\pt > 30$\GeV in the $\eta$ region
between these two leading jets.
Signal candidates are further divided into same-flavor (SF) dilepton ($\mumu$ or $\Pep\Pem$) and
different-flavor (DF) dilepton ($\Pgm^\pm\Pe^\mp$) categories.
The bulk of the signal arises through direct $\PW$ boson decays to muons or electrons, with the
small contribution from $\PW \to \tau\nu \to \ell+X$ decays implicitly included.

{\tolerance=500 In addition to high-$\pt$ isolated leptons and minimal jet activity, significant $\MET$ is expected to be
present in signal events, as opposed to none to moderate $\MET$ in background.
For this channel, an $\METpr$ variable is employed,
defined
as (i) the magnitude  of the $\vecEtm$
 component transverse to the closest lepton, if $\Delta \phi(\ell, \vecEtm ) < \pi/2$,
or (ii) the magnitude of $\vecEtm$ otherwise.
This observable more efficiently distinguishes $\dytt$ background events in which
the $\vecEtm$
is preferentially aligned with the leptons,
and
$\dyll$ events with mismeasured
$\vecEtm$.
Since the $\METpr$ resolution is degraded as pileup increases,
the minimum of two different observables is used: the first includes all particle candidates in the
event, while the second uses only the charged
particle candidates associated with the primary vertex.
Events with $\METpr > 20$\GeV are selected for this analysis.\par}

The backgrounds are suppressed using techniques described in Refs.~\cite{CMSobservation125,CMSlongpaper}.
Top quark background is controlled with a selection based on the presence of a soft muon and b-jet tagging~\cite{BTV}.
Rejection of events with
a third lepton passing the same requirements as the two selected leptons reduces both $\WZ$ and $\wgamma^*$ backgrounds.

The Drell--Yan (DY) process produces SF lepton pairs ($\Pgmp\Pgmm$ and $\Pep\Pem$) and therefore
additional requirements are
applied for the SF final state. Firstly, the resonant component of the DY background
is rejected by requiring a dilepton mass outside a 30\GeV window centered on the $\cPZ$ boson mass. The remaining
off-peak contribution is further suppressed by requiring significant missing transverse energy in the event.
For the $\sqrt{s}$ = 7\TeV data, we require $\METpr > (37 + {N_\text{vtx}}/2)$\GeV directly, while for $\sqrt{s}$ = 8\TeV data,
a boosted decision tree (BDT) multivariate discriminant including
$\METpr$ is used.
For events with two jets, the dominant source of misreconstructed \MET is the mismeasurement of
the hadronic recoil, and optimal performance is obtained by requiring $\MET > 45$\GeV.
Finally, the momenta of the dilepton system and of the most energetic jet must not be back-to-back in the transverse plane.
For the $\sqrt{s}$ = 7\TeV data, we require the angle to be less than 165 degrees , while for $\sqrt{s}$ = 8\TeV data, this information is included in the BDT.
These selections
reduce the DY background by three orders of magnitude, while rejecting less than 50\% of the signal.

The final analysis in the DF and in the SF 0-jet or 1-jet categories is performed using a two-dimensional fit in two observables,
the dilepton invariant mass $\mll$ and the transverse mass $\mt$ determined between the transverse dilepton system, $\vecPtell$, and
the $\vecEtm$:
\begin{equation}
\mt = \sqrt{2 \pt^{\ell\ell}\MET (1-\cos\Delta\Phi(\vecPtell,\vecEtm))},
\end{equation}
where $\Delta\Phi(\vecPtell,\vecEtm)$ is the azimuthal angle between
the dilepton transverse momentum and the \vecEtm.

\begin{figure}[h!t]
\centering
   \includegraphics[width=0.48\textwidth]{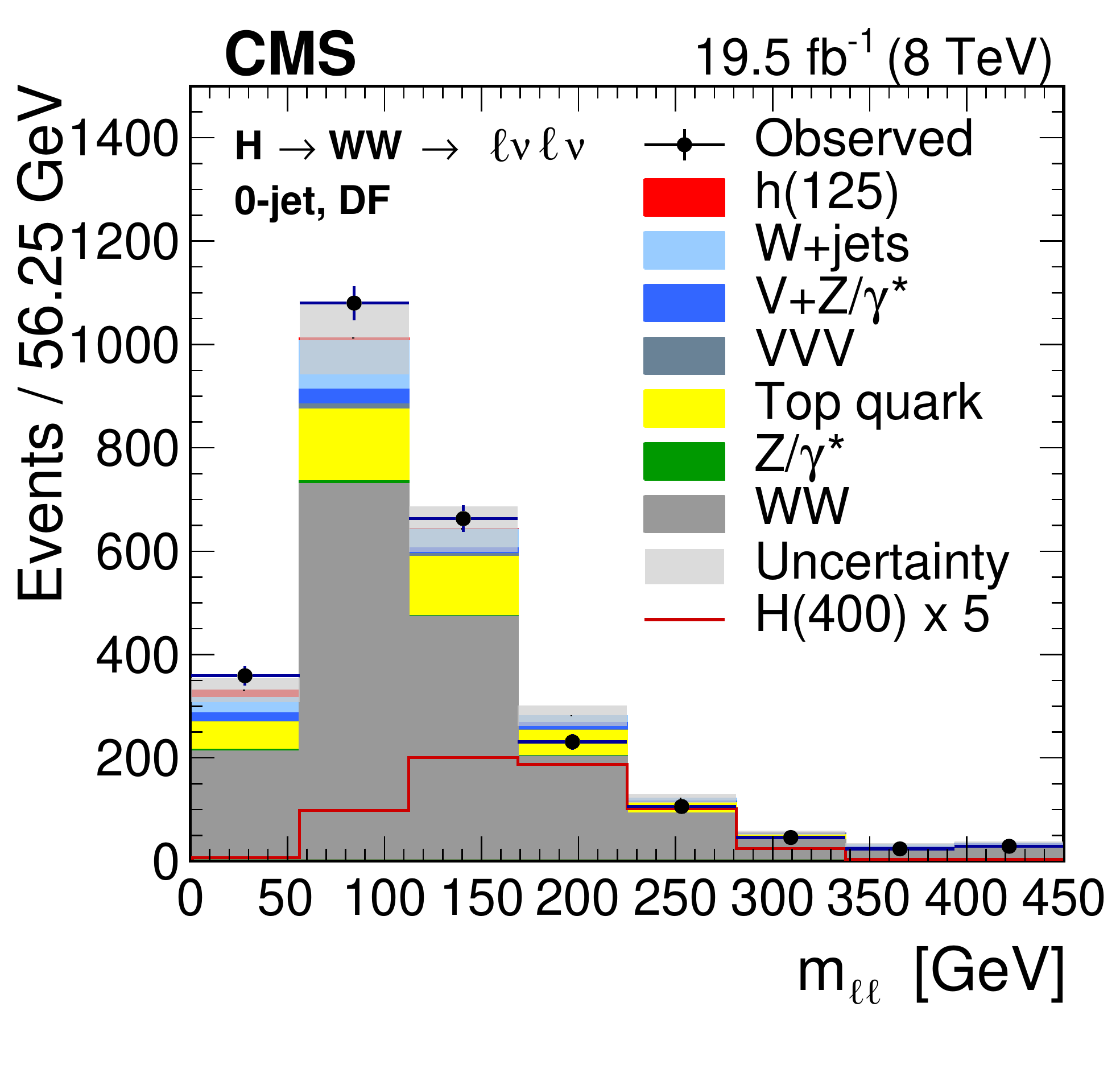}
   \includegraphics[width=0.48\textwidth]{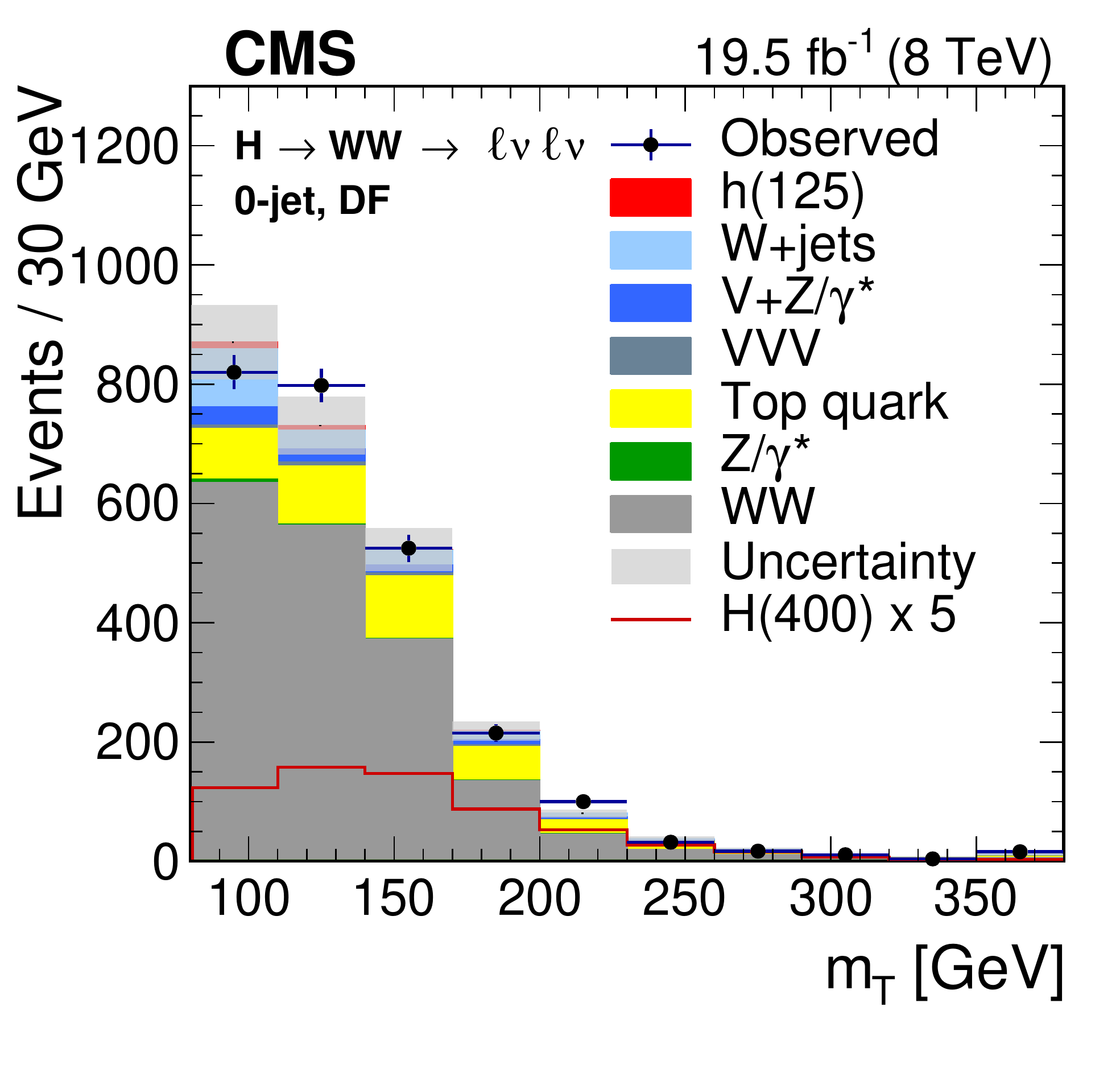}\\
   \caption{Distributions of $m_{\ell \ell}$ (left) and $\mt$ (right) for the 0-jet DF category of
   the $\PH \to \PW\PW \to \ell\nu\ell\nu$ search. The uncertainty in the background histograms includes the systematic
   uncertainties on all background estimates and is centered on the sum of all backgrounds, including the \Ph{}(125) in red. 
   The $\PW+\text{jets}$ distributions include the contributions from QCD multijet processes as well.
   The red open histogram shows five times the expectation for a $\mH = 400$\GeV SM-like Higgs boson. The selection has been optimized to suppress the \Ph{}(125) contribution as explained in the text. }
\label{fig:hww2l2n_2d}
\end{figure}

Figure~\ref{fig:hww2l2n_2d} shows the distributions of these two observables for the 0-jet DF category.
The fit ranges in $m_{\ell\ell}$ and $\mt$ are dependent on the $\mH$ hypothesis.
For $\mH > 200$\GeV, additional selections requiring $\pt  >50$\GeV for the leading lepton and $\mt >80$\GeV are imposed
to suppress the \Ph{}(125) contribution.
A cross-check counting analysis is performed by applying selection criteria
to several kinematic observables including $\mll$ and $\mt$.

For the VBF production mode~\cite{Ciccolini:2007jr,Ciccolini:2007ec,Arnold:2008rz,Cahn:1987},
the cross section is roughly ten times smaller than for ggF at lower $\mH$ hypotheses
and is roughly three times smaller at the highest $\mH$ hypothesis. We optimize the selection in the 2-jet category to tag these
VBF-type events by requiring the mass of the dijet system to fulfill $\mjj > 500$\GeV,
and the angular separation of the two jets to pass $\abs{\Delta \eta_{\mathrm{jj}}} > 3.5$.
Given the small event yield in this category, the signal extraction in the DF category is performed using
a one-dimensional fit in $\mll$ where an $\mH$ dependent requirement on the transverse mass
is imposed.
A counting analysis is performed in the SF category and is used as a cross-check in the DF category.

The normalization of the background contributions relies on observed events rather than simulation whenever possible, and exploits a combination
of techniques~\cite{CMSobservation125,CMSlongpaper}. The $\ttbar$ background is estimated by extrapolation from the observed
number of events with the b-tagging requirement inverted. The DY background measurement is based on extrapolation
from the observed number of $\Pgmp \Pgmm$ and $\Pep\Pem$ events with the $\cPZ$ boson veto requirement inverted. The background
of $\PW+\text{jets}$ and Quantum Chromodynamics (QCD) multijet events is estimated by measuring the number of events with one loosely isolated lepton.
The probability for such loosely isolated nongenuine leptons to pass the tight isolation criteria is measured in
observed data using multijet events.
In the 0-jet and 1-jet bins, the nonresonant $\WW$ contribution is estimated from a fit to the data
while in the 2-jet bin it is estimated from simulation.
Other backgrounds, such as
V+$\cPZ/\gamma^*$ and triple boson production (VVV) are estimated from simulation and are small.

Experimental effects, theoretical predictions, and the choice of event generators (\POWHEG, \textsc{gg2ww}, \MADGRAPH, \textsc{Phantom}) are considered as sources of
systematic uncertainty, and their impact on the signal efficiency is assessed.
The overall signal normalization uncertainty is estimated to be about 20\%, and is
dominated by the theoretical uncertainty associated with missing higher-order QCD corrections and PDF uncertainties,
estimated following the PDF4LHC recommendations~\cite{Alekhin:2011sk,Botje:2011sn,Lai:2010vv,Martin:2009iq,Ball:2011mu}.
The total uncertainty in the background estimation in the $\Hww$ signal region is about 15\% and is dominated by the
statistical uncertainty in the observed number of events in the background control regions.

\subsection{\texorpdfstring{$\PH \to \PW\PW \to \ell\nu \cPq\cPq$}{H to WW to l nu qq}}
\subsubsection{Unmerged-jet category}
\label{sec:hwwlnujj}
In the $\PH \to \PW\PW \to \ell\nu \cPq\cPq$ channel we search for a Higgs boson decaying to WW,
where one W decays leptonically, thus providing a trigger handle for the event, while the other decays hadronically. This channel has a
larger branching fraction than the two-lepton final state and allows one to
reconstruct the Higgs boson candidate invariant mass~\cite{intro2}.
The final state consists of an isolated electron or muon, \MET, and two separated jets.
The main experimental challenge is to control the large W+jets background.

We use data collected with a suite of single-lepton triggers mostly
using \pt thresholds of 24 (27)\GeV for muons (electrons).
Basic kinematic selections are applied to the final-state objects to reduce the background contribution.
The muon (electron) is required to be isolated and have $\pt >$ 25 (35)\GeV.
A veto is imposed on additional muons and electrons in the event to reduce backgrounds from DY events.
The events are required to have $\MET > 25\,(30)$\GeV in case of muons (electrons).
The transverse mass $\mt = \sqrt{\smash[b]{2\pt\MET ( 1 - \cos\Delta\Phi )}}$, with $\Delta\Phi$ being the azimuthal angle between the lepton
$\pt$ and \vecEtm, needs to be greater than 30\GeV.
The two leading-\pt jets in the event each must have $\pt > 30$\GeV and must together have an invariant mass, $m_{\mathrm{jj}}$,
between 66 and 98\GeV.
An unbinned maximum likelihood fit is performed on $m_{\mathrm{jj}}$ to determine
the background normalizations in the signal
region for each Higgs mass hypothesis independently.
Events that contain a b-tagged jet are vetoed in order to reduce the background from top quark decays.

To further exploit the differences in kinematics between the signal and
background, a likelihood discriminant is constructed using
angles between the Higgs boson decay products that fully describe the Higgs boson
production kinematics~\cite{Gao:2010qx,Chatrchyan:2013mxa} and the lepton charge.
The lepton charge provides discrimination power because of the charge asymmetry in the background, which is not present in the signal.
This approach improves the expected sensitivity to
a Higgs boson across the entire mass range.

The background normalization in the signal region is extracted for
each Higgs boson mass hypothesis independently with an unbinned maximum
likelihood fit to the dijet invariant mass distribution, $m_{\mathrm{jj}}$, of
the two leading jets. The signal region corresponding to the W boson mass
window, $66 < m_{\mathrm{jj}} < 98$\GeV, is excluded from the fit.  The
background is overwhelmingly due to events from W+jets production.
The normalization of the W+jets component is a free parameter and is determined in
the fit.  The electroweak diboson, \ttbar, and single top shapes are
based on simulation and their normalizations are constrained to the
theoretical predictions with associated uncertainties.  In the case of diboson backgrounds, both WW
and WZ normalizations come from NLO predictions.
The uncertainties in the
normalization of the W+jets component obtained from the fit, as well as those from all other backgrounds, are
included in the limit calculation as systematic uncertainties.

\begin{figure}[h!t]
  \centering
  \includegraphics[width=0.48\textwidth]{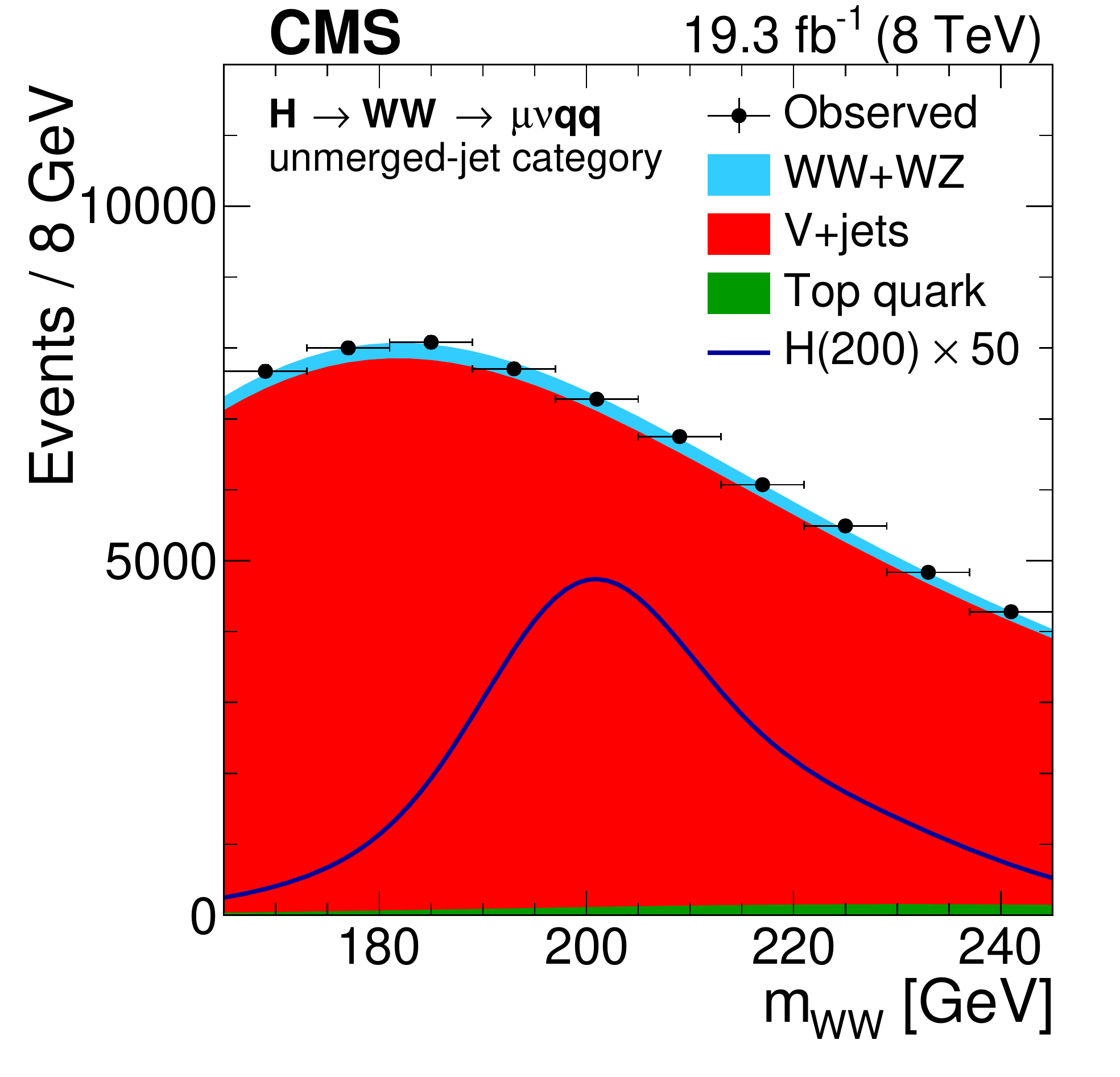}
  \includegraphics[width=0.48\textwidth]{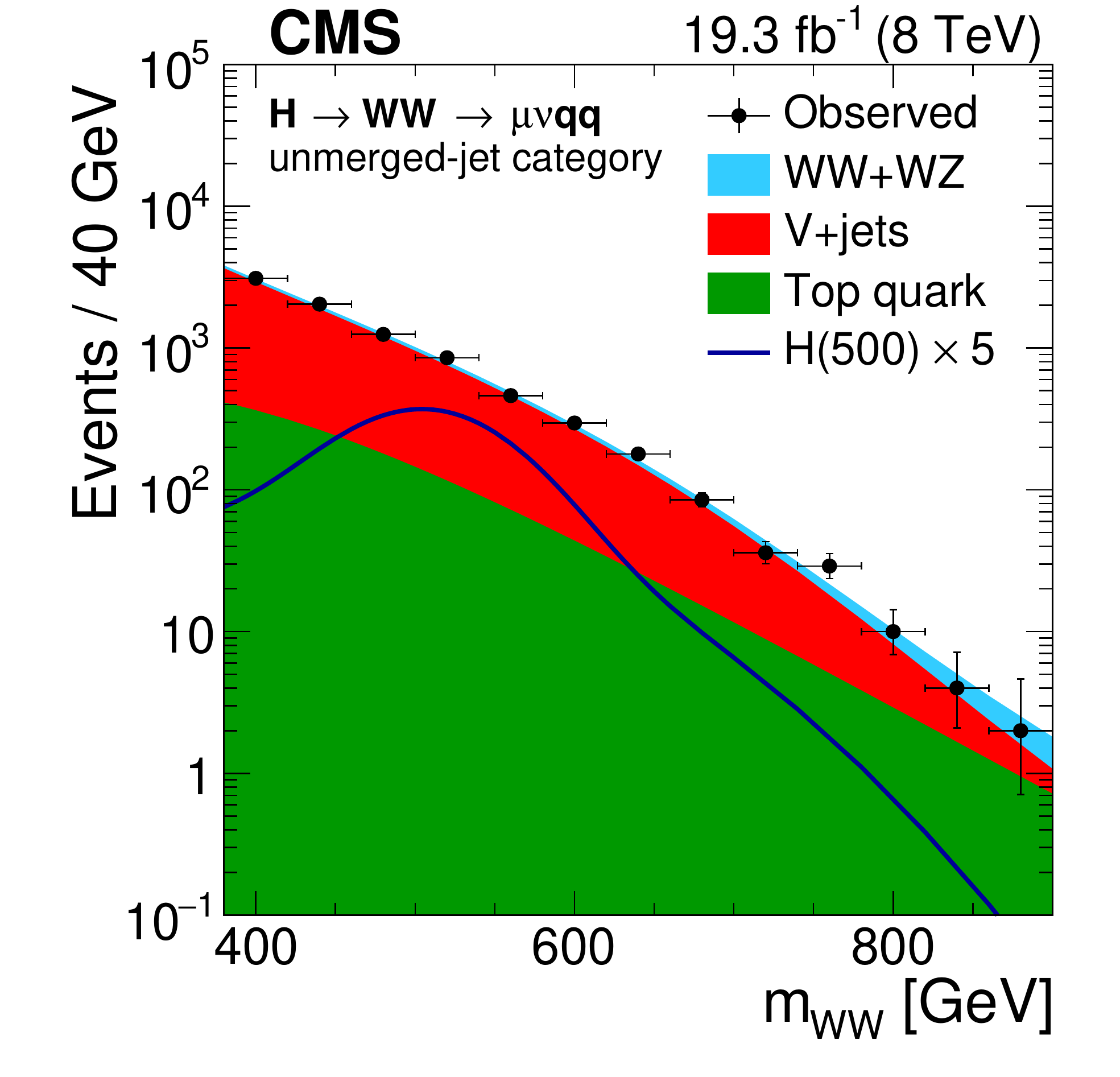}
  \caption{\label{fig:mlvjj_mH350} The WW invariant mass distribution
    with the fit projections in the signal region $66 < m_{\mathrm{jj}} <
    98$\GeV, for the muon channel of the unmerged-jet category. The V+jets background includes both, the large contribution from
    W+jets production, and the much smaller component of Z+jets. 
    The blue curve on the left (right) shows 50 (5) times the expectation for a $\mH = 200$ (500)\GeV
    SM-like Higgs boson.
  }
\end{figure}

For the signal search, we use the binned distribution of $m_{\PW\PW}$,
which is computed using a neutrino longitudinal momentum, $p_{z}$, that is determined from the constraint
that the leptonically decaying $\PW$ boson is produced on-shell.
The ambiguity of two possible solutions is resolved by taking the solution that yields the smaller $\abs{p_{z}}$ value for the neutrino.
According to simulation over 85\% of signal events are assigned the correct $p_z$ value, thus improving the mass resolution,
especially at low Higgs mass.
The contribution of each background in the
signal region ($66 < m_{\mathrm{jj}} < 98$\GeV) is described by an
expected shape and normalization.
The distribution of the
W+jets contribution is parameterized with a polynomial functional form
determined through simulation.
The parameters of the function are
determined from the observed data spectrum in the binned likelihood fit, which
is also used to determine the exclusion limits.
The other background shape distributions are parameterized using simulation, and their normalizations and uncertainties are included as nuisance parameters.
Figure~\ref{fig:mlvjj_mH350} depicts the $m_{\PW\PW}$ distribution for the muon category for
two different Higgs boson mass hypotheses. The observed data are compared to background and signal expectations.

Experimental effects, theoretical predictions, and uncertainties due to the choice of fit functions are considered as sources of
systematic uncertainty.
Because of the large background, the
dominant source of systematic uncertainty is the shape uncertainty of
the W+jets $m_{\PW\PW}$ distribution, followed by the
normalization uncertainty that is extracted from the $m_{\mathrm{jj}}$ fit.
The main uncertainty in the signal normalization stems from the uncertainty in the efficiency of the likelihood discriminant selection.
This effect occurs because of mismodeling of the likelihood discriminant and is studied in a signal-depleted region of the analysis.

\subsubsection{Merged-jet category}
\label{sec:hwwlnuJ}
In the highest mass range of this search, from 600 to 1000\GeV, the \pt of the decaying
$\PW$ bosons is typically greater than 200\GeV.
At this \pt, the daughter quarks of the hadronically decaying $\PW$ are often merged into a single jet
to the point where traditional dijet searches cannot be performed.
For a signal mass of 600\GeV (1\TeV), and signal events falling in the detector acceptance,
approximately 65\% (82\%) of the hadronic $\PW$ decay
products are contained in a cone of $\Delta R < 0.8$; alternatively,
approximately 10\% (42\%) of the hadronic $\PW$ decay products are separated by a distance of $\Delta R < 0.5$
and would not be reconstructed by the standard CMS AK5
jet finding algorithm. The larger cone CA8 jet affords more signal acceptance in the single
jet signature while not losing events when the decay products are separated by $\Delta R < 0.5$.
In this case, we use jet substructure techniques for identifying jets that have originated from a
hadronically decaying $\PW$ boson with high Lorentz boost.

As in the unmerged case, we use data collected with the single electron or muon trigger.
The dominant background is $\PW$+jets with a smaller background contribution from \ttbar.
Remaining backgrounds arise from $\PW\PW$, $\PW\cPZ$, $\cPZ\cPZ$, and single top quark production.

The hadronic $\PW$ boson candidate is reconstructed with CA8 jets to increase acceptance.
To reduce the contribution of quark- and gluon-initiated jets from QCD processes, a selection is made on the pruned jet mass
of the CA8 jet of $65 < m_{\mathrm{J}} < 105$\GeV and the N-subjettiness ratio $\tau_2/\tau_1 < 0.5$.

The kinematic selections of muons and electrons are slightly more stringent than those for the unmerged case,
with the \pt thresholds of isolated muons and electrons being 30 and 35\GeV, respectively.
The \MET requirement is increased to be greater than 50 (70)\GeV in the muon (electron) channel.
The \pt of both, leptonically and hadronically decaying $\PW$  boson
candidates, is required to be greater than 200\GeV in order to select events with a large Lorentz boost.
Additional selections are made to ensure that the $\PW$ boson candidates are sufficiently separated in
a back-to-back topology.
As in the unmerged category, there is an additional requirement to veto events with a b tagged jet
in order to reduce the background from top quark decays.

To increase the sensitivity of the analysis, the analysis is split into exclusive jet multiplicity categories: 0+1-jet and 2-jets.
These additional AK5 jets must pass a \pt threshold of 30\GeV and be separated from the hadronic W candidate in the event by a distance $
\Delta R > 0.8$.
Additionally, the 2-jet category has further requirements to better identify a topology consistent with the VBF
production mode.
The two highest \pt AK5 jets in the event are required to satisfy $\abs{\Delta\eta_{\mathrm{jj}}} > 2.5$ and
$m_{\mathrm{jj}} > 250$\GeV.
There is also a requirement on the invariant mass of the $\PW$ boson candidates and the nearest AK5 jet to be greater than 200\GeV
in order to reject other top quark background.
The final discriminating distribution is the $m_{\PW\PW}$ shape reconstructed from the lepton + \MET + CA8 jet system.
Because the number of events in the 2-jet category is limited, the muon and electron datasets are merged.

The background is estimated using observed data rather than simulation wherever possible.
The dominant $\PW$+jets background normalization is determined from a fit to the pruned jet mass sideband
where other backgrounds contribute less significantly.
The $\PW$+jets shape is determined from extrapolating the pruned jet mass sideband region into the signal region.
The $\ttbar$ background is estimated by inverting the veto on AK5 b-tagged jets, which yields a high-purity $\ttbar$ control sample.
The normalization of $\ttbar$ in the signal region is then corrected
by a data-to-MC scale factor determined from the $\ttbar$ control sample.
Finally the $\ttbar$ control sample is also a good source of $\PW$ bosons with high Lorentz boost, which are used to calibrate the
jet substructure selection efficiency.

\begin{figure}[h!t]
  \centering
  \includegraphics[width=0.48\textwidth]{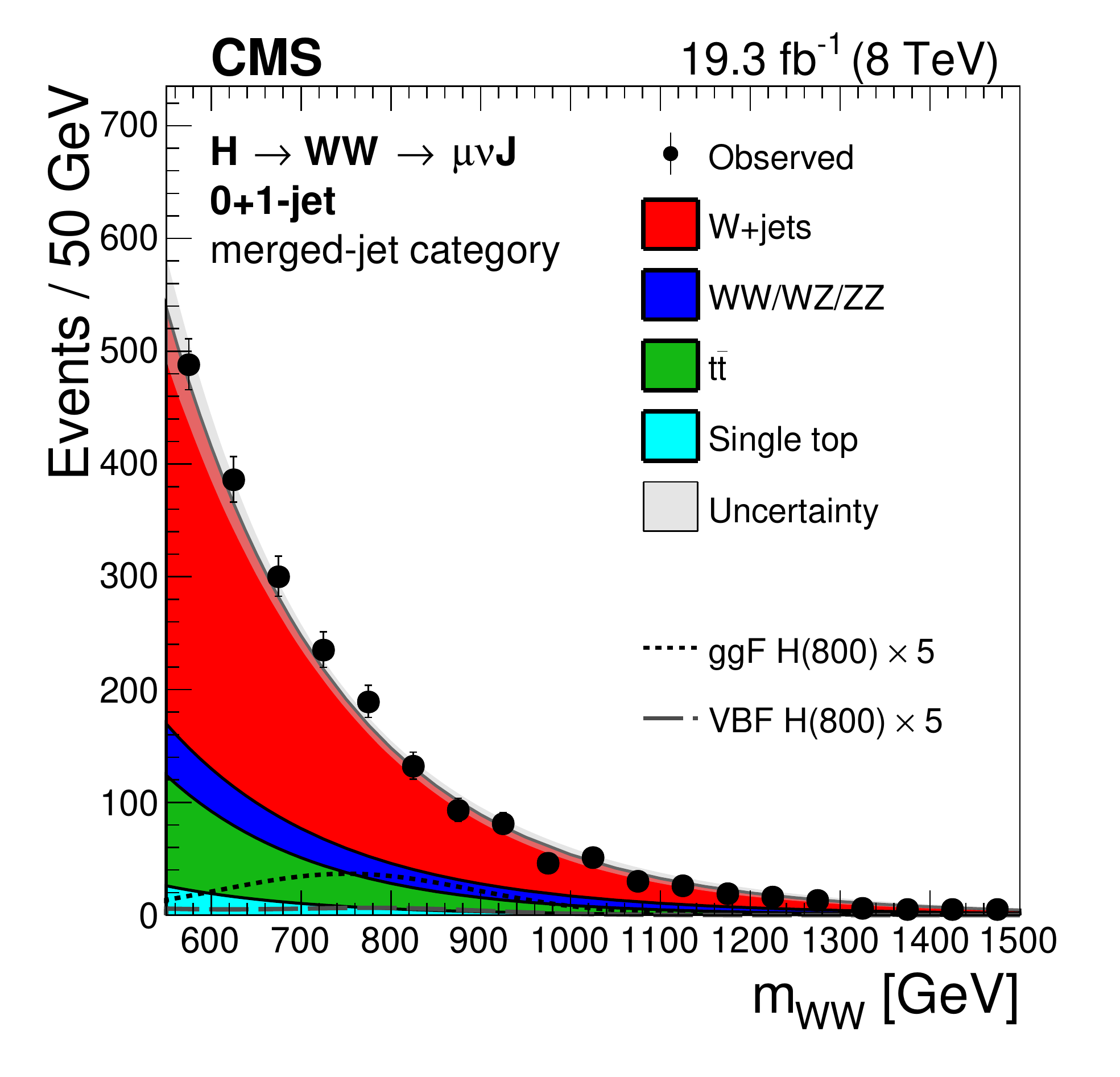}
  \includegraphics[width=0.48\textwidth]{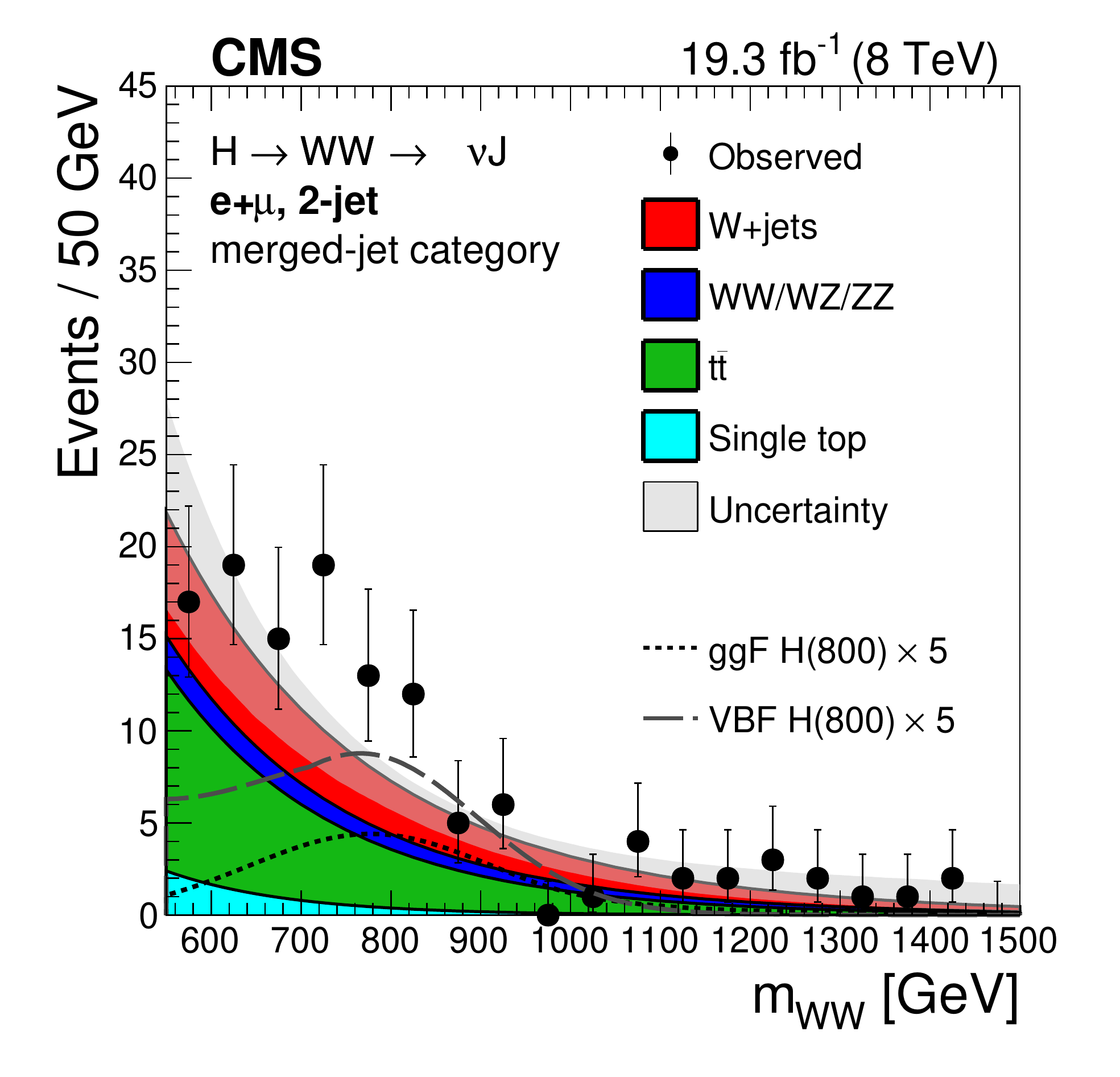}\\
  \hspace*{178pt}\raisebox{191pt}[0pt][0pt]{\small $\ell$}\vspace*{-13pt}
  \caption{\label{fig:mlvjJ_mH600} The final WW invariant mass distribution is shown for the 0+1-jet bin category for the muon channel
  only (left) and for the 2-jet bin category (right). Points represent the observed data, shaded
  graphs represent the background and dashed graphs represent five times the expectation for a $\mH = 800$\GeV SM-like Higgs boson
  from ggF and VBF production, separately.}
\end{figure}

Figure~\ref{fig:mlvjJ_mH600} displays the final $m_{\PW\PW}$ distributions after all selections and background
estimations for the 0+1-jet category for the muon channel only on the left and for the 2-jet category on the right.
In the 2-jet category, there is an excess observed around 750\GeV with local significances of $2.6\,\sigma$ at 700\GeV{} and
$2.1\,\sigma$ at 800\GeV. No such excess is observed in the 0+1-jet multiplicity bin.

Sources of systematic uncertainty are similar to the unmerged-jet category.
The dominant experimental uncertainties are the background normalization and shape uncertainty, particularly in the 2-jet category,
where the number of events is small, as well as the uncertainty in the
signal cross section due to the hadronic $\PW$ candidate selection.
The dominant theoretical uncertainty comes from the uncertainty in the cross section when dividing the analysis
into exclusive jet multiplicity bins.

\subsection{\texorpdfstring{$\PH \to \ZZ \to 2\ell 2\ell`$}{H to ZZ to 2 l 2 l`}}
\label{sec:hzz4l}
This analysis seeks to identify Higgs boson decays to a pair of $\cPZ$ bosons, with one $\cPZ$ decaying to a pair of muons or electrons
($\cPZ\to 2\ell$, with $\ell = \mu$ or e), and the second decaying to electrons, muons or taus ($\cPZ\to 2\ell`$,
with $\ell` = \Pe,\mu$ or $\tau$)~\cite{CMSobservation125,Chatrchyan:2013mxa,Chatrchyan:2012hr}.
This channel has extremely low background, and the presence of four leptons in the
final state allows reconstruction and isolation requirements to be very loose, leading to a high selection efficiency.
This channel is one of the most sensitive channels across the entire mass range.

For this analysis, we require triggers with two high-\pt muons or electrons.
The dilepton trigger \pt thresholds for the leading and trailing leptons were 17 and 8\GeV, respectively.
Events included in the analysis contain $\cPZ$ boson candidates formed from a pair of leptons of the same flavor and
opposite charge. Decay muons or electrons are required to be isolated, and to originate from the primary vertex.
Muons (electrons) are required to have $\pt > 5 (7)$\GeV and $\abs{\eta} < 2.1 (2.5)$, while taus are required to have a visible transverse momentum
$\pt > 20$\GeV and $\abs{\eta} < 2.3$. We reconstruct the $\cPZ\to\tau\tau$ in the following decay modes:
$\cPZ\to\tau_{\mathrm{h}}\tau_{\mathrm{h}}$, $\cPZ\to\tau_{\mathrm{e}}\tau_{\mathrm{h}}$,
$\cPZ\to\tau_\mu\tau_{\mathrm{h}}$, and $\cPZ\to\tau_{\mathrm{e}}\tau_\mu$. Overlap with the $2\ell 2\ell$ channel
is avoided by excluding events with both taus decaying to electrons or muons.

For the $2\ell2\ell$ final state, the lepton pair with invariant mass closest to the nominal $\cPZ$ boson mass, denoted
$\cPZ_1$, is identified and retained if $40 < m_{\cPZ_1} < 120$\GeV. The second $\cPZ$ boson, denoted $\cPZ_2$, is then
constructed from the remaining leptons in the event, and is required to satisfy $12 < m_{\cPZ_2} < 120$\GeV. If
more than one $\cPZ_2$ candidate remains, the ambiguity is resolved by choosing the leptons of highest $\pt$. Amongst
the four candidate decay leptons, at least one should have $\pt > 20$\GeV, and another
should have $\pt > 10$\GeV. This requirement ensures that selected events correspond to the high-efficiency
plateau of the trigger.

For the $2\ell2\Pgt$ final state, events are required to have one $\cPZ_1 \to 2\ell$ candidate, with one
lepton having $\pt > 20\GeV$ and the other $\pt > 10\GeV$. The leptons from leptonic decays of the tau are required to
have $\pt > 10$\GeV. The invariant
mass of the reconstructed $\cPZ_1$ candidate is required to satisfy $60 < m_{\ell\ell} < 120$\GeV. The $\cPZ_2$ candidate mass reconstructed
from the visible tau decay products (visible mass), $m_{\Pgt\Pgt}$, must satisfy $30 < m_{\Pgt\Pgt} < 90$\GeV for events with at least
one hadronically decaying tau,
and $20 < m_{\Pgt\Pgt} < 90$\GeV for events with two leptonically decaying taus. Events with both taus decaying to muons or electrons are excluded
in order to avoid overlap with the $2\ell2\ell$ channel.
The selections on the reconstructed $\cPZ_1$ mass for the $2\ell2\Pgt$ final state are tighter than for the $2\ell2\ell$ channel,
because the search range starts at $\mH = 200$\GeV rather than $\mH = 145$\GeV.

In order to further separate signal from background and to distinguish different signal production mechanisms, we use
a matrix element likelihood approach, which relies on probabilities for an event
to come either from signal or background using a set of observables, such as angular and mass information,
which fully characterize the event topology in its center-of-mass
frame~\cite{Gao:2010qx}.

The events are categorized according to their jet multiplicity, counting jets with $\pt > 30$\GeV.
In the 0- or 1-jet category the $\pt$ spectrum of the four-lepton system
is exploited to distinguish between ggF and vector boson induced production modes, such as VBF and associated production with a vector
boson (VH). The events selected with two or more jets potentially come from several sources:
background (predominantly $\cPZ\cPZ+$2 jets),
VBF signal, V$\PH$ signal (with V$\to$2 jets), and ggF signal $\Pg\Pg\to \PH+$2 jets.
In order to isolate the production mechanism, we use a matrix-element-based probability~\cite{Gao:2010qx,Anderson:2014}
for the kinematics of the two jets and the Higgs boson
to come either from the VBF process or from the $\Pg\Pg\to \PH+$2 jets process.
This discriminant is equally efficient to separate VBF from either $\PH+$2 jets signal
or $\cPZ\cPZ+$2 jets background, because both processes show very similar jet kinematics.

To allow estimation of the $\ttbar$, $\cPZ$+jets, and $\PW\cPZ$+jets backgrounds, a $\cPZ_1$+$X$ control region
is defined, well separated from the expected signal region.
In addition, a sample of $\cPZ_1 + \ell$ events, with at least
one reconstructed lepton in addition to a $\cPZ$, is
defined in order to estimate the lepton misidentification probability, which is the probability for non-prompt leptons and other
particles, which are not leptons, to be reconstructed as leptons and to pass the isolation and identification requirements.
The contamination of the sample from $\PW\cPZ$ events
containing a
genuine additional lepton is suppressed by requiring the energy imbalance measured in the transverse
plane to be below $25 (20)$\GeV for the $2\ell 2\ell$ ($2\ell2\Pgt$) channel. The event rates measured in the background control region are extrapolated to the
signal region. The systematic uncertainties associated with the
reducible background estimate vary from 30\% to 100\% and are combined in quadrature with
the statistical uncertainties.

The cross section for \ZZ\ production at NLO is
calculated with \textsc{mcfm}. The theoretical uncertainty in the
cross section is evaluated as a function of $m_{2\ell2\ell'}$, varying both the renormalization and factorization
scales and the PDF set, following the PDF4LHC recommendations~\cite{Botje:2011sn,Alekhin:2011sk,Lai:2010vv,Martin:2009iq,Ball:2011mu}.

\begin{figure}[h!]
\centering
\includegraphics[width=0.48\textwidth]{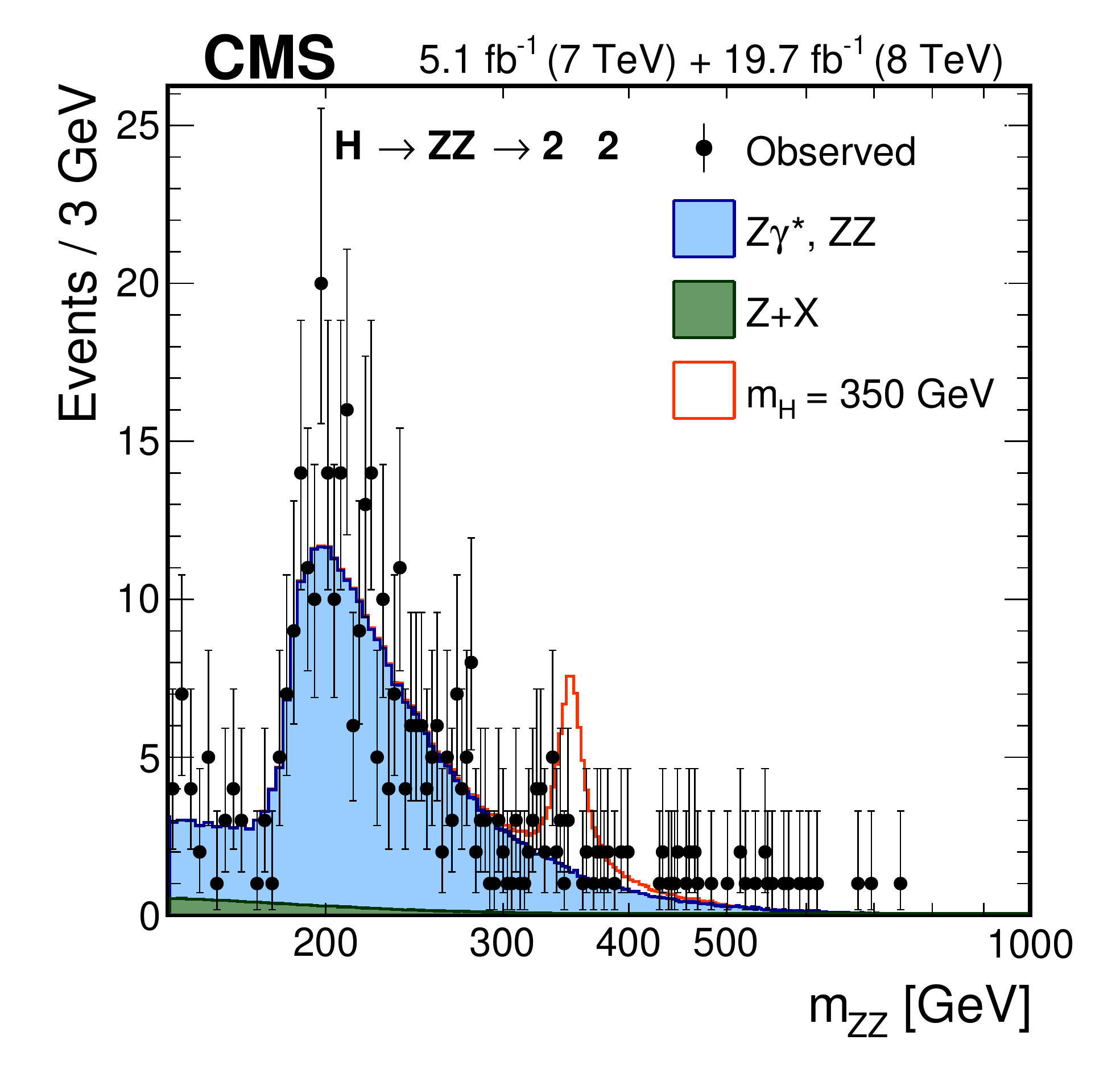}
\includegraphics[width=0.48\textwidth]{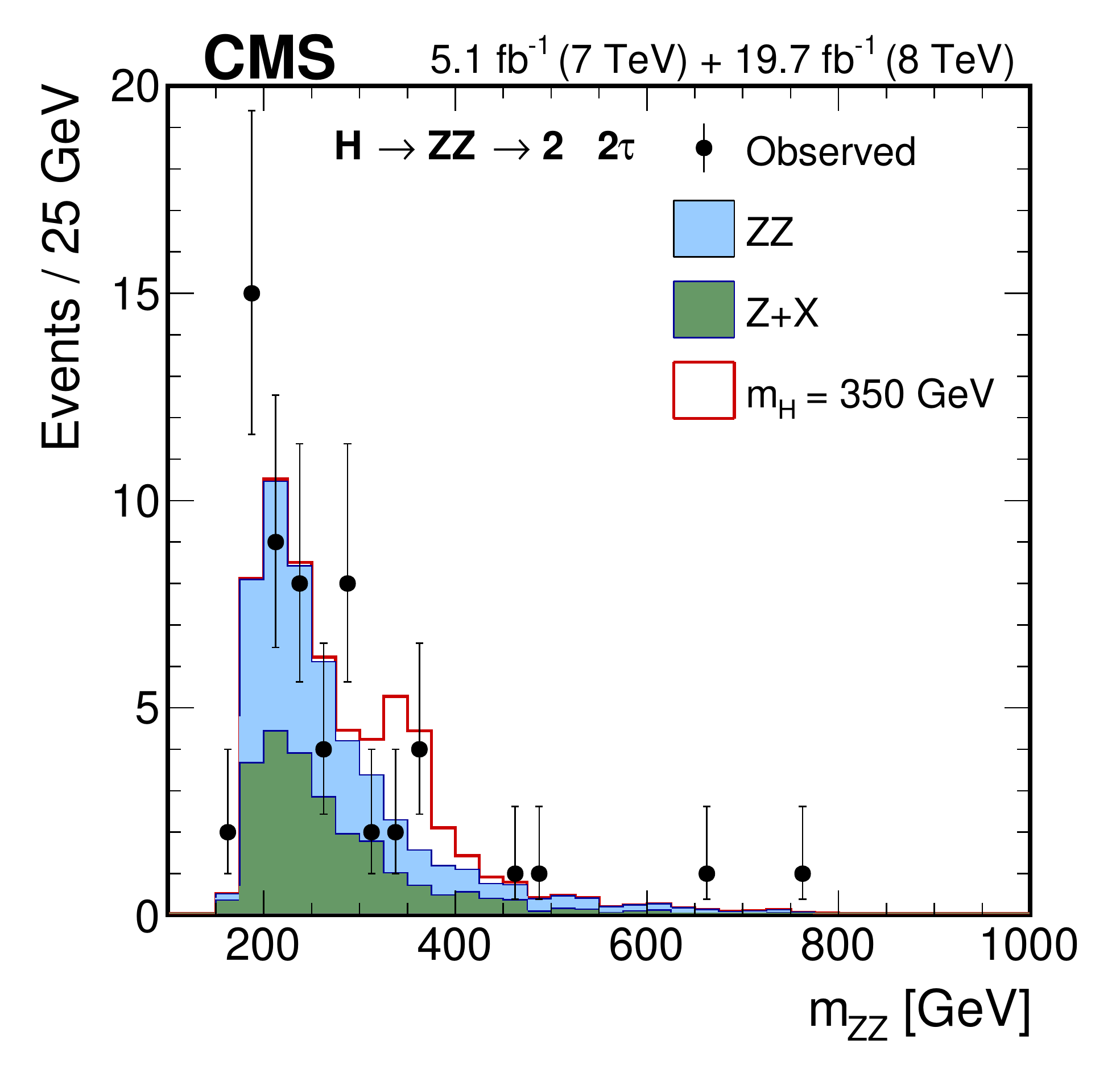}\\
\hspace*{8.5pt}\raisebox{192pt}[0pt][0pt]{\small $\ell$\hspace{6pt}$\ell$\hspace{205pt}$\ell$}\vspace*{-13pt}
\caption{
Distribution of the four-lepton reconstructed mass for the sum of
the $4\Pgm$, $2\Pgm2\Pe$, and $4\Pe$ channels (left), and for the sum over all
$2\ell2\tau$ channels (right).
Points represent the observed data, shaded histograms represent the background,
and the red open histogram shows the expectation for a $\mH = 350$\GeV SM-like Higgs boson.
}
\label{fig:Mass4l-2l2tau}

\end{figure}

The reconstructed invariant mass distribution for $2\ell 2\ell$, combining the $4\Pgm$, $2\Pgm2\Pe$, and $4\Pgm$
channels, is shown in Fig.~\ref{fig:Mass4l-2l2tau} (left), compared with the expectation from SM background processes.
Figure~\ref{fig:Mass4l-2l2tau} (right) plots the reconstructed visible mass distribution for the $2\ell2\Pgt$ selection,
combining all $2\ell2\tau$ final states.

The background shapes are taken from simulation, with rates normalized to the observed data.
The measured $\cPZ$ boson and $\cPZ\cPZ$ masses are underestimated due to the undetected neutrinos in tau decays.
To compensate for this effect we correct the momenta of the visible tau decay products to constrain the $\cPZ$ boson mass to 91.2\GeV.
The correction is applied on the selected events and only affects the final mass shape of the $\cPZ\cPZ$ system.
The observed mass distributions are well-matched to the SM background expectation.

In the $2\ell2\ell$ channel limits on the production of heavy Higgs bosons are extracted using the unbinned four lepton mass distribution
and the correlation of the kinematic discriminant with the Higgs boson mass.
In the case of the tagged 2-jet category a third dimension is introduced using the correlation of the aforementioned VBF discriminant with the four-lepton mass.
In the untagged 0/1-jet category the transverse momentum of the four lepton system is used in place of the VBF discriminant.
For the $2\ell2\tau$ final state, limits are set using the $m_{2\ell2\tau}$ distribution.

\subsection{\texorpdfstring{$\PH \to \ZZ \to 2\ell 2\nu$}{H to ZZ to 2 l 2 nu}}
\label{sec:hzz2l2nu}
This analysis seeks to identify Higgs boson decays to a pair of $\cPZ$ bosons, with one $\cPZ$ boson decaying to neutrinos and the other decaying to leptons. The
analysis strategy is based on selection requirements in the (\MET, \mt) phase space, with selections adjusted for
different $\mH$ hypotheses~\cite{Chatrchyan:2012ft}.
Here, the transverse mass is determined between the transverse dilepton system, $\vecPtell$, and the  $\vecEtm$.

As in the previous Section, we use data collected with the trigger requiring two high-\pt electrons or muons.
Events are required to have a pair of well-identified, isolated leptons of same flavor ($\Pgmp\Pgmm$ or $\Pep\Pem$),
each lepton with $\pt > 20$\GeV, with an invariant mass within a $30$\GeV window centered on the $\cPZ$ boson mass. The $\pt$
of the dilepton system is required to be greater than $55\GeV$. The presence of large \MET (70\GeV or more, depending on \mH) in the
event is also required.

To suppress the $\cPZ$+jets background, events are rejected if the angle in the transverse plane between
the $\vecEtm$ and the closest jet with $\pt > 30$\GeV is smaller than 0.5 radians. Events where the lepton is mismeasured
are rejected if \MET$ > $ 60\GeV and $\Delta\phi$($\ell, \vecEtm$) $ < 0.2$.
The top quark background is suppressed by applying a veto on events having a b tagged jet with $\pt > 30$\GeV and $\abs{\eta} < 2.5$.
To further suppress this background, a veto is applied on events containing an additional, soft muon
with $\pt > 3$\GeV, typically produced in the leptonic decay of a b quark. To reduce the $\PW\cPZ$ background,
in which both bosons decay leptonically, any event with a third lepton ($\Pgm$ or $\Pe$) with $\pt > 10$\GeV, and passing the identification and
isolation requirements, is rejected.

The search is carried out in two mutually exclusive categories. The VBF category contains events with two or more jets
in the forward region, with a $\abs{\Delta\eta_{\mathrm{jj}}}>4$ requirement between the two leading jets, and with the invariant mass of those two
jets greater than $500$\GeV. In addition, the two leptons forming the $\cPZ$ candidate are required to lie between
these two jets in $\eta$, while no other selected jets with $\pt > 30$\GeV are allowed in this region.
The ggF category includes all events failing the VBF selection, and is subdivided into subsamples according to the presence or absence of
reconstructed jets with $\pt > $ 30\GeV. The event categories are chosen in order to maximize the expected cross section
limit. In the case of the VBF category, a constant $\MET>70$\GeV and no $\mt$ requirement are used, as no gain in sensitivity is obtained with a
selection dependent on the Higgs boson mass hypothesis. In the case of ggF, we apply an $\mH$-dependent lower limit on \MET ranging from 80 to 100\GeV over the
search range.

The background composition is expected to vary with the $\mH$ hypothesis. At low-$\mH$, $\cPZ$+jets
and $\ttbar$ processes are the largest contributions, while at $\mH > 400$\GeV the irreducible $\cPZ\cPZ$ and
$\PW\cPZ$ backgrounds dominate. The $\cPZ\cPZ$ and $\PW\cPZ$ backgrounds are estimated from simulation and are
normalized to their respective NLO cross sections. The $\cPZ$+jets background is modeled from a control
sample of events with a single photon produced in association with jets. This procedure yields a good model of
the \MET distribution in $\cPZ$+jets events.

The uncertainty associated with the $\cPZ$+jets background is affected by residual contamination of the
$\Pgg$+jets control sample from processes involving a photon and genuine $\MET$.
We do not explicitly subtract this contamination, but include a shape uncertainty in the final fit, which is allowed to
vary this small residual background between 0 and 100\% of this estimate.

Background processes that do not involve a $\cPZ$ boson (nonresonant background), include $\ttbar$, single top quark production, $\PW$+jets and $\PW\PW$,
and are estimated with a control
sample of DF dilepton events ($\Pgm^{\mp}\Pe^{\pm}$) that pass the full event selection.
This method cannot distinguish between the nonresonant background and a possible contribution from
$\PH \to \PW\PW \to 2\ell 2\cPgn$ events, which are treated as part of the nonresonant background
estimate. The nonresonant background in the $\Pgmp\Pgmm$ or $\Pep\Pem$ final states is estimated by applying a
scale factor to the selected $\Pgm^{\pm}\Pe^{\mp}$ events, obtained from the events in the sidebands of the $\cPZ$ boson peak
($40 < m_{\ell\ell} < 70$\GeV and $110 < m_{\ell\ell} < 200$\GeV). The uncertainty associated with the estimate of this
background is determined to be 25\%. No significant excess of events is observed over
the SM background expectation.

\begin{figure}[ht]
\centering
\includegraphics[width=0.328\textwidth]{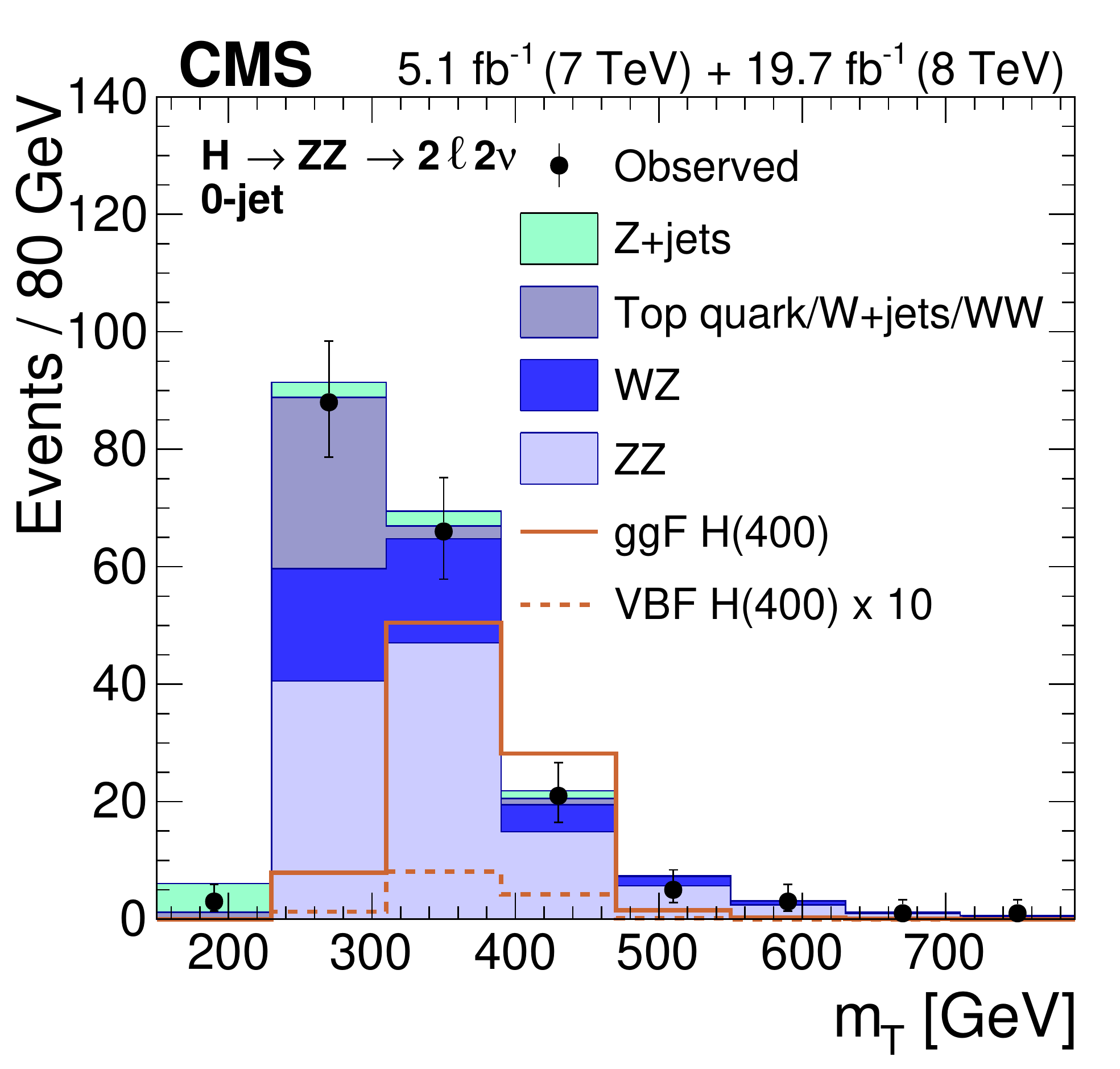}
\includegraphics[width=0.328\textwidth]{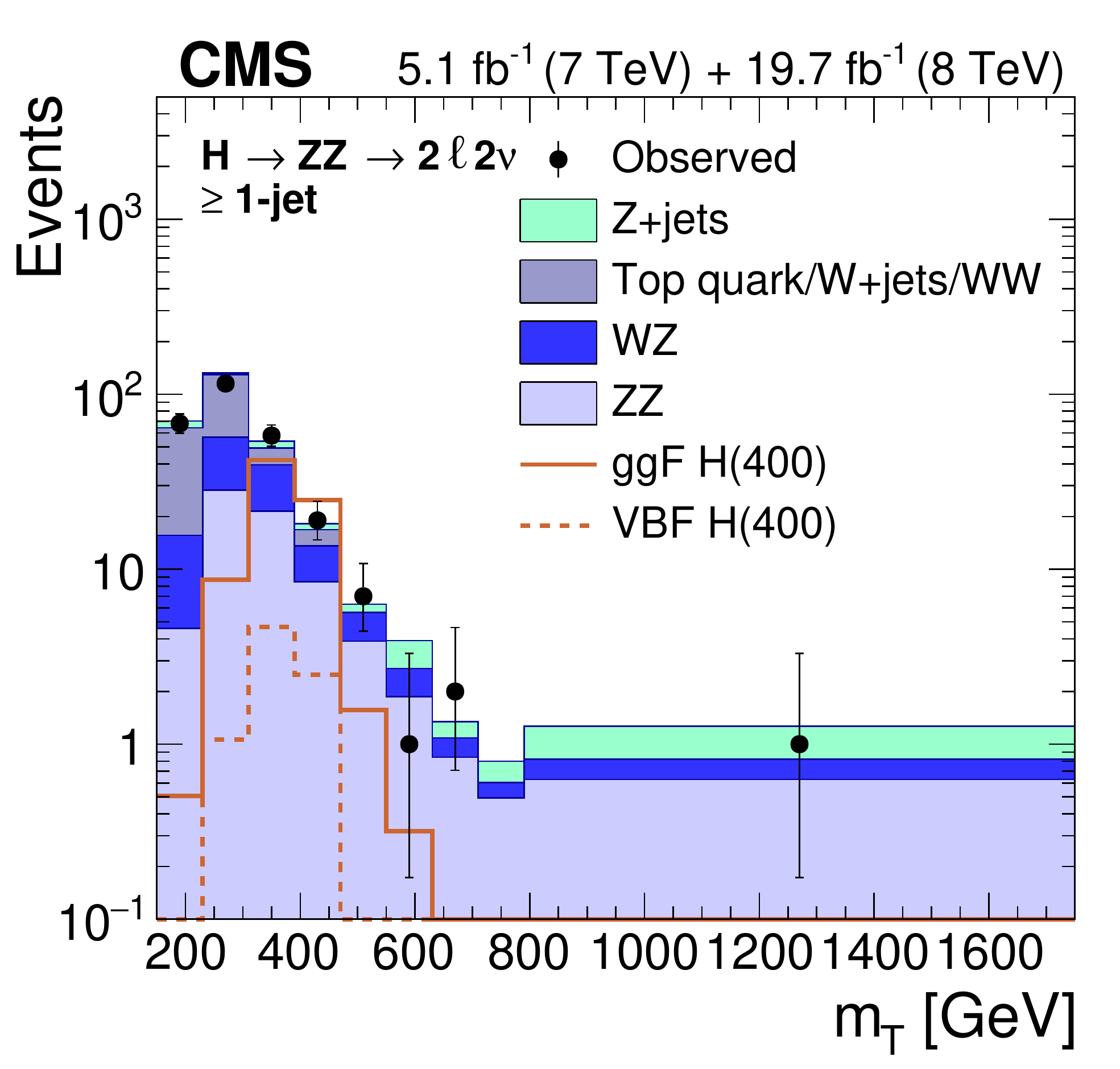}
\includegraphics[width=0.328\textwidth]{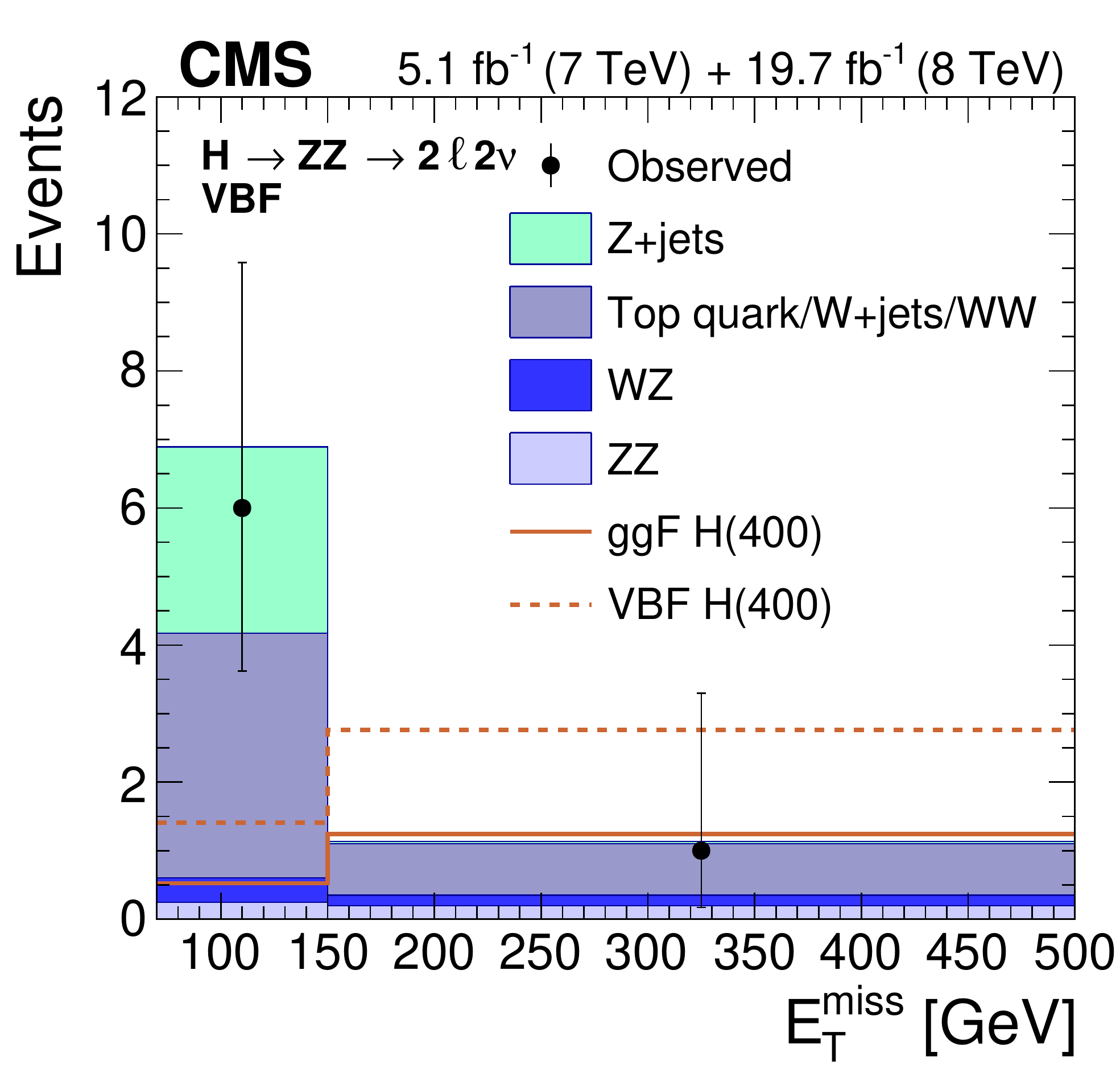}
\caption{The final transverse mass (left, center) and missing
  transverse energy (right) distributions are shown for all three event categories of the
  $\PH\to\cPZ\cPZ\to 2\ell 2\nu$ channel: (left) events with zero jets,
  (center) events with at least one jet, but not passing the VBF selection, (right) VBF events.
  The expected  distributions from the different background processes
  are stacked on top of each other. The red open histograms show the expectation for a $\mH = 400$\GeV SM-like Higgs boson separating the ggF and VBF contributions.
  In the case of the 0-jet category, the VBF contribution is multiplied by a factor of 10 to increase visibility.
\label{fig:hzz2l2nu-final_shapes}
}

\end{figure}

The $\mt$ and \MET\ distributions used to extract the results are shown
in Fig.~\ref{fig:hzz2l2nu-final_shapes} for all three event categories with the muon and electron channels combined.

\subsection{\texorpdfstring{$\PH \to \ZZ \to 2\ell 2\cPq$}{H to ZZ to 2 l 2q}}
\label{sec:hzz2l2q}
The $\PH \to \ZZ \to 2\ell 2\cPq$ channel has the largest branching fraction of all $\mathrm{H} \to \ZZ$ channels under consideration, but also a large background
contribution from $\cPZ$+jets production. Furthermore, the reconstruction of the
hadronically decaying $\cPZ$ boson is more difficult than the fully leptonic final states.
The dominant background is due to the DY process and can be reduced using b-tagging requirements on the selected jets, as will be described later in
this section.

As in the other channels with a \cPZ $ $ boson in the final state, we use data collected with the trigger requiring two high-\pt muons or electrons.
Reconstructed muons and electrons are required to have $\pt > 40 (20)$\GeV for the leading (second) lepton.
Muons are required to have $\abs{\eta} < 2.4$, and electrons $\abs{\eta} < 2.5$. Jets are required to have
$\pt > 30$\GeV and $\abs{\eta} < 2.4$. Each pair of oppositely charged leptons of the same flavor and each pair of jets
are considered as $\cPZ$ boson candidates.

To increase the sensitivity to a possible signal, the main analysis is
complemented with dedicated selections for the VBF signature. VBF events are identified requiring two additional jets with $\pt > 30$\GeV,
$m_{\mathrm{jj}} > 500$\GeV, and $|\Delta\eta_{\mathrm{jj}}| > 3.5$.

Analogously to the $\PH \to \WW \to \ell\nu\Pq\Pq$ channel, the analysis is modified for very high
Higgs boson masses ($\mH > 600$\GeV) to account for the fact that the two jets originate from a Lorentz-boosted
$\cPZ$ boson and therefore they may be reconstructed as a single, merged
jet (Sections~\ref{sec:reconstruction} and~\ref{sec:hwwlnuJ}).
Information about the internal structure of this kind of jet~\cite{CMS-PAS-JME-13-006} is used in
order to gain some insight about the origin of the jet, distinguishing the
DY background from jets produced from boosted $\cPZ$ bosons.
To reduce the contamination from nonboosted $\cPZ+X$ backgrounds, the selection for the merged-jet topology requires the hadronically decaying
$\cPZ$ boson to have $\pt > 100$\GeV and $\abs{\eta} < 2.4$, and the leptonically decaying $\cPZ$ boson to have $\pt > 200$\GeV.

In order to exploit the different jet composition of signal and background, events are classified into three
mutually exclusive categories according to the number of selected
$\cPqb$ tagged jets: 0 b tag, 1 b tag, and 2 b tag.
This distinction is not done in the VBF oriented selection where the main
discrimination is given by a multivariate discriminator for VBF topology based on angular and energy information of the two VBF tag jets.

Background contributions are reduced by requiring $71 < \mjj < 111$\GeV and
$76 < \mll < 106$\GeV in the selected events. The presence of $\cPZ$ bosons
decaying to leptons and dijets makes this selection very efficient for signal, whereas the
continuous background gets largely reduced. In case of the merged-jet analysis, the dijet requirement is applied
on the merged-jet mass, after applying the pruning procedure.

An angular likelihood discriminant is used to separate signal-like from background-like events in each category~\cite{Gao:2010qx,2l2qpaper}.
In order to suppress the substantial expected $\ttbar$ background in the 2 b tag category, a discriminant
is used, defined as the logarithm of the likelihood ratio of the hypothesis that the $\MET$ is equal to the value measured by
the PF algorithm and the null hypothesis $\MET = 0$~\cite{Chatrchyan:2011tn}. This discriminant provides a measure of
whether the event contains genuine missing transverse energy.
When an event contains multiple $\cPZ$ boson candidates passing the selection requirements,
those with jets in the highest b tag category are retained for analysis. If
multiple candidates are still
present, the ones with $\mjj$ and $\mll$ values closest to the $\cPZ$ boson mass are
retained. In the case of the merged-jet category all the requirements on jets with respect to b tagging and the
likelihood discriminant are applied to the two subjets reconstructed inside
the merged jet. In case of events with a merged-jet $\cPZ$ boson candidate in addition to a
dijet one, the candidate from the merged jet gets selected.

The dominant $\cPZ$+jets background is estimated using simulated
events properly corrected to reproduce the yield of the observed data in the control regions, defined
as $60 < \mjj < 71$\GeV and $111 < \mjj < 130$\GeV. Simulated events are weighted to reproduce the $\pt$
spectrum of the $\ell\ell$jj system in these control regions. In case of the merged-jet category both the $\pt$ of the $\ell\ell$ and $\ell\ell$J
systems have weights applied. The normalization of this distribution is taken from observed data
and used as an additional constraint for this background.

Other backgrounds are estimated using simulated
events. These include diboson and top quark events. In the
VBF selection, the background from top quark events is small. In the ggF analysis it is estimated from observed data using a
control sample of e$\mu$ events, invoking lepton flavor symmetry that is satisfied
by several background contributions but not in the signal events.

\begin{figure}[htbp]
\centering
\includegraphics[width=0.328\textwidth]{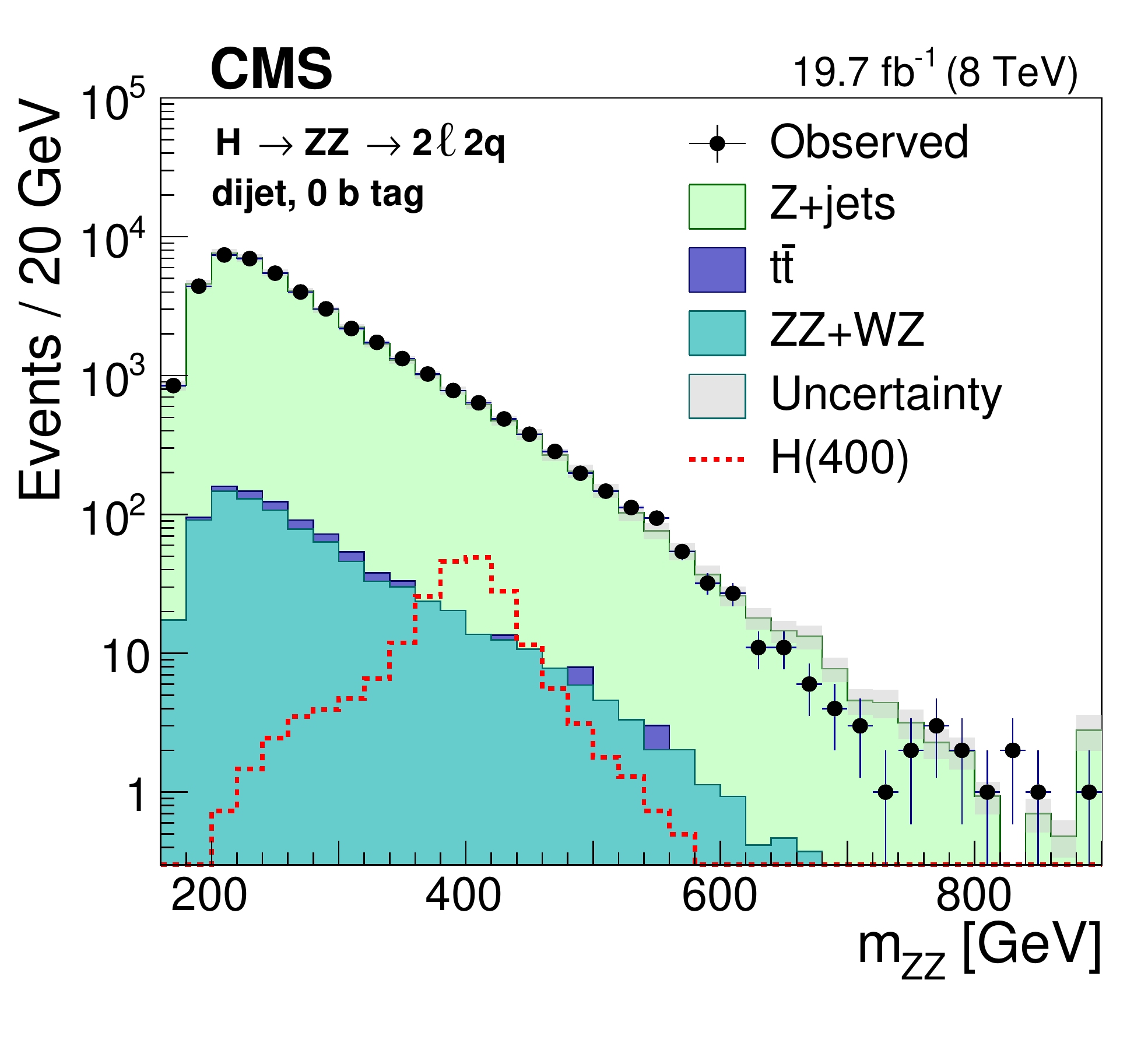}
\includegraphics[width=0.328\textwidth]{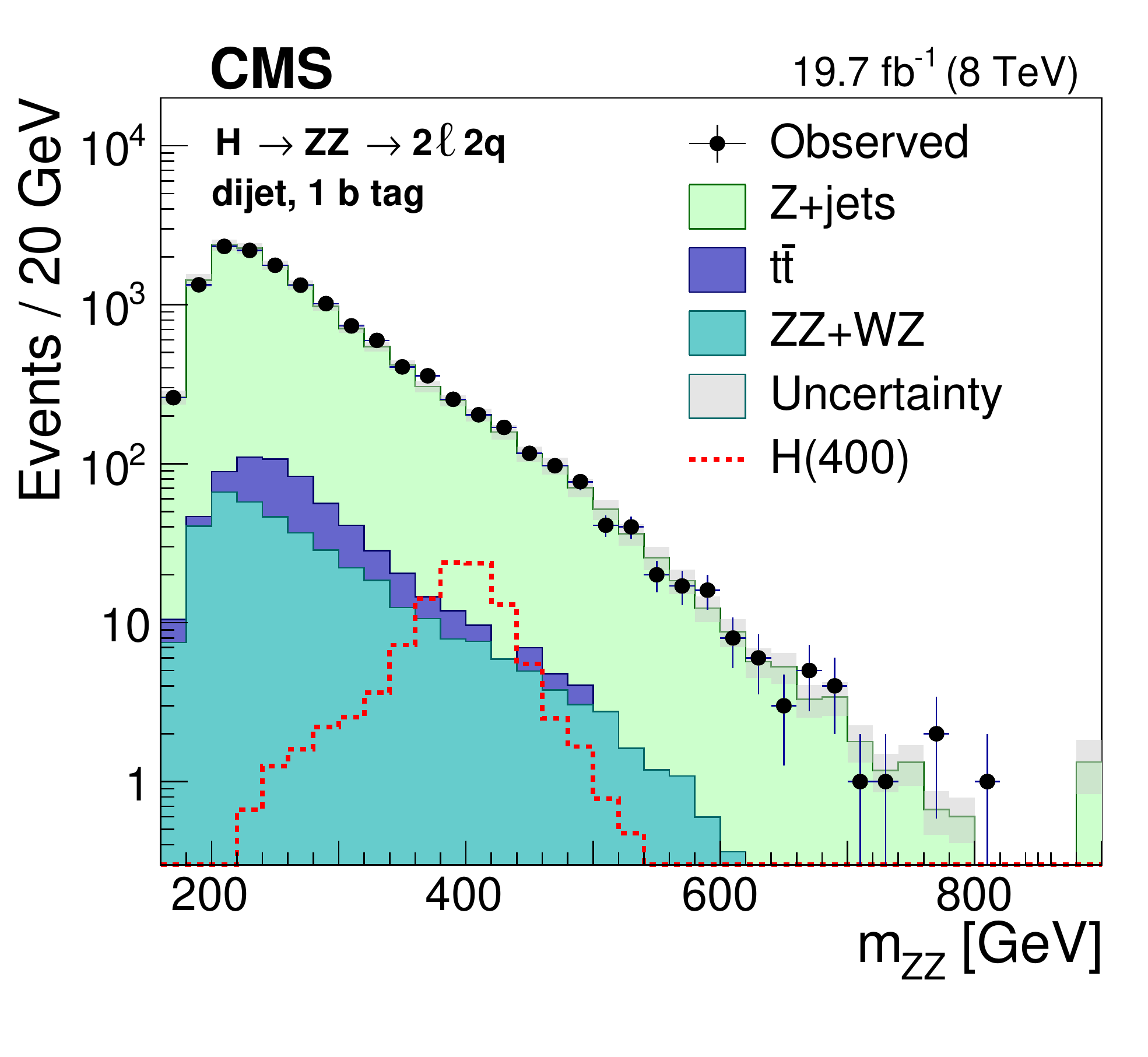}
\includegraphics[width=0.328\textwidth]{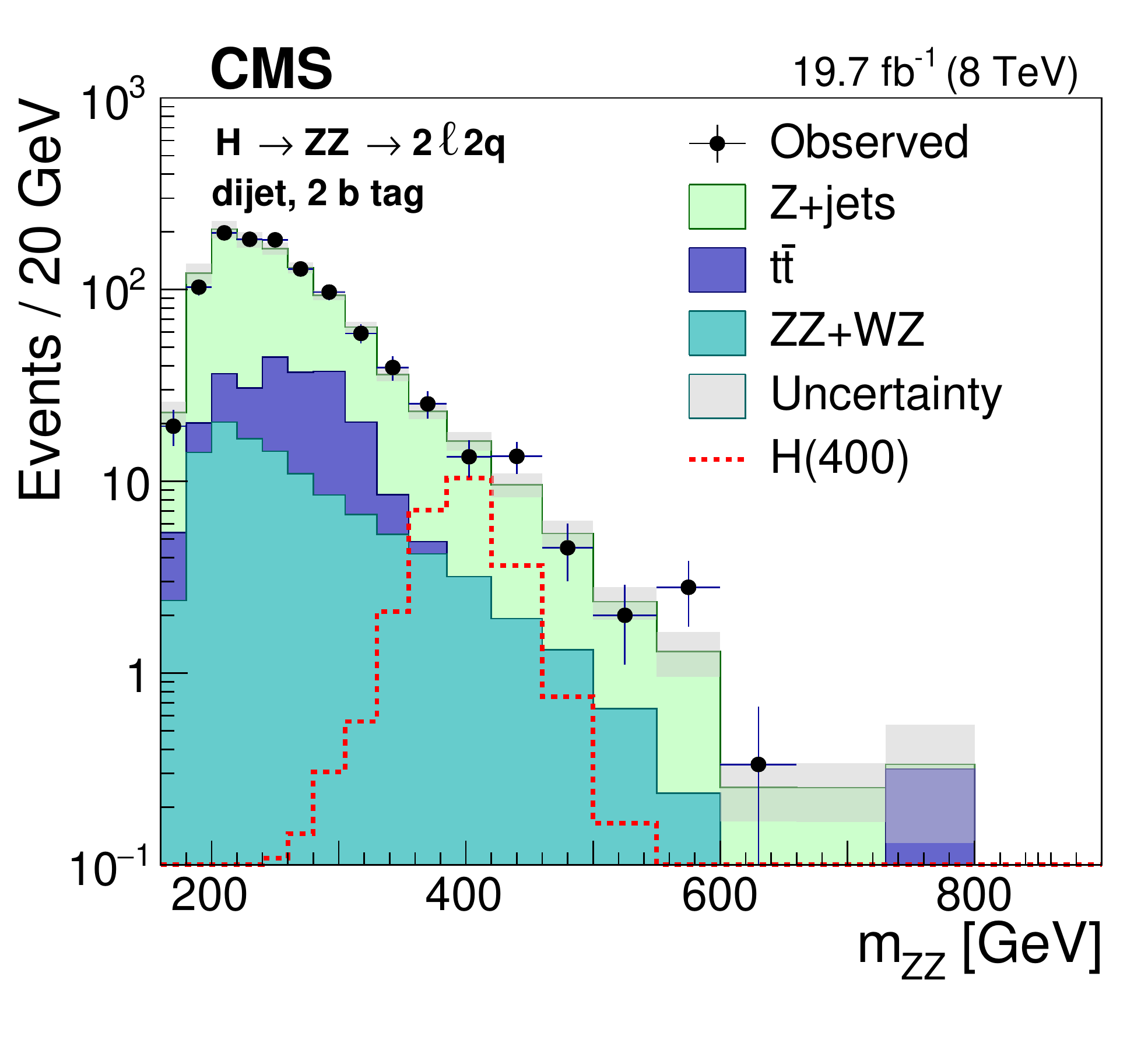}\\
\caption{
The $\mZZ$ invariant mass distribution after final selection in three categories of the $\PH\to\cPZ\cPZ\to 2\ell$2q dijet channel:
0 b tag (left),
1 b tag (center), and
2 b tag (right).
Points with error bars show distributions of observed data.
Solid histograms are depicting the background expectation from
simulated events for the different components.
The red open histogram shows the expectation for a $\mH = 400$\GeV SM-like Higgs boson.
}
\label{fig:mZZ_kinfit_hiMass}

\end{figure}

The distributions of $m_{\Z\Z}$ in the signal region are shown in Fig.~\ref{fig:mZZ_kinfit_hiMass}
for the three b tag
multiplicities of the dijet category comparing observed data with background expectations.
Good agreement is observed within the
uncertainties. For the merged and VBF categories good agreement is also observed.
The dominant systematic uncertainties are due to the b-tagging performance and the JES~\cite{CMS-PAS-JME-10-003}. Further systematic effects are due to the uncertainty in the predicted signal
and background shapes used in the analysis.
The distributions of $\mZZ$ in the signal region are used to extract the final
signal and background yields.

\subsection{Systematic uncertainties}
\label{sec:syst_unc}

Systematic uncertainties for the various final states come from common treatment of the signal model assumptions,
reconstructed objects used in the analysis and a few common experimental effects.

Uncertainties on the cross section for the production of heavy Higgs bosons arise from
uncertainties in the combined choice of the PDFs and $\alpha_s$, as well as in the renormalization and factorization
scales~\cite{Heinemeyer:2013tqa},
which are typically 6--7\% and 7--12\%, respectively, for the ggF
production mechanism, and 1--2\% and 2--5\%, respectively, for production via VBF.
Additionally, we add an uncertainty in the background coming from off-shell \Ph{}(125) production, which we estimate
with \textsc{gg2zz} (\textsc{Phantom}) for the ggF (VBF) case.
We find that at the largest  $\mH$ values, the size of the effect is approximately 3\% of the total background.
Uncertainties on the signal lineshape reweighting with interference varies for the ggF and VBF modes.
For ggF, we follow the prescription in Ref.~\cite{Heinemeyer:2013tqa}, which considers the NNLO contribution to
the signal interfering with the $\Pg\Pg \to \cPZ\cPZ$ background process.
For VBF, without a full prescription, we assign systematic uncertainties coming from renormalization and factorization scale
variations in the \textsc{Phantom} generator.

Other common systematic effects come from the luminosity uncertainty, which is 2.2\% (2.6\%) for the 7 (8)\TeV data.
Uncertainties on the muon and electron reconstruction efficiencies, and JES and jet energy resolution (JER) are
correlated among the various final states, where all these effects are subdominant.
The lepton fake rate is largest for the $\PH\to\cPZ\cPZ\to 2\ell 2\ell'$ channel, in which we consider
fake leptons at relatively lower \pt
than in the other channels.
A summary of the systematic uncertainties per channel is
given in Table~\ref{tab:syst}.

 \begin{table}[h!t]
   \caption{Sources of systematic uncertainties considered in each of the channels included in this
   analysis. Uncertainties are given in percent. Most uncertainties are affecting the normalisation of the
   observed data or simulated events, but some are uncertainties on the shape of kinematic
   distributions. Wherever ranges of uncertainties are given, they are either ranges in $\mH$, jet
   multiplicity categories, or dependent on the production mode.}
   \label{tab:syst}
\small
   \centering
   \begin{tabular}{l|c|c|c|c|c|c}
  \hline
  Source of uncertainty & $\PH\to\PW\PW$ & $\PH\to\PW\PW$ & $\PH\to\PW\PW$ & $\PH\to\cPZ\cPZ$ & $\PH\to\cPZ\cPZ$ & $\PH\to\cPZ\cPZ$ \\
  & $\to\ell\nu\ell\nu$ & $\to\ell\nu$jj & $\to\ell\nu$J & $\to 2\ell 2\ell'$ & $\to 2\ell 2\nu$ & $\to 2\ell$2q \\
  \hline
  Experimental sources \\
  \hline
   Luminosity, 7 (8)\TeV & 2.2 (2.6) & 2.2 (2.6) & 2.2 (2.6) & 2.2 (2.6) & 2.2 (2.6) & 2.2 (2.6) \\
   $\ell$ trigger, reco, id, iso & 1--4 & 1--2 & 1--2 & 0.5--7 & 2--3 & 1.8--2 \\
   $\ell$ mom./energy scale & 2--4 & & & 0.5--30 & 1--2 & 0.1--0.4 \\
   $\ell$ misid. rate & & & & 30 & & \\
   JES, JER, $\MET$ & 2--35 & $<$1 & 2 & 5--30 & 1 & 1--13 \\
   Pileup & & $<$1 & & & 1--3 & 1 \\
   b-tag/mistag &  & & 2.5 & & 1--3 & 1--6 \\
   $\PW$-tag/$\cPZ$-tag & & & 7.5 & & & 0--9.3 \\
   Signal selection eff. & & 10 & 2 & & \\
   Monte Carlo statistics & 1--20 & & & & 1--2 & 0--6 \\
  \hline
  Background estimates \\
  \hline
   $\ttbar$, $\PQt\PW$ & 20 & 7 & 6--30 & & 25 & 0--15 \\
   $\cPZ$+jets & 40--100 & & & 20--42 & 100 & 16 \\
   $\cPZ\cPZ$ & 3 & & & 13--14 & 12 & \\
   $\PW$+jets & 40 & 0.6 & 8 & & 25 & \\
   $\PW\PW$ & 8--30 & 10 & 30 & & 25 & \\
   $\PW\cPZ$, $\PW\gamma^*$ & 3--50 & & 30 & & 5.8--8.5 & \\
  \hline
  Theoretical sources \\
  \hline
   $\sigma(\Pg\Pg\to\PH)$ & 10--13 & 10--11 & 11--13 & 10--13 & 10--13 & 10--13 \\
   $\sigma(\PQq\PQq\to\PH)$ & 2.6--5.8 & 2.6--3.6 & 3.6--5.8 & 2.6--5.8 & 2.6--5.8 & 2.6--5.8 \\
   $\PH$ lineshape & & & & 5 & 2--8 & 0--7 \\
   $\PH$--$\PW\PW$ (ZZ) interference & 1--27 & & 10--50 & & 10--50 & \\
   Jet binning & 7--35 & & 7--35 & & 30 & \\
  \hline
   \end{tabular}
   
 \end{table}

\section{Statistical interpretation}
\label{sec:results}
The combination of the measurements in the different channels presented in this paper requires the simultaneous analysis
of the data selected by all
individual analyses, accounting for all statistical and systematic uncertainties, as well as their correlations.
The statistical methodology used in this
combination was developed by the ATLAS and CMS Collaborations in the context of the LHC Higgs Combination
Group~\cite{CMSlongpaper,LHC-HCG-Report,CMScombFeb2012}.
Upper limits on the model parameters are set for different $\mH$ hypotheses
using a modified frequentist method referred to as the CL$_\mathrm{s}$ method~\cite{Read:451614,Junk:1999kv},
where a likelihood ratio test statistic is used in which the nuisance parameters are profiled.
In the likelihood ratio, the total number of observed events is compared to the signal and background predictions
by means of a product of Poisson probabilities.
The predictions are subject to the multiple uncertainties described in
the previous section, and are handled by introducing nuisance
parameters with probability density functions.
The nuisance parameters modify parametrically the expectations for both signal and
background processes.
Furthermore, a signal strength modifier ($\mu$) is used
to scale the Higgs boson cross sections of all production
mechanisms by the same factor with respect to their SM predictions
while keeping the decay branching fractions unchanged.

\subsection{SM-like Higgs boson search}
\label{sec:SM_results}
\begin{figure}[ht]
{\centering
\includegraphics[width=0.98\textwidth]{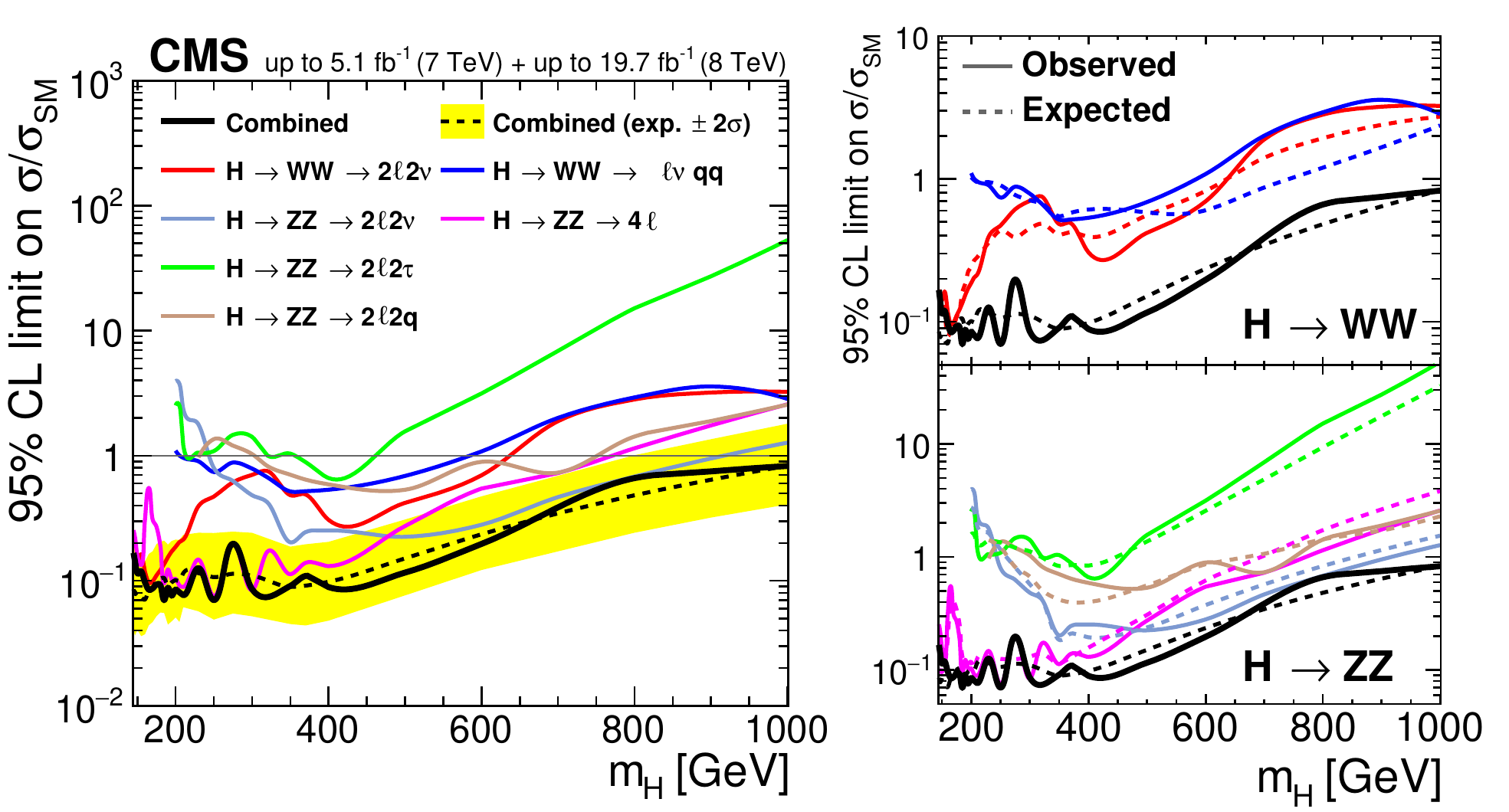}\\
\caption{
Upper limits at the 95\% CL for each of the contributing final states and their combination.
The theoretical cross section, $\sigma_\mathrm{SM}$, is
computed in Ref.~\cite{Heinemeyer:2013tqa}. The observed and expected limits of the six individual channels are compared with each other and with the combined results (right),
for $\PH\to\PW\PW$ channels (top right panel) and $\PH\to\cPZ\cPZ$ channels (bottom right panel) separately.}
\label{fig:abslimits_SM}
}
\end{figure}

The combined results obtained for a heavy Higgs boson with SM-like couplings
for all the different contributing final states are displayed in Fig.~\ref{fig:abslimits_SM}.
On the left, the observed 95\% CL limit is shown for each final state.
The expected combined 95\% CL limit of the six channels is plotted as a dashed black line, while the yellow shaded region is the
${\pm}2\sigma$ uncertainty in the expected limit.
On the right, the expected and observed limits are displayed for each of the individual channels as well as the combined result.
The top right panel shows the \PW\PW~final states, while the \cPZ\cPZ~final states are displayed in the bottom right panel.
In the lower mass region of the search range, the most sensitive channels are $\PH\to\cPZ\cPZ\to 4\ell$
and $\PH\to\PW\PW\to 2\ell 2\nu$.
At the highest masses, the $\PH\to\cPZ\cPZ\to 2\ell 2\nu$ channel has the best sensitivity, while
$\PH\to\cPZ\cPZ\to 4\ell$, $\PH\to\PW\PW\to 2\ell 2\nu$, $\PH\to\PW\PW\to\ell\nu$qq,
and $\PH\to\cPZ\cPZ\to 2\ell$2q contribute significantly.

Features in the combined observed limit can be traced to corresponding features in the limits of the individual channels.
At lower masses below 400\GeV, there are oscillations in the observed limit due to the high resolution channel, $\PH\to\cPZ\cPZ \to 4\ell$, and the narrow
width of the heavy Higgs boson in this mass range.
An excess in the combined limit at around 280\GeV is related to a small excess in the channels $\PH\to\cPZ\cPZ \to 4\ell$ and $\PH\to\PW\PW\to2\ell2\nu$.
The small excess of observed events seen around 700\GeV in the
$\PH\to\PW\PW\to\ell\nu$J merged-jet category is not supported by the other channels and is reduced to less than $0.5\,\sigma$ in the combination.
The combined upper limits at the 95\% CL on the product of the cross section and branching
fractions exclude a Higgs boson with SM-like couplings in the full search range $145 < \mH <  1000$\GeV.

\subsection{EW singlet Higgs boson search}
\label{sec:BSM_results}
\begin{figure}[ht]
\centering
\includegraphics[width=0.98\textwidth]{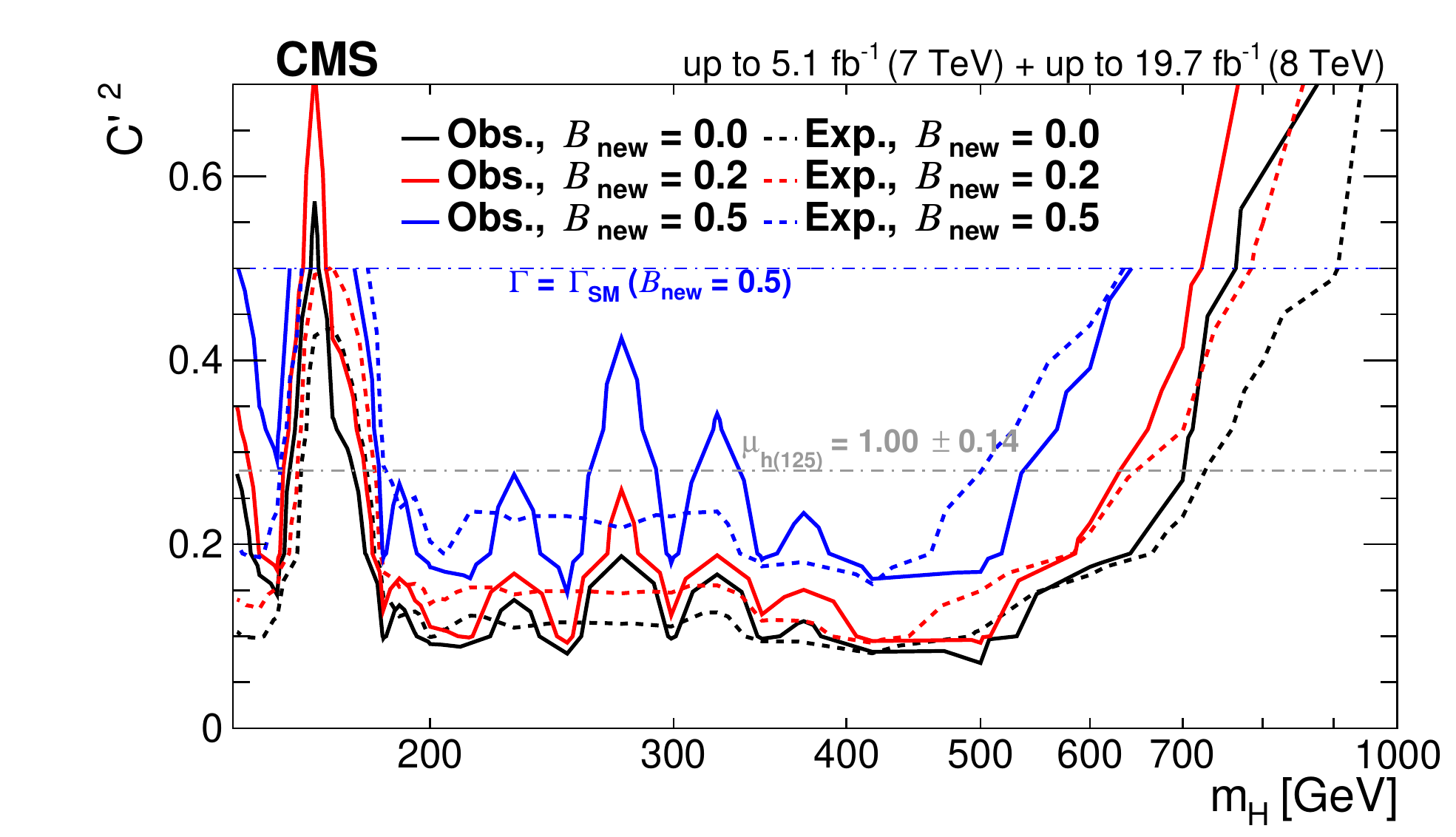}
\caption{
Upper limits at the 95\% CL on the EW singlet extension.
Upper limits are displayed as a function of the heavy Higgs boson mass and the model parameter $C'^2$ for
different values of $\mathcal{B}_\text{new}$. The upper dash-dotted line indicates where, for $\mathcal{B}_\text{new} = 0.5$, the variable width of
the heavy Higgs boson reaches the width of a SM-like Higgs boson. The lower dash-dotted line displays the indirect limit at 95\% CL on $C'^2$ from the
measurement of \Ph{}(125).
}
\label{fig:cpsqVmass}
\end{figure}

We interpret the search results in terms of a heavy Higgs boson in the EW singlet extension of the SM.
The parameters of the model are $C'^2$, the heavy Higgs boson contribution to electroweak symmetry breaking,
and $\mathcal{B}_\text{new}$, the contribution to the Higgs boson width of non-SM decays, and are defined in
Section~\ref{sec:sim}.
Figure~\ref{fig:cpsqVmass} shows the expected and observed upper limits at 95\% CL on the singlet scalar cross
section with respect to its expected cross section as a function of its mass.

The region above the curves shows the parameter space that is expected to be excluded at 95\% CL.
We show the exclusion region for various values of $\mathcal{B}_\text{new}$.
We find a large region of $C'^2$ versus mass parameters to be excluded for various values of $\mathcal{B}_\text{new}$.
We also plot the $\mu_{\Ph{}(125)} = 1 \pm 0.14$~\cite{CMS-PAS-HIG-14-009} indirect constraint $C'^2 < 0.28$ at 95\% CL for $\mathcal{B}_\text{new}$=0.
The upper dash-dotted line shows the cutoff of the allowed region for $\mathcal{B}_\text{new} = 0.5$ where the width of the
heavy Higgs boson becomes larger than the SM width at  that mass hypothesis.

\begin{figure}[ht]
{\centering
\includegraphics[width=0.49\textwidth]{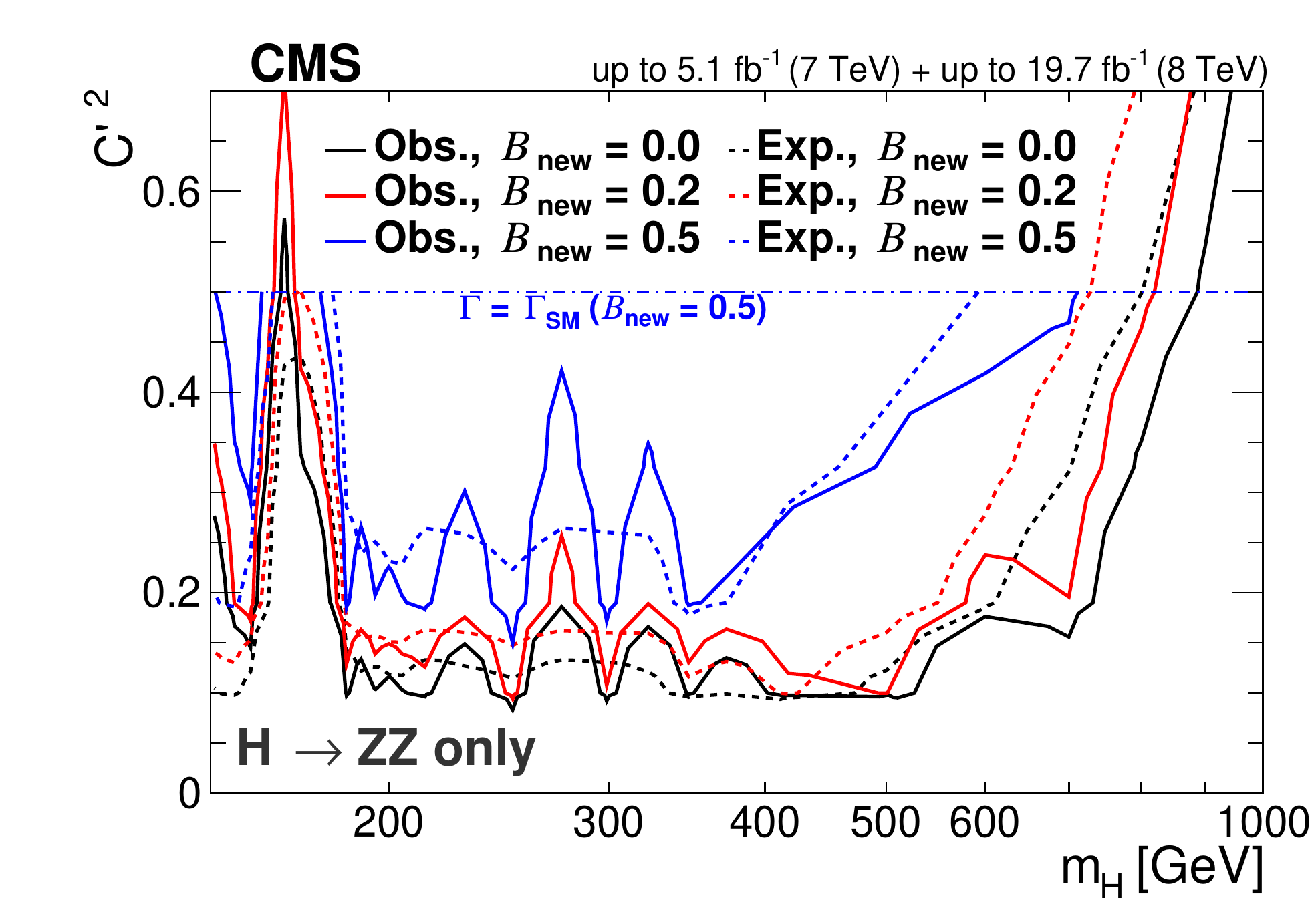}
\includegraphics[width=0.49\textwidth]{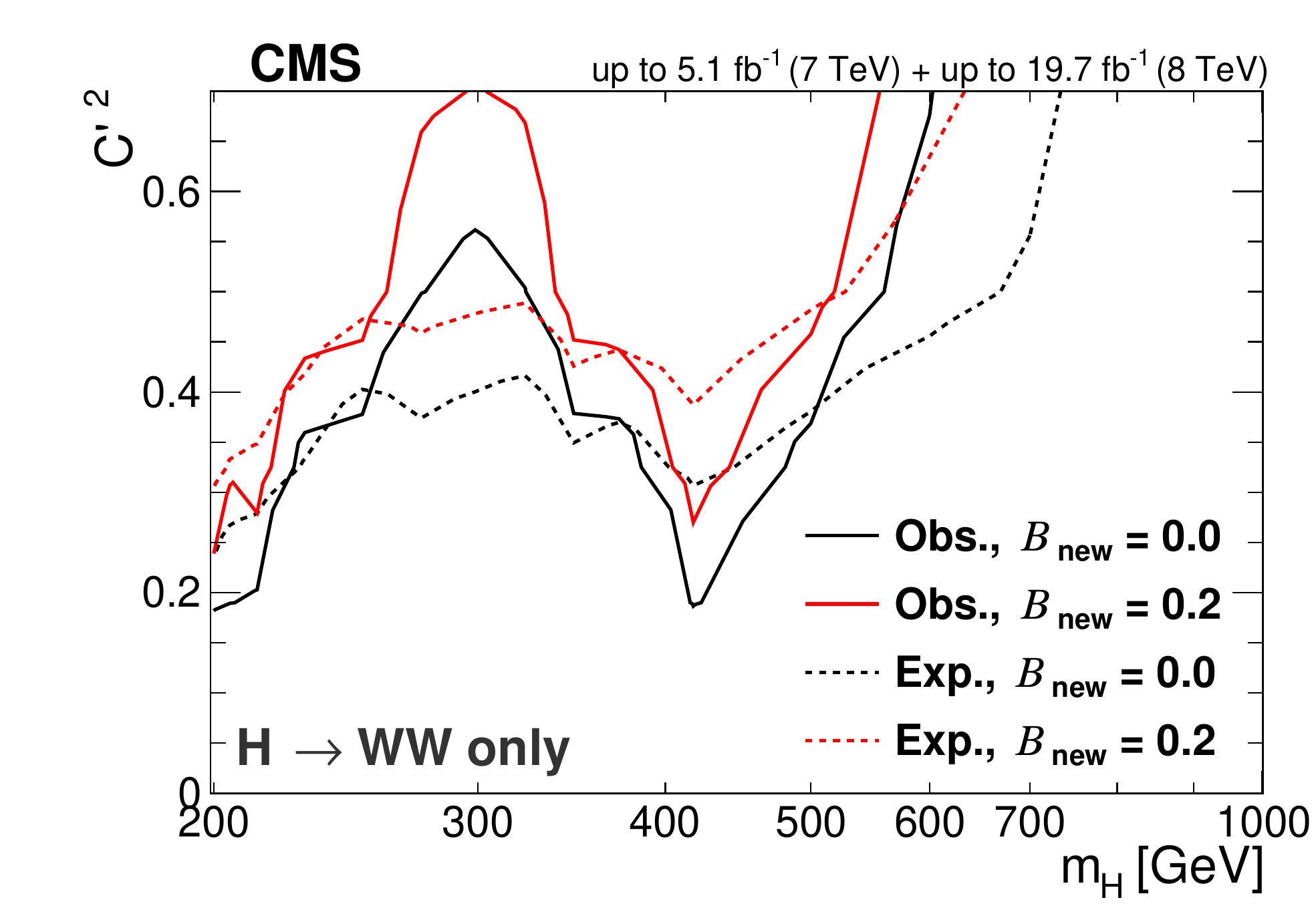}\\
\includegraphics[width=0.49\textwidth]{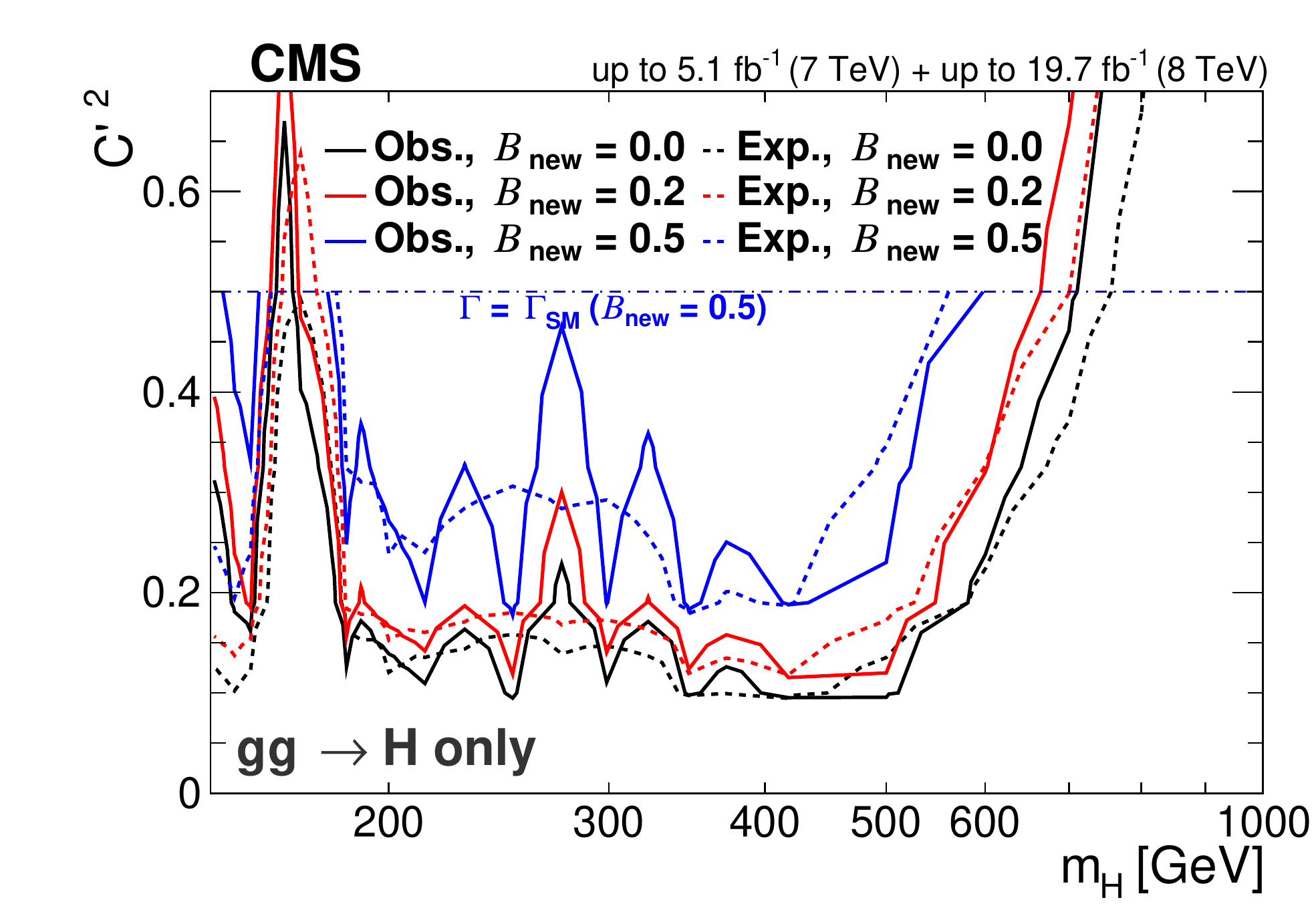}
\includegraphics[width=0.49\textwidth]{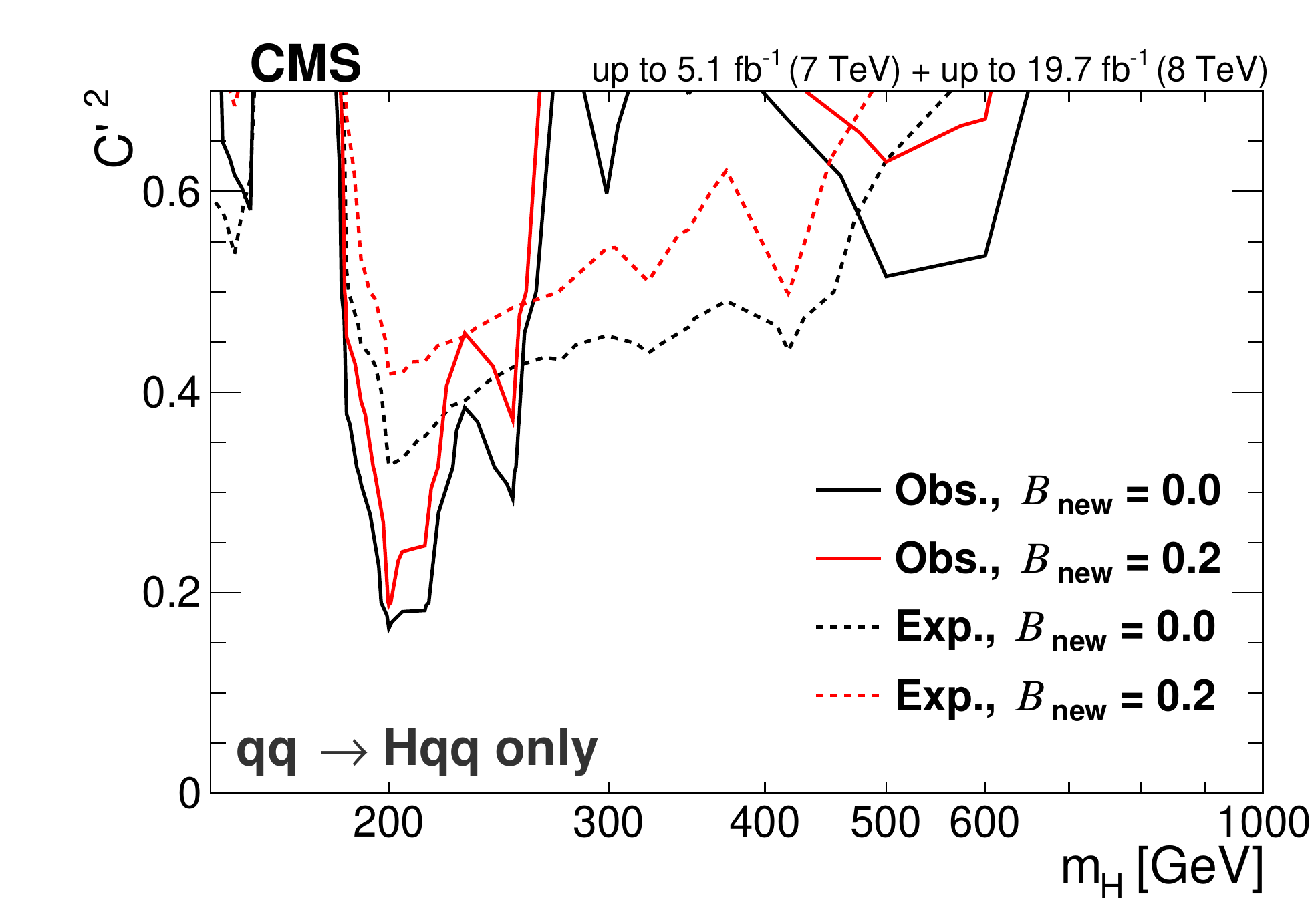}\\
\caption{
Upper limits at the 95\% CL on the EW singlet extension.
Upper limits are displayed as a function of the heavy Higgs boson mass and the model parameter $C'^2$ for
different values of $\mathcal{B}_\text{new}$.
All considered $\cPZ\cPZ$ decay channels combined (top left). All considered $\PW\PW$ decay channels combined (top right).
Limits for the ggF production mode only (bottom left). Limits for VBF production only (bottom right).
The dash-dotted line in the two left plots indicates where,
for $\mathcal{B}_\text{new} = 0.5$, the variable width of
the heavy Higgs boson reaches the width of a SM-like Higgs boson.
}
\label{fig:cpsqVmass_split}
}
\end{figure}

In order to understand the constraints of these results in a model-independent approach, we further subdivide the
results into categories.
In Fig.~\ref{fig:cpsqVmass_split} we show the limits in various configurations.
At the top of Fig.~\ref{fig:cpsqVmass_split} are the limits we obtain when we combine the ZZ (top left) and WW (top right)
channels separately.
Since the ZZ channels are more sensitive in the search for a Higgs boson with SM-like couplings, they better constrain the BSM case as well.
The bottom of Fig.~\ref{fig:cpsqVmass_split} shows the combined 95\% CL for all final states but only the
ggF or VBF production mechanism for the heavy Higgs boson.
In the heavy Higgs boson with SM-like couplings scenario, we assume the ratio of the cross sections for various production mechanisms
to be the same as in the SM case.

\clearpage
\section{Summary}
\label{sec:summary}

Combined results are presented from searches for a heavy Higgs boson in $\PH \to \WW$ and $\PH \to \ZZ$
decay channels, for Higgs boson mass hypotheses in the range $145 < \mH < 1000$\GeV.
In the case of a Higgs boson decaying into a pair of $\PW$ bosons, the
fully leptonic ($\PH \to \WW \to \ell\nu\ell\nu$) and semileptonic ($\PH \to \WW \to \ell \nu
\cPq\cPq$) final states are considered in the analysis. For a Higgs boson decaying into two $\cPZ$
bosons, final states containing four charged leptons ($\PH \to \ZZ \to 2\ell 2\ell'$), two charged leptons and two
quarks  ($\PH \to \ZZ \to 2\ell 2\cPq$), and two charged leptons and two neutrinos
($\PH \to \ZZ \to 2\ell 2\nu$) are considered, where $\ell = \Pe$ or $\Pgm$ and $\ell' = \Pe$, $\Pgm$,
or $\Pgt$.

The observed data are interpreted both in the context of a heavy Higgs boson with SM-like couplings and decays,
as well as a search for a heavy, narrow resonance as an EW singlet partner of the SM Higgs
boson at 125\GeV.
No significant excess over the expected SM background has been observed and exclusion limits have been set.
In the case of the search for a heavy Higgs boson with SM-like couplings and decays, we exclude the existence of such
a heavy Higgs boson over the entire search range of $145 < \mH < 1000$\GeV.
For the EW singlet partner of the SM Higgs, the parameters of the model are $C'^2$,
the heavy Higgs boson contribution to EW symmetry breaking,
and $\mathcal{B}_\text{new}$, the contribution to the Higgs boson width of non-SM decays.
We find that a large part of the $C'^2$ versus mass parameter space is excluded for various values of $\mathcal{B}_\text{new}$.
Additionally, we present limits for the EW singlet model for different production mechanisms, ggF and VBF, and
$\WW$ and $\ZZ$ decay modes separately.

\begin{acknowledgments}
\hyphenation{Bundes-ministerium Forschungs-gemeinschaft Forschungs-zentren} We congratulate our colleagues in the CERN accelerator departments for the excellent performance of the LHC and thank the technical and administrative staffs at CERN and at other CMS institutes for their contributions to the success of the CMS effort. In addition, we gratefully acknowledge the computing centers and personnel of the Worldwide LHC Computing Grid for delivering so effectively the computing infrastructure essential to our analyses. Finally, we acknowledge the enduring support for the construction and operation of the LHC and the CMS detector provided by the following funding agencies: the Austrian Federal Ministry of Science, Research and Economy and the Austrian Science Fund; the Belgian Fonds de la Recherche Scientifique, and Fonds voor Wetenschappelijk Onderzoek; the Brazilian Funding Agencies (CNPq, CAPES, FAPERJ, and FAPESP); the Bulgarian Ministry of Education and Science; CERN; the Chinese Academy of Sciences, Ministry of Science and Technology, and National Natural Science Foundation of China; the Colombian Funding Agency (COLCIENCIAS); the Croatian Ministry of Science, Education and Sport, and the Croatian Science Foundation; the Research Promotion Foundation, Cyprus; the Ministry of Education and Research, Estonian Research Council via IUT23-4 and IUT23-6 and European Regional Development Fund, Estonia; the Academy of Finland, Finnish Ministry of Education and Culture, and Helsinki Institute of Physics; the Institut National de Physique Nucl\'eaire et de Physique des Particules~/~CNRS, and Commissariat \`a l'\'Energie Atomique et aux \'Energies Alternatives~/~CEA, France; the Bundesministerium f\"ur Bildung und Forschung, Deutsche Forschungsgemeinschaft, and Helmholtz-Gemeinschaft Deutscher Forschungszentren, Germany; the General Secretariat for Research and Technology, Greece; the National Scientific Research Foundation, and National Innovation Office, Hungary; the Department of Atomic Energy and the Department of Science and Technology, India; the Institute for Studies in Theoretical Physics and Mathematics, Iran; the Science Foundation, Ireland; the Istituto Nazionale di Fisica Nucleare, Italy; the Ministry of Science, ICT and Future Planning, and National Research Foundation (NRF), Republic of Korea; the Lithuanian Academy of Sciences; the Ministry of Education, and University of Malaya (Malaysia); the Mexican Funding Agencies (CINVESTAV, CONACYT, SEP, and UASLP-FAI); the Ministry of Business, Innovation and Employment, New Zealand; the Pakistan Atomic Energy Commission; the Ministry of Science and Higher Education and the National Science Centre, Poland; the Funda\c{c}\~ao para a Ci\^encia e a Tecnologia, Portugal; JINR, Dubna; the Ministry of Education and Science of the Russian Federation, the Federal Agency of Atomic Energy of the Russian Federation, Russian Academy of Sciences, and the Russian Foundation for Basic Research; the Ministry of Education, Science and Technological Development of Serbia; the Secretar\'{\i}a de Estado de Investigaci\'on, Desarrollo e Innovaci\'on and Programa Consolider-Ingenio 2010, Spain; the Swiss Funding Agencies (ETH Board, ETH Zurich, PSI, SNF, UniZH, Canton Zurich, and SER); the Ministry of Science and Technology, Taipei; the Thailand Center of Excellence in Physics, the Institute for the Promotion of Teaching Science and Technology of Thailand, Special Task Force for Activating Research and the National Science and Technology Development Agency of Thailand; the Scientific and Technical Research Council of Turkey, and Turkish Atomic Energy Authority; the National Academy of Sciences of Ukraine, and State Fund for Fundamental Researches, Ukraine; the Science and Technology Facilities Council, UK; the US Department of Energy, and the US National Science Foundation.

Individuals have received support from the Marie-Curie program and the European Research Council and EPLANET (European Union); the Leventis Foundation; the A. P. Sloan Foundation; the Alexander von Humboldt Foundation; the Belgian Federal Science Policy Office; the Fonds pour la Formation \`a la Recherche dans l'Industrie et dans l'Agriculture (FRIA-Belgium); the Agentschap voor Innovatie door Wetenschap en Technologie (IWT-Belgium); the Ministry of Education, Youth and Sports (MEYS) of the Czech Republic; the Council of Science and Industrial Research, India; the HOMING PLUS program of the Foundation for Polish Science, cofinanced from European Union, Regional Development Fund; the Compagnia di San Paolo (Torino); the Consorzio per la Fisica (Trieste); MIUR project 20108T4XTM (Italy); the Thalis and Aristeia programs cofinanced by EU-ESF and the Greek NSRF; and the National Priorities Research Program by Qatar National Research Fund.
\end{acknowledgments}

\bibliography{auto_generated}

\providecommand{\href}[2]{#2}\begingroup\raggedright\begin{thebibliography}{100}%
\makeatletter
\providecommand{\hrefCMSnoop }[0]{\@secondoftwo}%
\makeatother
\providecommand{\doi}{\texttt{doi:}\begingroup \urlstyle{tt}\Url}

\bibitem{StandardModel67_1}
\hrefCMSnoop {}{S.~L. Glashow, ``{Partial-symmetries of weak interactions}'',}
  \textit{ Nucl. Phys.} \textbf{ 22} (1961) 579,
\href{http://dx.doi.org/10.1016/0029-5582(61)90469-2}{\doi{10.1016/0029-5582(61)90469-2}}.

\bibitem{StandardModel67_2}
\hrefCMSnoop {}{S.~Weinberg, ``{A Model of Leptons}'',} \textit{ Phys. Rev.
  Lett.} \textbf{ 19} (1967) 1264,
\href{http://dx.doi.org/10.1103/PhysRevLett.19.1264}{\doi{10.1103/PhysRevLett.19.1264}}.

\bibitem{StandardModel67_3}
\hrefCMSnoop {}{A.~Salam, ``Weak and electromagnetic interactions'',} in
  \textit{ Elementary particle physics: relativistic groups and analyticity},
  N.~Svartholm, ed., p.~367.
\newblock Almqvist \& Wiksell, 1968.
\newblock Proceedings of the eighth Nobel symposium.

\bibitem{Englert:1964et}
\hrefCMSnoop {}{F.~Englert and R.~Brout, ``{Broken Symmetry and the Mass of
  Gauge Vector Mesons}'',} \textit{ Phys. Rev. Lett.} \textbf{ 13} (1964) 321,
  \href{http://dx.doi.org/10.1103/PhysRevLett.13.321}{\doi{10.1103/PhysRevLett.13.321}}.

\bibitem{Higgs:1964ia}
\hrefCMSnoop {}{P.~W. Higgs, ``{Broken symmetries, massless particles and gauge
  fields}'',} \textit{ Phys. Lett.} \textbf{ 12} (1964) 132,
  \href{http://dx.doi.org/10.1016/0031-9163(64)91136-9}{\doi{10.1016/0031-9163(64)91136-9}}.

\bibitem{Higgs:1964pj}
\hrefCMSnoop {}{P.~W. Higgs, ``{Broken Symmetries and the Masses of Gauge
  Bosons}'',} \textit{ Phys. Rev. Lett.} \textbf{ 13} (1964) 508,
  \href{http://dx.doi.org/10.1103/PhysRevLett.13.508}{\doi{10.1103/PhysRevLett.13.508}}.

\bibitem{Guralnik:1964eu}
\hrefCMSnoop {}{G.~S. Guralnik, C.~R. Hagen, and T.~W.~B. Kibble, ``{Global
  Conservation Laws and Massless Particles}'',} \textit{ Phys. Rev. Lett.}
  \textbf{ 13} (1964) 585,
  \href{http://dx.doi.org/10.1103/PhysRevLett.13.585}{\doi{10.1103/PhysRevLett.13.585}}.

\bibitem{Higgs:1966ev}
\hrefCMSnoop {}{P.~W. Higgs, ``{Spontaneous Symmetry Breakdown without Massless
  Bosons}'',} \textit{ Phys. Rev.} \textbf{ 145} (1966) 1156,
  \href{http://dx.doi.org/10.1103/PhysRev.145.1156}{\doi{10.1103/PhysRev.145.1156}}.

\bibitem{Kibble:1967sv}
\hrefCMSnoop {}{T.~W.~B. Kibble, ``{Symmetry Breaking in Non-Abelian Gauge
  Theories}'',} \textit{ Phys. Rev.} \textbf{ 155} (1967) 1554,
  \href{http://dx.doi.org/10.1103/PhysRev.155.1554}{\doi{10.1103/PhysRev.155.1554}}.

\bibitem{ATLASobservation125}
\hrefCMSnoop {}{{ATLAS Collaboration}, ``{Observation of a new particle in the
  search for the Standard Model Higgs boson with the ATLAS detector at the
  LHC}'',} \textit{ Phys. Lett. B} \textbf{ 716} (2012) 1,
  \href{http://dx.doi.org/10.1016/j.physletb.2012.08.020}{\doi{10.1016/j.physletb.2012.08.020}},
\href{http://www.arXiv.org/abs/1207.7214}{\texttt{arXiv:1207.7214}}.

\bibitem{CMSobservation125}
\hrefCMSnoop {}{{CMS Collaboration}, ``Observation of a new boson at a mass of
  125 GeV with the CMS experiment at the LHC'',} \textit{ Phys. Lett. B}
  \textbf{ 716} (2012) 30,
  \href{http://dx.doi.org/10.1016/j.physletb.2012.08.021}{\doi{10.1016/j.physletb.2012.08.021}},
\href{http://www.arXiv.org/abs/1207.7235}{\texttt{arXiv:1207.7235}}.

\bibitem{CMSlongpaper}
\hrefCMSnoop {}{{CMS Collaboration}, ``Observation of a new boson with mass
  near 125 GeV in pp collisions at $\sqrt{s}$ = 7 and 8 TeV'',} \textit{ JHEP}
  \textbf{ 06} (2013) 081,
  \href{http://dx.doi.org/10.1007/JHEP06(2013)081}{\doi{10.1007/JHEP06(2013)081}},
\href{http://www.arXiv.org/abs/1303.4571}{\texttt{arXiv:1303.4571}}.

\bibitem{ATLASlongpaper}
\hrefCMSnoop {}{{ATLAS Collaboration}, ``Measurement of the Higgs boson mass
  from the $H\rightarrow\gamma\gamma$ and $H\rightarrow ZZ^*\rightarrow 4\ell$
  channels in pp collisions at center-of-mass energies of 7 and 8 TeV with the
  ATLAS detector'',} \textit{ Phys. Rev. D} \textbf{ 90} (2014) 052004,
  \href{http://dx.doi.org/10.1103/PhysRevD.90.052004}{\doi{10.1103/PhysRevD.90.052004}},
\href{http://www.arXiv.org/abs/1406.3827}{\texttt{arXiv:1406.3827}}.

\bibitem{LHCmasscombination:2015}
\hrefCMSnoop {}{{ATLAS, CMS} Collaboration, ``{Combined Measurement of the
  Higgs Boson Mass in $pp$ Collisions at $\sqrt{s}=7$ and 8 TeV with the ATLAS
  and CMS Experiments}'',} \textit{ Phys. Rev. Lett.} \textbf{ 114} (2015)
  191803,
  \href{http://dx.doi.org/10.1103/PhysRevLett.114.191803}{\doi{10.1103/PhysRevLett.114.191803}},
\href{http://www.arXiv.org/abs/1503.07589}{\texttt{arXiv:1503.07589}}.

\bibitem{CMSttHbb7TeV}
\hrefCMSnoop {}{{CMS Collaboration}, ``{Search for the standard model Higgs
  boson produced in association with a top-quark pair in pp collisions at the
  LHC}'',} \textit{ JHEP} \textbf{ 05} (2013) 145,
  \href{http://dx.doi.org/10.1007/JHEP05(2013)145}{\doi{10.1007/JHEP05(2013)145}},
\href{http://www.arXiv.org/abs/1303.0763}{\texttt{arXiv:1303.0763}}.

\bibitem{ATLASDiboson2013}
\hrefCMSnoop {}{{ATLAS Collaboration}, ``{Measurements of Higgs boson
  production and couplings in diboson final states with the ATLAS detector at
  the LHC}'',} \textit{ Phys. Lett. B} \textbf{ 726} (2013) 88,
  \href{http://dx.doi.org/10.1016/j.physletb.2013.08.010}{\doi{10.1016/j.physletb.2013.08.010}},
\href{http://www.arXiv.org/abs/1307.1427}{\texttt{arXiv:1307.1427}}.

\bibitem{Chatrchyan:2013zna}
\hrefCMSnoop {}{{CMS Collaboration}, ``{Search for the standard model Higgs
  boson produced in association with a W or a Z boson and decaying to bottom
  quarks}'',} \textit{ Phys. Rev. D} \textbf{ 89} (2014) 012003,
  \href{http://dx.doi.org/10.1103/PhysRevD.89.012003}{\doi{10.1103/PhysRevD.89.012003}},
\href{http://www.arXiv.org/abs/1310.3687}{\texttt{arXiv:1310.3687}}.

\bibitem{CMSHWWlvlv2014}
\hrefCMSnoop {}{{CMS Collaboration}, ``{Measurement of Higgs boson production
  and properties in the WW decay channel with leptonic final states}'',}
  \textit{ JHEP} \textbf{ 01} (2014) 096,
  \href{http://dx.doi.org/10.1007/JHEP01(2014)096}{\doi{10.1007/JHEP01(2014)096}},
\href{http://www.arXiv.org/abs/1312.1129}{\texttt{arXiv:1312.1129}}.

\bibitem{Chatrchyan:2013mxa}
\hrefCMSnoop {}{{CMS Collaboration}, ``{Measurement of the properties of a
  Higgs boson in the four-lepton final state}'',} \textit{ Phys. Rev. D}
  \textbf{ 89} (2014) 092007,
  \href{http://dx.doi.org/10.1103/PhysRevD.89.092007}{\doi{10.1103/PhysRevD.89.092007}},
\href{http://www.arXiv.org/abs/1312.5353}{\texttt{arXiv:1312.5353}}.

\bibitem{Chatrchyan:2014nva}
\hrefCMSnoop {}{{CMS Collaboration}, ``{Evidence for the 125 GeV Higgs boson
  decaying to a pair of $\tau$ leptons}'',} \textit{ JHEP} \textbf{ 05} (2014)
  104,
  \href{http://dx.doi.org/10.1007/JHEP05(2014)104}{\doi{10.1007/JHEP05(2014)104}},
\href{http://www.arXiv.org/abs/1401.5041}{\texttt{arXiv:1401.5041}}.

\bibitem{CMSHggLegacyRun1-arxiv}
\hrefCMSnoop {}{{CMS Collaboration}, ``{Observation of the diphoton decay of
  the Higgs boson and measurement of its properties}'',} \textit{ Eur. Phys. J.
  C} \textbf{ 74} (2014) 3076,
  \href{http://dx.doi.org/10.1140/epjc/s10052-014-3076-z}{\doi{10.1140/epjc/s10052-014-3076-z}},
\href{http://www.arXiv.org/abs/1407.0558}{\texttt{arXiv:1407.0558}}.

\bibitem{CMSHzg2013}
\hrefCMSnoop {}{{CMS Collaboration}, ``{Search for a Higgs boson decaying into
  a Z and a photon in pp collisions at $\sqrt{s} = 7$ and 8 TeV}'',} \textit{
  Phys. Lett. B} \textbf{ 726} (2013) 587,
  \href{http://dx.doi.org/10.1016/j.physletb.2013.09.057}{\doi{10.1016/j.physletb.2013.09.057}},
\href{http://www.arXiv.org/abs/1307.5515}{\texttt{arXiv:1307.5515}}.

\bibitem{ATLASHzg2014}
\hrefCMSnoop {}{{ATLAS Collaboration}, ``{Search for Higgs boson decays to a
  photon and a Z boson in pp collisions at $\sqrt{s}=7$ and 8 TeV with the
  ATLAS detector}'',} \textit{ Phys. Lett. B} \textbf{ 732} (2014) 8,
  \href{http://dx.doi.org/10.1016/j.physletb.2014.03.015}{\doi{10.1016/j.physletb.2014.03.015}},
\href{http://www.arXiv.org/abs/1402.3051}{\texttt{arXiv:1402.3051}}.

\bibitem{ATLASInvisible2014}
\hrefCMSnoop {}{{ATLAS Collaboration}, ``Search for Invisible Decays of a Higgs
  Boson Produced in Association with a $Z$ Boson in ATLAS'',} \textit{ Phys.
  Rev. Lett.} \textbf{ 112} (2014) 201802,
  \href{http://dx.doi.org/10.1103/PhysRevLett.112.201802}{\doi{10.1103/PhysRevLett.112.201802}},
\href{http://www.arXiv.org/abs/1402.3244}{\texttt{arXiv:1402.3244}}.

\bibitem{CMSInvisible2014}
\hrefCMSnoop {}{{CMS Collaboration}, ``{Search for invisible decays of Higgs
  bosons in the vector boson fusion and associated ZH production modes}'',}
  \textit{ Eur. Phys. J. C} \textbf{ 74} (2014) 2980,
  \href{http://dx.doi.org/10.1140/epjc/s10052-014-2980-6}{\doi{10.1140/epjc/s10052-014-2980-6}},
\href{http://www.arXiv.org/abs/1404.1344}{\texttt{arXiv:1404.1344}}.

\bibitem{CMSMassParity2012}
\hrefCMSnoop {}{{CMS Collaboration}, ``{Study of the Mass and Spin-Parity of
  the Higgs Boson Candidate via Its Decays to Z Boson Pairs}'',} \textit{ Phys.
  Rev. Lett.} \textbf{ 110} (2013) 081803,
  \href{http://dx.doi.org/10.1103/PhysRevLett.110.081803}{\doi{10.1103/PhysRevLett.110.081803}},
\href{http://www.arXiv.org/abs/1212.6639}{\texttt{arXiv:1212.6639}}.

\bibitem{Aad:2013xqa}
\hrefCMSnoop {}{{ATLAS Collaboration}, ``{Evidence for the spin-0 nature of the
  Higgs boson using ATLAS data}'',} \textit{ Phys. Lett. B} \textbf{ 726}
  (2013) 120,
  \href{http://dx.doi.org/10.1016/j.physletb.2013.08.026}{\doi{10.1016/j.physletb.2013.08.026}},
\href{http://www.arXiv.org/abs/1307.1432}{\texttt{arXiv:1307.1432}}.

\bibitem{Dicus:1992vj}
\hrefCMSnoop {}{D.~A. Dicus and V.~S. Mathur, ``Upper Bounds on the Values of
  Masses in Unified Gauge Theories'',} \textit{ Phys. Rev. D} \textbf{ 7}
  (1973) 3111,
\href{http://dx.doi.org/10.1103/PhysRevD.7.3111}{\doi{10.1103/PhysRevD.7.3111}}.

\bibitem{Veltman:1976rt}
\hrefCMSnoop {}{M.~J.~G. Veltman, ``{Second Threshold in Weak Interactions}'',}
  \textit{ Acta Phys. Polon. B} \textbf{ 8} (1977)
475.

\bibitem{Lee:1977eg}
\hrefCMSnoop {}{B.~W. Lee, C.~Quigg, and H.~B. Thacker, ``Weak interactions at
  very high energies: The role of the Higgs-boson mass'',} \textit{ Phys. Rev.
  D} \textbf{ 16} (1977) 1519,
\href{http://dx.doi.org/10.1103/PhysRevD.16.1519}{\doi{10.1103/PhysRevD.16.1519}}.

\bibitem{Lee:1977yc}
\hrefCMSnoop {}{B.~W. Lee, C.~Quigg, and H.~B. Thacker, ``{Strength of Weak
  Interactions at Very High Energies and the Higgs Boson Mass}'',} \textit{
  Phys. Rev. Lett.} \textbf{ 38} (1977) 883,
\href{http://dx.doi.org/10.1103/PhysRevLett.38.883}{\doi{10.1103/PhysRevLett.38.883}}.

\bibitem{Passarino:1990hk}
\hrefCMSnoop {}{G.~Passarino, ``WW scattering and perturbative unitarity'',}
  \textit{ Nucl. Phys. B} \textbf{ 343} (1990) 31,
\href{http://dx.doi.org/10.1016/0550-3213(90)90593-3}{\doi{10.1016/0550-3213(90)90593-3}}.

\bibitem{Chanowitz:1985hj}
\hrefCMSnoop {}{M.~S. Chanowitz and M.~K. Gaillard, ``The TeV physics of
  strongly interacting W's and Z's'',} \textit{ Nucl. Phys. B} \textbf{ 261}
  (1985) 379,
\href{http://dx.doi.org/10.1016/0550-3213(85)90580-2}{\doi{10.1016/0550-3213(85)90580-2}}.

\bibitem{Duncan:1985vj}
\hrefCMSnoop {}{M.~J. Duncan, G.~L. Kane, and W.~W. Repko, ``WW physics at
  future colliders'',} \textit{ Nucl. Phys. B} \textbf{ 272} (1986) 517,
\href{http://dx.doi.org/10.1016/0550-3213(86)90234-8}{\doi{10.1016/0550-3213(86)90234-8}}.

\bibitem{Dicus:1986jg}
\hrefCMSnoop {}{D.~A. Dicus and R.~Vega, ``WW Production from pp collisions'',}
  \textit{ Phys. Rev. Lett.} \textbf{ 57} (1986) 1110,
\href{http://dx.doi.org/10.1103/PhysRevLett.57.1110}{\doi{10.1103/PhysRevLett.57.1110}}.

\bibitem{Bagger:1995mk}
J.~Bagger\hrefCMSnoop {}{ {et~al.}, ``CERN LHC analysis of the strongly
  interacting WW system: Gold-plated modes'',} \textit{ Phys. Rev. D} \textbf{
  52} (1995) 3878,
  \href{http://dx.doi.org/10.1103/PhysRevD.52.3878}{\doi{10.1103/PhysRevD.52.3878}},
\href{http://www.arXiv.org/abs/hep-ph/9504426}{\texttt{arXiv:hep-ph/9504426}}.

\bibitem{Ballestrero:2009vw}
\hrefCMSnoop {}{A.~Ballestrero, G.~Bevilacqua, D.~B. Franzosi, and E.~Maina,
  ``How well can the LHC distinguish between the SM light Higgs scenario, a
  composite Higgs and the Higgsless case using VV scattering channels?'',}
  \textit{ JHEP} \textbf{ 11} (2009) 126,
  \href{http://dx.doi.org/10.1088/1126-6708/2009/11/126}{\doi{10.1088/1126-6708/2009/11/126}},
\href{http://www.arXiv.org/abs/0909.3838}{\texttt{arXiv:0909.3838}}.

\bibitem{Branco:2011iw}
G.~C. Branco\hrefCMSnoop {}{ {et~al.}, ``Theory and phenomenology of
  two-Higgs-doublet models'',} \textit{ Phys. Rept.} \textbf{ 516} (2012) 1,
  \href{http://dx.doi.org/10.1016/j.physrep.2012.02.002}{\doi{10.1016/j.physrep.2012.02.002}},
\href{http://www.arXiv.org/abs/1106.0034}{\texttt{arXiv:1106.0034}}.

\bibitem{craig}
\hrefCMSnoop {}{N.~Craig and T.~Scott, ``Exclusive signals of an extended Higgs
  sector'',} \textit{ JHEP} \textbf{ 11} (2012) 083,
  \href{http://dx.doi.org/10.1007/JHEP11(2012)083}{\doi{10.1007/JHEP11(2012)083}},
\href{http://www.arXiv.org/abs/1207.4835}{\texttt{arXiv:1207.4835}}.

\bibitem{Chacko:2005pe}
\hrefCMSnoop {}{Z.~Chacko, H.-S. Goh, and R.~Harnik, ``{The Twin Higgs: Natural
  electroweak breaking from mirror symmetry}'',} \textit{ Phys.Rev.Lett.}
  \textbf{ 96} (2006) 231802,
  \href{http://dx.doi.org/10.1103/PhysRevLett.96.231802}{\doi{10.1103/PhysRevLett.96.231802}},
\href{http://www.arXiv.org/abs/hep-ph/0506256}{\texttt{arXiv:hep-ph/0506256}}.

\bibitem{Patt:2006fw}
\hrefCMSnoop {}{B.~Patt and F.~Wilczek, ``Higgs-field portal into hidden
  sectors'',} (2006).
\href{http://www.arXiv.org/abs/hep-ph/0605188}{\texttt{arXiv:hep-ph/0605188}}.

\bibitem{Barger:2007}
V.~Barger\hrefCMSnoop {}{ {et~al.}, ``{CERN LHC phenomenology of an extended
  standard model with a real scalar singlet}'',} \textit{ Phys. Rev. D}
  \textbf{ 77} (2008) 035005,
  \href{http://dx.doi.org/10.1103/PhysRevD.77.035005}{\doi{10.1103/PhysRevD.77.035005}},
\href{http://www.arXiv.org/abs/0706.4311}{\texttt{arXiv:0706.4311}}.

\bibitem{Bowen:2007ia}
\hrefCMSnoop {}{M.~Bowen, Y.~Cui, and J.~Wells, ``Narrow trans-TeV Higgs bosons
  and $H\rightarrow hh$ decays: Two LHC search paths for a hidden sector Higgs
  boson'',} \textit{ JHEP} \textbf{ 03} (2007) 036,
  \href{http://dx.doi.org/10.1088/1126-6708/2007/03/036}{\doi{10.1088/1126-6708/2007/03/036}},
\href{http://www.arXiv.org/abs/hep-ph/0701035}{\texttt{arXiv:hep-ph/0701035}}.

\bibitem{Bhattacharyya:2007}
\hrefCMSnoop {}{G.~Bhattacharyya, C.~Branco, and S.~Nandi, ``{Universal
  Doublet-Singlet Higgs Couplings and phenomenology at the CERN Large Hadron
  Collider}'',} \textit{ Phys. Rev. D} \textbf{ 77} (2008) 117701,
  \href{http://dx.doi.org/10.1103/PhysRevD.77.117701}{\doi{10.1103/PhysRevD.77.117701}},
\href{http://www.arXiv.org/abs/0712.2693}{\texttt{arXiv:0712.2693}}.

\bibitem{Profumo:2007wc}
\hrefCMSnoop {}{S.~Profumo, M.~J. Ramsey-Musolf, and G.~Shaughnessy, ``{Singlet
  Higgs phenomenology and the electroweak phase transition}'',} \textit{ JHEP}
  \textbf{ 0708} (2007) 010,
  \href{http://dx.doi.org/10.1088/1126-6708/2007/08/010}{\doi{10.1088/1126-6708/2007/08/010}},
\href{http://www.arXiv.org/abs/0705.2425}{\texttt{arXiv:0705.2425}}.

\bibitem{Dawson:2009}
\hrefCMSnoop {}{S.~Dawson and Y.~Wenbin, ``{Hiding the Higgs boson with
  multiple scalars}'',} \textit{ Phys. Rev. D} \textbf{ 79} (2009) 095002,
  \href{http://dx.doi.org/10.1103/PhysRevD.79.095002}{\doi{10.1103/PhysRevD.79.095002}},
\href{http://www.arXiv.org/abs/0904.2005}{\texttt{arXiv:0904.2005}}.

\bibitem{Bock:2010}
S.~Bock\hrefCMSnoop {}{ {et~al.}, ``{Measuring hidden Higgs and
  strongly-interacting Higgs scenarios}'',} \textit{ Phys. Lett. B} \textbf{
  694} (2010) 44,
  \href{http://dx.doi.org/10.1016/j.physletb.2010.09.032}{\doi{10.1016/j.physletb.2010.09.032}},
\href{http://www.arXiv.org/abs/1007.2645}{\texttt{arXiv:1007.2645}}.

\bibitem{Baek:2011aa}
\hrefCMSnoop {}{S.~Baek, P.~Ko, and W.-I. Park, ``{Search for the Higgs portal
  to a singlet fermionic dark matter at the LHC}'',} \textit{ JHEP} \textbf{
  1202} (2012) 047,
  \href{http://dx.doi.org/10.1007/JHEP02(2012)047}{\doi{10.1007/JHEP02(2012)047}},
\href{http://www.arXiv.org/abs/1112.1847}{\texttt{arXiv:1112.1847}}.

\bibitem{Fox:2011}
\hrefCMSnoop {}{P.~Fox, D.~Tucker-Smith, and N.~Weiner, ``{Higgs friends and
  counterfeits at hadron colliders}'',} \textit{ JHEP} \textbf{ 06} (2011) 127,
  \href{http://dx.doi.org/10.1007/JHEP06(2011)127}{\doi{10.1007/JHEP06(2011)127}},
\href{http://www.arXiv.org/abs/1104.5450}{\texttt{arXiv:1104.5450}}.

\bibitem{Englert:2011}
\hrefCMSnoop {}{C.~Englert, T.~Plehn, D.~Zeerwas, and P.~M. Zerwas,
  ``{Exploring the Higgs portal}'',} \textit{ Phys. Lett. B} \textbf{ 703}
  (2011) 298,
  \href{http://dx.doi.org/10.1016/j.physletb.2011.08.002}{\doi{10.1016/j.physletb.2011.08.002}},
\href{http://www.arXiv.org/abs/1106.3097}{\texttt{arXiv:1106.3097}}.

\bibitem{Englert:2012}
\hrefCMSnoop {}{C.~Englert, J.~Jaeckel, E.~Re, and M.~Spannowsky, ``{Evasive
  Higgs boson maneuvers at the LHC}'',} \textit{ Phys. Rev. D} \textbf{ 85}
  (2012) 035008,
  \href{http://dx.doi.org/10.1103/PhysRevD.85.035008}{\doi{10.1103/PhysRevD.85.035008}},
  \href{http://www.arXiv.org/abs/1111.1719}{\texttt{arXiv:1111.1719}}.

\bibitem{Batell:2012}
\hrefCMSnoop {}{B.~Batell, S.~Gori, and L.-T. Wang, ``{Exploring the Higgs
  portal with 10 fb$^{-1}$ at the LHC}'',} \textit{ JHEP} \textbf{ 06} (2012)
  172,
  \href{http://dx.doi.org/10.1007/JHEP06(2012)172}{\doi{10.1007/JHEP06(2012)172}},
\href{http://www.arXiv.org/abs/1112.5180}{\texttt{arXiv:1112.5180}}.

\bibitem{Englert:2012a}
C.~Englert\hrefCMSnoop {}{ {et~al.}, ``{LHC: Standard Higgs and hidden
  Higgs}'',} \textit{ Phys. Lett. B} \textbf{ 707} (2012) 512,
  \href{http://dx.doi.org/10.1016/j.physletb.2011.12.067}{\doi{10.1016/j.physletb.2011.12.067}},
\href{http://www.arXiv.org/abs/1112.3007}{\texttt{arXiv:1112.3007}}.

\bibitem{Gupta:2012}
\hrefCMSnoop {}{R.~Gupta and J.~Wells, ``{Higgs boson search significance
  deformations due to mixed-in scalars}'',} \textit{ Phys. Lett. B} \textbf{
  710} (2012) 154,
  \href{http://dx.doi.org/10.1016/j.physletb.2012.02.056}{\doi{10.1016/j.physletb.2012.02.056}},
\href{http://www.arXiv.org/abs/1110.0824}{\texttt{arXiv:1110.0824}}.

\bibitem{Dolan:2012}
\hrefCMSnoop {}{M.~Dolan, C.~Englert, and M.~Spannowsky, ``{New physics in LHC
  Higgs boson pair production}'',} \textit{ Phys. Rev. D} \textbf{ 87} (2012)
  055002,
  \href{http://dx.doi.org/10.1103/PhysRevD.87.055002}{\doi{10.1103/PhysRevD.87.055002}},
\href{http://www.arXiv.org/abs/1210.8166}{\texttt{arXiv:1210.8166}}.

\bibitem{Bertolini:2012}
\hrefCMSnoop {}{D.~Bertolini and M.~McCullough, ``{The social Higgs}'',}
  \textit{ JHEP} \textbf{ 12} (2012) 118,
  \href{http://dx.doi.org/10.1007/JHEP12(2012)118}{\doi{10.1007/JHEP12(2012)118}},
\href{http://www.arXiv.org/abs/1207.4209}{\texttt{arXiv:1207.4209}}.

\bibitem{Batell:2012a}
\hrefCMSnoop {}{B.~Batell, D.~McKeen, and M.~Pospelov, ``{Singlet neighbors of
  the Higgs boson}'',} \textit{ JHEP} \textbf{ 10} (2012) 104,
  \href{http://dx.doi.org/10.1007/JHEP10(2012)104}{\doi{10.1007/JHEP10(2012)104}},
\href{http://www.arXiv.org/abs/1207.6252}{\texttt{arXiv:1207.6252}}.

\bibitem{Chpoi:2013wga}
\hrefCMSnoop {}{S.~Choi, S.~Jung, and P.~Ko, ``{Implications of LHC data on 125
  GeV Higgs-like boson for the Standard Model and its various extensions}'',}
  \textit{ JHEP} \textbf{ 1310} (2013) 225,
  \href{http://dx.doi.org/10.1007/JHEP10(2013)225}{\doi{10.1007/JHEP10(2013)225}},
\href{http://www.arXiv.org/abs/1307.3948}{\texttt{arXiv:1307.3948}}.

\bibitem{Hambye:2013dgv}
\hrefCMSnoop {}{T.~Hambye and A.~Strumia, ``{Dynamical generation of the weak
  and Dark Matter scale}'',} \textit{ Phys.Rev.} \textbf{ D88} (2013) 055022,
  \href{http://dx.doi.org/10.1103/PhysRevD.88.055022}{\doi{10.1103/PhysRevD.88.055022}},
\href{http://www.arXiv.org/abs/1306.2329}{\texttt{arXiv:1306.2329}}.

\bibitem{Craig:2014lda}
\hrefCMSnoop {}{N.~Craig, H.~K. Lou, M.~McCullough, and A.~Thalapillil, ``{The
  Higgs Portal Above Threshold}'',}
\href{http://www.arXiv.org/abs/1412.0258}{\texttt{arXiv:1412.0258}}.

\bibitem{Curtin:2014jma}
\hrefCMSnoop {}{D.~Curtin, P.~Meade, and C.-T. Yu, ``{Testing Electroweak
  Baryogenesis with Future Colliders}'',} \textit{ JHEP} \textbf{ 1411} (2014)
  127,
  \href{http://dx.doi.org/10.1007/JHEP11(2014)127}{\doi{10.1007/JHEP11(2014)127}},
\href{http://www.arXiv.org/abs/1409.0005}{\texttt{arXiv:1409.0005}}.

\bibitem{Lopez-Val:2014jva}
\hrefCMSnoop {}{D.~Lopez-Val and T.~Robens, ``{$\Delta\mathrm{r}$ and the
  W-boson mass in the singlet extension of the standard model}'',} \textit{
  Phys. Rev. D} \textbf{ 90} (2014) 114018,
  \href{http://dx.doi.org/10.1103/PhysRevD.90.114018}{\doi{10.1103/PhysRevD.90.114018}},
\href{http://www.arXiv.org/abs/1406.1043}{\texttt{arXiv:1406.1043}}.

\bibitem{Robens:2015gla}
\hrefCMSnoop {}{T.~Robens and T.~Stefaniak, ``{Status of the Higgs singlet
  extension of the standard model after LHC run 1}'',} \textit{ Eur. Phys. J.
  C} \textbf{ 75} (2015) 105,
  \href{http://dx.doi.org/10.1140/epjc/s10052-015-3323-y}{\doi{10.1140/epjc/s10052-015-3323-y}},
\href{http://www.arXiv.org/abs/1501.02234}{\texttt{arXiv:1501.02234}}.

\bibitem{CMShighmass}
\hrefCMSnoop {}{{CMS Collaboration}, ``{Search for a standard-model-like Higgs
  boson with a mass in the range 145 to 1000 GeV at the LHC}'',} \textit{ Eur.
  Phys. J. C} \textbf{ 73} (2013) 27,
  \href{http://dx.doi.org/10.1140/epjc/s10052-013-2469-8}{\doi{10.1140/epjc/s10052-013-2469-8}},
\href{http://www.arXiv.org/abs/1304.0213}{\texttt{arXiv:1304.0213}}.

\bibitem{Chatrchyan:2008zzk}
\hrefCMSnoop {}{{CMS Collaboration}, ``The {CMS} experiment at the {CERN}
  {LHC}'',} \textit{ JINST} \textbf{ 3} (2008) S08004,
  \href{http://dx.doi.org/10.1088/1748-0221/3/08/S08004}{\doi{10.1088/1748-0221/3/08/S08004}}.

\bibitem{Bagnaschi:2012}
\hrefCMSnoop {}{E.~Bagnaschi, G.~Degrassi, P.~Slavich, and A.~Vicini, ``{Higgs
  production via gluon fusion in the POWHEG approach in the SM and in the
  MSSM}'',} \textit{ JHEP} \textbf{ 02} (2012) 088,
  \href{http://dx.doi.org/10.1007/JHEP02(2012)088}{\doi{10.1007/JHEP02(2012)088}},
\href{http://www.arXiv.org/abs/1111.2854}{\texttt{arXiv:1111.2854}}.

\bibitem{Nason:2004rx}
\hrefCMSnoop {}{P.~Nason, ``{A new method for combining NLO QCD with shower
  Monte Carlo algorithms}'',} \textit{ JHEP} \textbf{ 11} (2004) 040,
  \href{http://dx.doi.org/10.1088/1126-6708/2004/11/040}{\doi{10.1088/1126-6708/2004/11/040}},
\href{http://www.arXiv.org/abs/hep-ph/0409146}{\texttt{arXiv:hep-ph/0409146}}.

\bibitem{Frixione:2007vw}
\hrefCMSnoop {}{S.~Frixione, P.~Nason, and C.~Oleari, ``{Matching NLO QCD
  computations with Parton Shower simulations: the POWHEG method}'',} \textit{
  JHEP} \textbf{ 11} (2007) 070,
  \href{http://dx.doi.org/10.1088/1126-6708/2007/11/070}{\doi{10.1088/1126-6708/2007/11/070}},
\href{http://www.arXiv.org/abs/0709.2092}{\texttt{arXiv:0709.2092}}.

\bibitem{Alioli:2010xd}
\hrefCMSnoop {}{S.~Alioli, P.~Nason, C.~Oleari, and E.~Re, ``{A general
  framework for implementing NLO calculations in shower Monte Carlo programs:
  the POWHEG BOX}'',} \textit{ JHEP} \textbf{ 06} (2010) 043,
  \href{http://dx.doi.org/10.1007/JHEP06(2010)043}{\doi{10.1007/JHEP06(2010)043}},
\href{http://www.arXiv.org/abs/1002.2581}{\texttt{arXiv:1002.2581}}.

\bibitem{Nason:2010}
\hrefCMSnoop {}{P.~Nason and C.~Oleari, ``{NLO Higgs boson production via
  vector-boson fusion matched with shower in POWHEG}'',} \textit{ JHEP}
  \textbf{ 02} (2010) 037,
  \href{http://dx.doi.org/10.1007/JHEP02(2010)037}{\doi{10.1007/JHEP02(2010)037}},
\href{http://www.arXiv.org/abs/0911.5299}{\texttt{arXiv:0911.5299}}.

\bibitem{Gao:2010qx}
Y.~Gao\hrefCMSnoop {}{ {et~al.}, ``{Spin determination of single-produced
  resonances at hadron colliders}'',} \textit{ Phys. Rev. D} \textbf{ 81}
  (2010) 075022,
  \href{http://dx.doi.org/10.1103/PhysRevD.81.075022}{\doi{10.1103/PhysRevD.81.075022}},
\href{http://www.arXiv.org/abs/1001.3396}{\texttt{arXiv:1001.3396}}.

\bibitem{Sjostrand:2006za}
\hrefCMSnoop {}{T.~Sj{\"{o}}strand, S.~Mrenna, and P.~Z. Skands, ``{PYTHIA 6.4
  physics and manual}'',} \textit{ JHEP} \textbf{ 05} (2006) 026,
  \href{http://dx.doi.org/10.1088/1126-6708/2006/05/026}{\doi{10.1088/1126-6708/2006/05/026}},
\href{http://www.arXiv.org/abs/hep-ph/0603175}{\texttt{arXiv:hep-ph/0603175}}.

\bibitem{Heinemeyer:2013tqa}
\hrefCMSnoop {}{S.~Heinemeyer {et~al.}, ``{Handbook of LHC Higgs cross
  sections: 3. Higgs properties}'',} CERN Report CERN-2013-004, 2013.
\newblock
  \href{http://dx.doi.org/10.5170/CERN-2013-004}{\doi{10.5170/CERN-2013-004}},
  \href{http://www.arXiv.org/abs/1307.1347}{\texttt{arXiv:1307.1347}}.

\bibitem{Passarino:2010qk}
\hrefCMSnoop {}{G.~Passarino, C.~Sturm, and S.~Uccirati, ``{Higgs
  pseudo-observables, second Riemann sheet and all that}'',} \textit{ Nucl.
  Phys. B} \textbf{ 834} (2010) 77,
  \href{http://dx.doi.org/10.1016/j.nuclphysb.2010.03.013}{\doi{10.1016/j.nuclphysb.2010.03.013}},
\href{http://www.arXiv.org/abs/1001.3360}{\texttt{arXiv:1001.3360}}.

\bibitem{Goria:2011wa}
\hrefCMSnoop {}{S.~Goria, G.~Passarino, and D.~Rosco, ``{The Higgs-boson
  lineshape}'',} \textit{ Nucl. Phys. B} \textbf{ 864} (2012) 530,
  \href{http://dx.doi.org/10.1016/j.nuclphysb.2012.07.006}{\doi{10.1016/j.nuclphysb.2012.07.006}},
\href{http://www.arXiv.org/abs/1112.5517}{\texttt{arXiv:1112.5517}}.

\bibitem{Kauer:2012hd}
\hrefCMSnoop {}{N.~Kauer and G.~Passarino, ``{Inadequacy of zero-width
  approximation for a light Higgs boson signal}'',} \textit{ JHEP} \textbf{ 08}
  (2012) 116,
  \href{http://dx.doi.org/10.1007/JHEP08(2012)116}{\doi{10.1007/JHEP08(2012)116}},
\href{http://www.arXiv.org/abs/1206.4803}{\texttt{arXiv:1206.4803}}.

\bibitem{Alwall:2007st}
J.~Alwall\hrefCMSnoop {}{ {et~al.}, ``{MadGraph/MadEvent} v4: the new web
  generation'',} \textit{ JHEP} \textbf{ 09} (2007) 028,
  \href{http://dx.doi.org/10.1088/1126-6708/2007/09/028}{\doi{10.1088/1126-6708/2007/09/028}},
\href{http://www.arXiv.org/abs/0706.2334}{\texttt{arXiv:0706.2334}}.

\bibitem{ggww}
\hrefCMSnoop {}{T.~Binoth, M.~Ciccolini, N.~Kauer, and M.~Kr{\"a}mer,
  ``Gluon-induced {$W$}-boson pair production at the {LHC}'',} \textit{ JHEP}
  \textbf{ 12} (2006) 046,
  \href{http://dx.doi.org/10.1088/1126-6708/2006/12/046}{\doi{10.1088/1126-6708/2006/12/046}},
  \href{http://www.arXiv.org/abs/hep-ph/0611170}{\texttt{arXiv:hep-ph/0611170}}.

\bibitem{Binoth:2008pr}
\hrefCMSnoop {}{T.~Binoth, N.~Kauer, and P.~Mertsch, ``{Gluon-induced QCD
  corrections to $pp \to ZZ \to \ell\bar{\ell}\ell'\bar{\ell'}$}'',} in
  \textit{ Proceedings of the XVI Int. Workshop on Deep-Inelastic Scattering
  and Related Topics (DIS'07)}.
\newblock 2008.
\newblock \href{http://www.arXiv.org/abs/0807.0024}{\texttt{arXiv:0807.0024}}.
\newblock
\href{http://dx.doi.org/10.3360/dis.2008.142}{\doi{10.3360/dis.2008.142}}.

\bibitem{CTEQ66}
H.-L. Lai\hrefCMSnoop {}{ {et~al.}, ``Uncertainty induced by {QCD} coupling in
  the {CTEQ} global analysis of parton distributions'',} \textit{ Phys. Rev. D}
  \textbf{ 82} (2010) 054021,
  \href{http://dx.doi.org/10.1103/PhysRevD.82.054021}{\doi{10.1103/PhysRevD.82.054021}},
  \href{http://www.arXiv.org/abs/1004.4624}{\texttt{arXiv:1004.4624}}.

\bibitem{Lai:2010vv}
H.-L. Lai\hrefCMSnoop {}{ {et~al.}, ``{New parton distributions for collider
  physics}'',} \textit{ Phys. Rev. D} \textbf{ 82} (2010) 074024,
  \href{http://dx.doi.org/10.1103/PhysRevD.82.074024}{\doi{10.1103/PhysRevD.82.074024}},
  \href{http://www.arXiv.org/abs/1007.2241}{\texttt{arXiv:1007.2241}}.

\bibitem{Jadach:1993hs}
\hrefCMSnoop {}{S.~Jadach, Z.~W{\c{a}}s, R.~Decker, and J.~H. K{\"u}hn, ``{The
  $\tau$ decay library TAUOLA, version 2.4}'',} \textit{ Comput. Phys. Commun.}
  \textbf{ 76} (1993) 361,
\href{http://dx.doi.org/10.1016/0010-4655(93)90061-G}{\doi{10.1016/0010-4655(93)90061-G}}.

\bibitem{GEANT}
\hrefCMSnoop {}{{{GEANT4}} Collaboration, ``{Geant4---a simulation toolkit}'',}
  \textit{ Nucl. Instrum. Meth. A} \textbf{ 506} (2003) 250,
  \href{http://dx.doi.org/10.1016/S0168-9002(03)01368-8}{\doi{10.1016/S0168-9002(03)01368-8}}.

\bibitem{Chatrchyan:2011id}
\hrefCMSnoop {}{{CMS Collaboration}, ``Measurement of the underlying event
  activity at the LHC with $\sqrt{s}= 7$ TeV and comparison with $\sqrt{s}$ =
  0.9 TeV'',} \textit{ JHEP} \textbf{ 09} (2011) 109,
  \href{http://dx.doi.org/10.1007/JHEP09(2011)109}{\doi{10.1007/JHEP09(2011)109}},
\href{http://www.arXiv.org/abs/1107.0330}{\texttt{arXiv:1107.0330}}.

\bibitem{CMS-PAS-HIG-14-009}
\hrefCMSnoop {}{{CMS Collaboration}, ``{Precise determination of the mass of
  the Higgs boson and tests of compatibility of its couplings with the standard
  model predictions using proton collisions at 7 and 8 $\,\text {TeV}$}'',}
  \textit{ Eur. Phys. J.} \textbf{ C75} (2015), no.~5, 212,
  \href{http://dx.doi.org/10.1140/epjc/s10052-015-3351-7}{\doi{10.1140/epjc/s10052-015-3351-7}},
\href{http://www.arXiv.org/abs/1412.8662}{\texttt{arXiv:1412.8662}}.

\bibitem{MCFM}
\hrefCMSnoop {}{J.~M. Campbell and R.~K. Ellis, ``{MCFM for the Tevatron and
  the LHC}'',} \textit{ Nucl. Phys. Proc. Suppl.} \textbf{ 205} (2010) 10,
  \href{http://dx.doi.org/10.1016/j.nuclphysbps.2010.08.011}{\doi{10.1016/j.nuclphysbps.2010.08.011}},
\href{http://www.arXiv.org/abs/1007.3492}{\texttt{arXiv:1007.3492}}.

\bibitem{Ballestrero:2007xq}
A.~Ballestrero\hrefCMSnoop {}{ {et~al.}, ``{PHANTOM: A Monte Carlo event
  generator for six parton final states at high energy colliders}'',} \textit{
  Comput. Phys. Commun.} \textbf{ 180} (2009) 401,
  \href{http://dx.doi.org/10.1016/j.cpc.2008.10.005}{\doi{10.1016/j.cpc.2008.10.005}},
\href{http://www.arXiv.org/abs/0801.3359}{\texttt{arXiv:0801.3359}}.

\bibitem{Englert:2014}
\hrefCMSnoop {}{C.~Englert, Y.~Soreq, and M.~Spannowsky, ``{Off-shell Higgs
  coupling measurements in BSM scenarios}'',} \textit{ JHEP} \textbf{ 1505}
  (2015) 145,
  \href{http://dx.doi.org/10.1007/JHEP05(2015)145}{\doi{10.1007/JHEP05(2015)145}},
\href{http://www.arXiv.org/abs/1410.5440}{\texttt{arXiv:1410.5440}}.

\bibitem{Maina:2015ela}
\hrefCMSnoop {}{E.~Maina, ``{Interference effects in Heavy Higgs production via
  gluon fusion in the Singlet Extension of the Standard Model}'',} (2015).
\href{http://www.arXiv.org/abs/1501.02139}{\texttt{arXiv:1501.02139}}.

\bibitem{CMS-PAS-PFT-09-001}
\href {http://cdsweb.cern.ch/record/1194487}{{CMS Collaboration},
  ``Particle-Flow Event Reconstruction in CMS and Performance for Jets, Taus
  and MET'',} Technical Report CMS-PAS-PFT-09-001.

\bibitem{Chatrchyan:2011tn}
\hrefCMSnoop {}{{CMS Collaboration}, ``Missing transverse energy performance of
  the CMS detector'',} \textit{ JINST} \textbf{ 6} (2011) P09001,
  \href{http://dx.doi.org/10.1088/1748-0221/6/09/P09001}{\doi{10.1088/1748-0221/6/09/P09001}},
  \href{http://www.arXiv.org/abs/1106.5048}{\texttt{arXiv:1106.5048}}.

\bibitem{CMS-PAS-MUO-10-002}
\hrefCMSnoop {}{{CMS Collaboration}, ``{Performance of CMS muon reconstruction
  in pp collision events at $\sqrt{s}=7\TeV$}'',} \textit{ JINST} \textbf{ 7}
  (2012) P10002,
  \href{http://dx.doi.org/10.1088/1748-0221/7/10/P10002}{\doi{10.1088/1748-0221/7/10/P10002}},
\href{http://www.arXiv.org/abs/1206.4071}{\texttt{arXiv:1206.4071}}.

\bibitem{CMSTracking}
\hrefCMSnoop {}{{CMS Collaboration}, ``Description and performance of track and
  primary-vertex reconstruction with the CMS tracker'',} \textit{ JINST}
  \textbf{ 9} (2014) P10009,
  \href{http://dx.doi.org/10.1088/1748-0221/9/10/P10009}{\doi{10.1088/1748-0221/9/10/P10009}},
\href{http://www.arXiv.org/abs/1405.6569}{\texttt{arXiv:1405.6569}}.

\bibitem{ElePerformance:2015}
\hrefCMSnoop {}{{CMS Collaboration}, ``{Performance of electron reconstruction
  and selection with the CMS detector in proton-proton collisions at $\sqrt{s}$
  = 8 TeV}'',} \textit{ JINST} \textbf{ 10} (2015), no.~06, P06005,
  \href{http://dx.doi.org/10.1088/1748-0221/10/06/P06005}{\doi{10.1088/1748-0221/10/06/P06005}},
\href{http://www.arXiv.org/abs/1502.02701}{\texttt{arXiv:1502.02701}}.

\bibitem{Chatrchyan:2011xq}
\hrefCMSnoop {}{{CMS Collaboration}, ``{Performance of $\tau$-lepton
  reconstruction and identification in CMS}'',} \textit{ JINST} \textbf{ 7}
  (2012) P01001,
  \href{http://dx.doi.org/10.1088/1748-0221/7/01/P01001}{\doi{10.1088/1748-0221/7/01/P01001}},
\href{http://www.arXiv.org/abs/1109.6034}{\texttt{arXiv:1109.6034}}.

\bibitem{FASTJET}
\hrefCMSnoop {}{{M. Cacciari and G. P. Salam}, ``Pileup subtraction using jet
  areas'',} \textit{ Phys. Lett. B} \textbf{ 659} (2008) 119,
  \href{http://dx.doi.org/10.1016/j.physletb.2007.09.077}{\doi{10.1016/j.physletb.2007.09.077}},
  \href{http://www.arXiv.org/abs/0707.1378}{\texttt{arXiv:0707.1378}}.

\bibitem{CMS-PAS-JME-10-003}
\hrefCMSnoop {}{{CMS Collaboration}, ``{Determination of jet energy calibration
  and transverse momentum resolution in CMS}'',} \textit{ JINST} \textbf{ 6}
  (2011) 11002,
  \href{http://dx.doi.org/10.1088/1748-0221/6/11/P11002}{\doi{10.1088/1748-0221/6/11/P11002}},
  \href{http://www.arXiv.org/abs/1107.4277}{\texttt{arXiv:1107.4277}}.

\bibitem{Cacciari:2008gp}
\hrefCMSnoop {}{M.~Cacciari, G.~P. Salam, and G.~Soyez, ``The anti-$k_t$ jet
  clustering algorithm'',} \textit{ JHEP} \textbf{ 04} (2008) 063,
  \href{http://dx.doi.org/10.1088/1126-6708/2008/04/063}{\doi{10.1088/1126-6708/2008/04/063}},
\href{http://www.arXiv.org/abs/0802.1189}{\texttt{arXiv:0802.1189}}.

\bibitem{fastjetmanual}
\hrefCMSnoop {}{M.~Cacciari, G.~P. Salam, and G.~Soyez, ``FastJet user
  manual'',} \textit{ Eur. Phys. J. C} \textbf{ 72} (2012) 1896,
  \href{http://dx.doi.org/10.1140/epjc/s10052-012-1896-2}{\doi{10.1140/epjc/s10052-012-1896-2}},
\href{http://www.arXiv.org/abs/1111.6097}{\texttt{arXiv:1111.6097}}.

\bibitem{Dokshitzer:1997gp}
\hrefCMSnoop {}{Y.~L. Dokshitzer, G.~D. Leder, S.~Moretti, and B.~R. Webber,
  ``Better jet clustering algorithms'',} \textit{ JHEP} \textbf{ 08} (1997)
  001,
  \href{http://dx.doi.org/10.1088/1126-6708/1997/08/001}{\doi{10.1088/1126-6708/1997/08/001}},
  \href{http://www.arXiv.org/abs/hep-ph/9707323}{\texttt{arXiv:hep-ph/9707323}}.

\bibitem{Cacciari:2008gn}
\hrefCMSnoop {}{M.~Cacciari, G.~P. Salam, and G.~Soyez, ``The catchment area of
  jets'',} \textit{ JHEP} \textbf{ 04} (2008) 005,
  \href{http://dx.doi.org/10.1088/1126-6708/2008/04/005}{\doi{10.1088/1126-6708/2008/04/005}},
\href{http://www.arXiv.org/abs/0802.1188}{\texttt{arXiv:0802.1188}}.

\bibitem{CMS-PAS-JME-13-005}
\href {http://cdsweb.cern.ch/record/1581583}{{CMS Collaboration}, ``Pileup Jet
  Identification'',} Technical Report CMS-PAS-JME-13-005.

\bibitem{CMS-PAS-JME-13-006}
\hrefCMSnoop {}{{CMS Collaboration}, ``Identification techniques for highly
  boosted W bosons that decay into hadrons'',} \textit{ JHEP} \textbf{ 12}
  (2014) 017,
  \href{http://dx.doi.org/10.1007/JHEP12(2014)017}{\doi{10.1007/JHEP12(2014)017}},
  \href{http://www.arXiv.org/abs/1410.4227}{\texttt{arXiv:1410.4227}}.

\bibitem{Thaler:2011}
\hrefCMSnoop {}{J.~Thaler and K.~Van~Tilburg, ``Identifying boosted objects
  with N-subjettiness'',} \textit{ JHEP} \textbf{ 03} (2011) 015,
  \href{http://dx.doi.org/10.1007/JHEP03(2011)015}{\doi{10.1007/JHEP03(2011)015}},
  \href{http://www.arXiv.org/abs/1011.2268}{\texttt{arXiv:1011.2268}}.

\bibitem{BTV}
\hrefCMSnoop {}{{CMS Collaboration}, ``Identification of b-quark jets with the
  CMS experiment'',} \textit{ JINST} \textbf{ 8} (2013) P04013,
  \href{http://dx.doi.org/10.1088/1748-0221/8/04/P04013}{\doi{10.1088/1748-0221/8/04/P04013}},
\href{http://www.arXiv.org/abs/1211.4462}{\texttt{arXiv:1211.4462}}.

\bibitem{CMS-PAS-JME-10-005}
\href
  {http://cms-physics.web.cern.ch/cms-physics/public/JME-10-005-pas.pdf}{{CMS
  Collaboration}, ``CMS MET Performance in Events Containing Electroweak Bosons
  from pp Collisions at $\sqrt{s}$ = 7 {TeV}'',} Technical Report
  CMS-PAS-JME-10-005.

\bibitem{Ciccolini:2007jr}
\hrefCMSnoop {}{M.~Ciccolini, A.~Denner, and S.~Dittmaier, ``{Strong and
  Electroweak Corrections to the Production of a Higgs Boson + 2 Jets via Weak
  Interactions at the Large Hadron Collider}'',} \textit{ Phys. Rev. Lett.}
  \textbf{ 99} (2007) 161803,
  \href{http://dx.doi.org/10.1103/PhysRevLett.99.161803}{\doi{10.1103/PhysRevLett.99.161803}},
  \href{http://www.arXiv.org/abs/0707.0381}{\texttt{arXiv:0707.0381}}.

\bibitem{Ciccolini:2007ec}
\hrefCMSnoop {}{M.~Ciccolini, A.~Denner, and S.~Dittmaier, ``{Electroweak and
  QCD corrections to Higgs production via vector-boson fusion at the CERN
  LHC}'',} \textit{ Phys. Rev. D} \textbf{ 77} (2008) 013002,
  \href{http://dx.doi.org/10.1103/PhysRevD.77.013002}{\doi{10.1103/PhysRevD.77.013002}},
\href{http://www.arXiv.org/abs/0710.4749}{\texttt{arXiv:0710.4749}}.

\bibitem{Arnold:2008rz}
\hrefCMSnoop {}{K.~Arnold {et~al.}, ``{VBFNLO: A parton level Monte Carlo for
  processes with electroweak bosons}'',} \textit{ Comput. Phys. Commun.}
  \textbf{ 180} (2009) 1661,
  \href{http://dx.doi.org/10.1016/j.cpc.2009.03.006}{\doi{10.1016/j.cpc.2009.03.006}},
\href{http://www.arXiv.org/abs/0811.4559}{\texttt{arXiv:0811.4559}}.

\bibitem{Cahn:1987}
\hrefCMSnoop {}{R.~Cahn, S.~D. Ellis, R.~Kleiss, and W.~J. Stirling,
  ``Transverse-momentum signatures for heavy Higgs bosons'',} \textit{ Phys.
  Rev. D} \textbf{ 35} (1987) 1626,
  \href{http://dx.doi.org/10.1103/PhysRevD.35.1626}{\doi{10.1103/PhysRevD.35.1626}}.

\bibitem{Alekhin:2011sk}
\hrefCMSnoop {}{S.~Alekhin {et~al.}, ``{The PDF4LHC Working Group Interim
  Report}'',} (2011).
  \href{http://www.arXiv.org/abs/1101.0536}{\texttt{arXiv:1101.0536}}.

\bibitem{Botje:2011sn}
M.~Botje\hrefCMSnoop {}{ {et~al.}, ``{The PDF4LHC Working Group Interim
  Recommendations}'',} (2011).
  \href{http://www.arXiv.org/abs/1101.0538}{\texttt{arXiv:1101.0538}}.

\bibitem{Martin:2009iq}
\hrefCMSnoop {}{A.~D. Martin, W.~J. Stirling, R.~S. Thorne, and G.~Watt,
  ``{Parton distributions for the LHC}'',} \textit{ Eur. Phys. J. C} \textbf{
  63} (2009) 189,
  \href{http://dx.doi.org/10.1140/epjc/s10052-009-1072-5}{\doi{10.1140/epjc/s10052-009-1072-5}},
  \href{http://www.arXiv.org/abs/0901.0002}{\texttt{arXiv:0901.0002}}.

\bibitem{Ball:2011mu}
R.~D. Ball\hrefCMSnoop {}{ {et~al.}, ``{Impact of heavy quark masses on parton
  distributions and LHC phenomenology}'',} \textit{ Nucl. Phys. B} \textbf{
  849} (2011) 296,
  \href{http://dx.doi.org/10.1016/j.nuclphysb.2011.03.021}{\doi{10.1016/j.nuclphysb.2011.03.021}},
\href{http://www.arXiv.org/abs/1101.1300}{\texttt{arXiv:1101.1300}}.

\bibitem{intro2}
\hrefCMSnoop {}{B.~A. Dobrescu and J.~D. Lykken, ``{Semileptonic decays of the
  standard Higgs boson}'',} \textit{ JHEP} \textbf{ 04} (2010) 083,
  \href{http://dx.doi.org/10.1007/JHEP04(2010)083}{\doi{10.1007/JHEP04(2010)083}},
\href{http://www.arXiv.org/abs/0912.3543}{\texttt{arXiv:0912.3543}}.

\bibitem{Chatrchyan:2012hr}
\hrefCMSnoop {}{{CMS Collaboration}, ``Search for the standard model Higgs
  boson in the $\PH \to \ZZ \to \ell\ell\tau\tau$ decay channel in pp
  collisions at $\sqrt{s}=7\TeV$'',} \textit{ JHEP} \textbf{ 03} (2012) 081,
  \href{http://dx.doi.org/10.1007/JHEP03(2012)081}{\doi{10.1007/JHEP03(2012)081}},
\href{http://www.arXiv.org/abs/1202.3617}{\texttt{arXiv:1202.3617}}.

\bibitem{Anderson:2014}
I.~Anderson\hrefCMSnoop {}{ {et~al.}, ``{Constraining anomalous $HVV$
  interactions at proton and lepton colliders}'',} \textit{ Phys. Rev. D}
  \textbf{ 89} (2014) 035007,
  \href{http://dx.doi.org/10.1103/PhysRevD.89.035007}{\doi{10.1103/PhysRevD.89.035007}},
\href{http://www.arXiv.org/abs/1309.4819}{\texttt{arXiv:1309.4819}}.

\bibitem{Chatrchyan:2012ft}
\hrefCMSnoop {}{{CMS Collaboration}, ``Search for the standard model Higgs
  boson in the $\PH \to \ZZ \to 2\ell 2\nu$ channel in pp collisions at
  $\sqrt{s} = 7\TeV$'',} \textit{ JHEP} \textbf{ 03} (2012) 040,
  \href{http://dx.doi.org/10.1007/JHEP03(2012)040}{\doi{10.1007/JHEP03(2012)040}},
\href{http://www.arXiv.org/abs/1202.3478}{\texttt{arXiv:1202.3478}}.

\bibitem{2l2qpaper}
\hrefCMSnoop {}{{CMS Collaboration}, ``Search for a Higgs boson in the decay
  channel $\PH \to \ZZ^{*} \to {\rm{q\bar{q}}}\ell^+\ell^-$ in pp collisions at
  $\sqrt{s} = 7\TeV$'',} \textit{ JHEP} \textbf{ 04} (2012) 036,
  \href{http://dx.doi.org/10.1007/JHEP04(2012)036}{\doi{10.1007/JHEP04(2012)036}},
\href{http://www.arXiv.org/abs/1202.1416}{\texttt{arXiv:1202.1416}}.

\bibitem{LHC-HCG-Report}
\href {http://cdsweb.cern.ch/record/1379837}{{ATLAS and CMS Collaborations, LHC
  Higgs Combination Group}, ``Procedure for the {LHC} Higgs boson search
  combination in {S}ummer 2011'',} Technical Report ATL-PHYS-PUB 2011-11, CMS
  NOTE 2011/005, 2011.

\bibitem{CMScombFeb2012}
\hrefCMSnoop {}{{CMS Collaboration}, ``{Combined results of searches for the
  standard model Higgs boson in pp collisions at $\sqrt{s}$ = 7 TeV}'',}
  \textit{ Phys. Lett. B} \textbf{ 710} (2012) 26,
  \href{http://dx.doi.org/10.1016/j.physletb.2012.02.064}{\doi{10.1016/j.physletb.2012.02.064}},
\href{http://www.arXiv.org/abs/1202.1488}{\texttt{arXiv:1202.1488}}.

\bibitem{Read:451614}
\href {https://cdsweb.cern.ch/record/451614}{A.~L. Read, ``Modified frequentist
  analysis of search results (the $CL_{s}$ method)'',} CERN Report
  {CERN-OPEN-2000-005}, 2000.

\bibitem{Junk:1999kv}
\hrefCMSnoop {}{T.~Junk, ``{Confidence level computation for combining searches
  with small statistics}'',} \textit{ Nucl. Instrum. Meth. A} \textbf{ 434}
  (1999) 435,
  \href{http://dx.doi.org/10.1016/S0168-9002(99)00498-2}{\doi{10.1016/S0168-9002(99)00498-2}},
\href{http://www.arXiv.org/abs/hep-ex/9902006}{\texttt{arXiv:hep-ex/9902006}}.

\end{thebibliography}\endgroup

\cleardoublepage \appendix\section{The CMS Collaboration \label{app:collab}}\begin{sloppypar}\hyphenpenalty=5000\widowpenalty=500\clubpenalty=5000\textbf{Yerevan Physics Institute,  Yerevan,  Armenia}\\*[0pt]
V.~Khachatryan, A.M.~Sirunyan, A.~Tumasyan
\vskip\cmsinstskip
\textbf{Institut f\"{u}r Hochenergiephysik der OeAW,  Wien,  Austria}\\*[0pt]
W.~Adam, E.~Asilar, T.~Bergauer, J.~Brandstetter, E.~Brondolin, M.~Dragicevic, J.~Er\"{o}, M.~Flechl, M.~Friedl, R.~Fr\"{u}hwirth\cmsAuthorMark{1}, V.M.~Ghete, C.~Hartl, N.~H\"{o}rmann, J.~Hrubec, M.~Jeitler\cmsAuthorMark{1}, V.~Kn\"{u}nz, A.~K\"{o}nig, M.~Krammer\cmsAuthorMark{1}, I.~Kr\"{a}tschmer, D.~Liko, T.~Matsushita, I.~Mikulec, D.~Rabady\cmsAuthorMark{2}, B.~Rahbaran, H.~Rohringer, J.~Schieck\cmsAuthorMark{1}, R.~Sch\"{o}fbeck, J.~Strauss, W.~Treberer-Treberspurg, W.~Waltenberger, C.-E.~Wulz\cmsAuthorMark{1}
% ...
\vskip\cmsinstskip
% ...
1:~~Also at Vienna University of Technology, Vienna, Austria\\
2:~~Also at CERN, European Organization for Nuclear Research, Geneva, Switzerland\\
% ...

\end{sloppypar}
\end{document}